\newif\iffigs
 \figstrue
\documentclass[12pt]{article}
\usepackage{latexsym,amssymb,lscape}
\iffigs
    \usepackage{graphicx}
    \input{epsf}
\else
\fi
\newtheorem{definizione}{Definition}[section]

\newtheorem{principio}{Principle}[section]
\newtheorem{statement}{Statement}[section]

\newtheorem{property}{Property}[section]

\def\IC{\relax\,\hbox{$\inbar\kern-.3em{\rm C}$}}
\def\IG{\relax\,\hbox{$\inbar\kern-.3em{\rm G}$}}
\def\IB{\relax{\rm I\kern-.18em B}}
\def\ID{\relax{\rm I\kern-.18em D}}
\def\IL{\relax{\rm I\kern-.18em L}}
\def\IF{\relax{\rm I\kern-.18em F}}
\def\IH{\relax{\rm I\kern-.18em H}}
\def\II{\relax{\rm I\kern-.17em I}}
\def\IN{\relax{\rm I\kern-.18em N}}
\def\IP{\relax{\rm I\kern-.18em P}}
\def\IQ{\relax\,\hbox{$\inbar\kern-.3em{\rm Q}$}}
\def\bfzero{\relax\,\hbox{$\inbar\kern-.3em{\rm 0}$}}
\def\IK{\relax{\rm I\kern-.18em K}}
\def\IG{\relax\,\hbox{$\inbar\kern-.3em{\rm G}$}}
 \font\cmss=cmss10 \font\cmsss=cmss10 at 7pt
\def\IR{\relax{\rm I\kern-.18em R}}
\def\ZZ{\relax\ifmmode\mathchoice
{\hbox{\cmss Z\kern-.4em Z}}{\hbox{\cmss Z\kern-.4em Z}}
{\lower.9pt\hbox{\cmsss Z\kern-.4em Z}} {\lower1.2pt\hbox{\cmsss
Z\kern-.4em Z}}\else{\cmss Z\kern-.4em Z}\fi}
\def\bfone{\relax{\rm 1\kern-.35em 1}}

\def\inbar{\vrule height1.5ex width.4pt depth0pt}
\def\bfzero{\relax{\rm I\kern-.18em 0}}
\def\bfone{\relax{\rm 1\kern-.35em 1}}

\newcommand{\ft}[2]{{\textstyle\frac{#1}{#2}}}

\def\1bar{1\hskip -.275cm -}
\def\2bar{2\hskip -.275cm -}
\def\3bar{3\hskip -.275cm -}

\newsavebox{\uuunit}
\sbox{\uuunit}
                 {\setlength{\unitlength}{0.825em}
                      \begin{picture}(0.6,0.7)
                                      \thinlines
                                      \put(0,0){\line(1,0){0.5}}
                                      \put(0.15,0){\line(0,1){0.7}}
                                      \put(0.35,0){\line(0,1){0.8}}
                                     \multiput(0.3,0.8)(-0.04,-0.02){10}{\rule{0.5pt}{0.5pt}}
                      \end {picture}}

\makeatletter \@addtoreset{equation}{section} \makeatother

\newcommand{\be}{\begin{equation}}
\newcommand{\ee}{\end{equation}}
\newcommand{\ba}{\begin{eqnarray}}
\newcommand{\ea}{\end{eqnarray}}
\def\bfone{\relax{\rm 1\kern-.35em 1}}

\def\bfone{\relax{\rm 1\kern-.35em 1}}
\font\cmss=cmss10 \font\cmsss=cmss10 at 7pt
\newcommand{\p}[1]{(\ref{#1})}
\newcommand{\nn}{\nonumber}
\newcommand{\so}{\mathfrak{so}}

\newcommand{\uu}{\mathfrak{u}}
\newcommand{\sym}{\mathfrak{sp}}
\newcommand{\slal}{\mathfrak{sl}}
\begin{document}
\begin{titlepage}
\begin{flushright}
DFTT07/23\\
JINR-E2-2007-145
\end{flushright}
\vskip 0.5cm
\begin{center}
{\LARGE {\bf The arrow of time and the Weyl group:
}}\\[0.4cm]
{\LARGE {\bf
 all supergravity billiards are integrable$^\dagger$
}}\\[1cm]
{\large Pietro Fr\'e$^{a}$  and Alexander S. Sorin$^{b}$}
{}~\\
\quad \\
{{\em $^{a}$ Dipartimento di Fisica Teorica, Universit\'a di Torino,}}
\\
{{\em $\&$ INFN - Sezione di Torino}}\\
{\em via P. Giuria 1, I-10125 Torino, Italy}~\quad\\
{\tt fre@to.infn.it}
{}~
{}~
{}~\\
\quad \\
{{\em $^{b}$ Bogoliubov Laboratory of Theoretical Physics,}}\\
{{\em Joint Institute for Nuclear Research,}}\\
{\em 141980 Dubna, Moscow Region, Russia}~\quad\\
{\tt sorin@theor.jinr.ru}
{}~
{}~
\quad \\
\end{center}
~{}
\begin{abstract}
In this paper we show that all supergravity billiards corresponding
to $\sigma$-models on any $\mathrm{U/H}$ non compact-symmetric space and obtained
by compactifying supergravity  to $D=3$ are fully integrable. The key
point in establishing the integration algorithm is provided by an
upper triangular embedding of the solvable Lie algebra associated
with $\mathrm{U/H}$ into $\slal(\mathrm{N},\mathbb{R})$ which always
exists. In this context we establish a remarkable relation between
the arrow of time and the properties of the Weyl group. The
asymptotic states of the developing Universe are in one-to-one
correspondence with the elements of the Weyl group which is a
property of the Tits Satake universality classes and not of their
single representatives. Furthermore the Weyl group admits a natural
ordering in terms of $\ell_T$, the number of reflections with respect
to the simple roots and the direction of time flows is always towards
increasing $\ell_T$, which plays the unexpected role of an entropy.

\end{abstract}
\vfill
\vspace{1.5cm}
\vfill
\vspace{1.5cm}
\vspace{2mm} \vfill \hrule width 3.cm {\footnotesize $^ \dagger $
This work is supported in part by the European Union RTN contract
MRTN-CT-2004-005104 and by the Italian Ministry of University (MIUR) under
contracts PRIN 2005-024045 and PRIN 2005-023102. Furthermore the work of A.S. was partially supported by the RFBR Grant No. 06-01-00627-a,
RFBR-DFG Grant No. 06-02-04012-a, DFG Grant 436 RUS 113/669-3, the
Program for Supporting Leading Scientific Schools (Grant No. NSh-5332.2006.2),
and the Heisenberg-Landau Program.}
\end{titlepage}
\section{Foreword}
Notwithstanding its length and its somewhat pedagogical
organization, the present one is a research article and not a review. All the
presented material is, up to our knowledge, new. Due to the
combination of several different mathematical results and techniques
necessary  to make our point, which is instead physical in spirit and relevant
to basic questions in supergravity and superstring cosmology, we
considered it appropriate to choose the present somewhat unconventional format for our paper. After
the theoretical statement of our result, we have illustrated it with the
detailed study of a few examples. These case-studies were essential for us in order
to understand the main point which we have formalized in mathematical terms in part I and we think that they will be similarly
essential for the physicist reader. The table of
contents helps  the reader to get a comprehensive view of the
article and of its structure.
\tableofcontents
\part{Theory: Stating the principles}
\section{Supergravity billiards: a paradigm for cosmology}
Cosmological implications of superstring theory have been under
attentive consideration in the last few years from various
viewpoints \cite{cosmicstringliteraturegeneral}.
This involves the classification and the study of possible time-evolving string
backgrounds which amounts to the construction, classification and analysis of supergravity solutions
depending only on time or, more generally, on a low number of coordinates including time.
\par
In this context a quite challenging and potentially
highly relevant phenomenon for the overall interpretation of extra--dimensions and string dynamics is provided by the so
named \textit{cosmic billiard} phenomenon
\cite{cosmicbilliardliterature1}, \cite{cosmicbilliardliterature2},
\cite{cosmicbilliardliterature3}, \cite{cosmicbilliardliterature4}.
This is based on the relation between the  cosmological scale factors
and the  duality groups $\mathrm{U}$ of  string theory. The group
$\mathrm{U}$ appears as isometry group of the scalar
manifold $\mathcal{M}_{scalar}$ emerging in compactifications of
$10$--dimensional supergravity to lower dimensions $D<10$ and
 depends both on the geometry of the compact dimensions and on
the number of preserved supersymmetries $\mathcal{N}_Q \le 32$.  For $\mathcal{N}_Q > 8$ the scalar manifold is always a homogeneous
space $\mathrm{U/H}$. The cosmological scale factors $a_i(t)$ associated with the
various  dimensions of supergravity  are interpreted as exponentials of those
scalar fields $h_i(t)$ which lie in the Cartan subalgebra of
$\mathbb{U}$, while the other scalar fields in $\mathrm{U/H}$ correspond to
positive roots $\alpha >0$ of the Lie algebra $\mathbb{U}$.
The cosmological evolution is described by  a
\textit{fictitious ball} that moves in the CSA of $\mathbb{U}$ and occasionally bounces on the hyperplanes orthogonal
to the various roots: the billiard walls. Such bounces  represent
inversions in the time evolution of scale factors.
Such a scenario was introduced by Damour, Henneaux, Julia and Nicolai in
\cite{cosmicbilliardliterature1}, \cite{cosmicbilliardliterature2},
\cite{cosmicbilliardliterature3}, \cite{cosmicbilliardliterature4},
generalizing classical results obtained in the context of pure
General Relativity \cite{Kassner}. In these papers the billiard
phenomenon was mainly considered as an asymptotic regime near
singularieties.
\par
In a series of papers \cite{noiconsasha},
\cite{Weylnashpaper, noipaintgroup, noiKacmodpaper} involving both
the present authors and other collaborators it was started and developed what
can be described as the \textit{smooth cosmic billiard programme}. This
amounts to the study of the \textit{billiard features} within the
framework of exact analytic solutions of supergravity rather than in
asymptotic regimes. Crucial starting point in this programme
was the observation \cite{noiconsasha} that the fundamental
mathematical setup underlying the appearance of the billiard
phenomenon is the so named \textit{Solvable Lie algebra
parametrization} of supergravity scalar manifolds, pioneered in \cite{primisolvi}
and later applied to the solution of a large variety of superstring/supergravity
problems \cite{noie7blackholes}, \cite{otherBHpape}, \cite{gaugedsugrapot}, \cite{mario1},
\cite{mario2} (for a comprehensive review see \cite{myparis}).
\par
Thanks to the solvable parametrization, one can
establish a precise algorithm to implement the following programme:
\begin{description}
  \item [a] Reduce the original supergravity in higher  dimensions $D \ge 4$ (for instance $D=10,11$) to
  a gravity-coupled $\sigma$--model in $D \le 3$ where gravity is
  non--dynamical and can be eliminated. The target manifold is the non compact coset
  $\mathrm{U/H} \cong \exp \left[ Solv\left(\mathrm{U/H} \right)
  \right]$ metrically equivalent to a solvable group manifold.
  \item [b] Utilize various group theoretical techniques  in order to integrate analytically the
  $\sigma$--model equations.
  \item [c] Dimensionally oxide the solutions obtained in this way to
  extract  time dependent solutions of  $D \ge 4$ supergravity.
\end{description}
In view of the above observation we will use the following definition
of supergravity billiards:
\begin{definizione}\label{defibigliardi}
$<<$
A supergravity billiard is a one-dimensional $\sigma$-model whose
target space is a non-compact coset manifold  $\mathrm{U/H}$, metrically
equivalent, in force of a general theorem, to a solvable group
manifold $\exp \left[ Solv\left(\mathrm{U/H} \right)
  \right]$.
$>>$
\end{definizione}
There exists a complete classification \cite{myparis,contoine,noiKacmodpaper} of all non-compact coset
manifolds $\mathrm{U/H}$ relevant to the various instances of supergravities
in all space-time dimensions $D$ and for all numbers
$\mathcal{N}_Q$ of supercharges. A general important feature is that
maximal supersymmetry $\mathcal{N}_Q=32$ corresponds to \textbf{maximally
split} symmetric cosets.
\begin{definizione}\label{maxsplit}
$<<$
A symmetric coset manifold $\mathrm{U/H}$ is maximally split when the
Lie algebra $\mathbb{U}$ of $\mathrm{U}$ is the maximally non compact
real section of its own complexification and $\mathbb{H}\, \subset \, \mathbb{U}$ is the unique maximal compact subalgebra.
In this case the Cartan subalgebra $\mathcal{C}$ is completely
non-compact, namely the non-compact rank $r_{n.c.}=r$ equals the rank
and the solvable Lie algebra $Solv\left(\mathrm{U/H} \right)$ is made
by all the  Cartan generators $\mathcal{H}_i$ plus the step operators
$E^{\alpha}$ for all the positive roots $\alpha >0$.
$>>$
\end{definizione}
In \cite{sahaedio} the present authors shew that for maximally split
cosets the one-dimensional $\sigma$-model is fully integrable and the
general integral can be constructed using a well established
algorithm endowed with a series of distinctive and quite inspiring
features.
\par
In the present paper we demonstrate that the algorithm of \cite{sahaedio}
can be actually extended to all the other cases, also those not maximally
split, so that all supergravity billiards are in fact completely
integrable as claimed in the title.
\par
Besides demonstrating the integrability we will illustrate the main
features of the general integral which reveal a very rich and highly
interesting geometrical structure of the parameter space. In this
context it will emerge a challenging new concept. The time flows
appearing as exact analytical solutions of supergravity billiards
have a preferred orientation which is intrinsically determined in
group theoretical terms. There emerges a similarity between the
second law of thermodynamics and the properties of cosmological evolutions just as
there is such a similarity in the case of black-hole dynamics.
We establish the following principle
\begin{principio}\label{secondoprincipio}
$<<$
The asymptotic states of the cosmic billiard at past and future
infinity $t=\pm \infty$ are in one-to-one correspondence with the elements $w_i$ of
the duality algebra Weyl group $\mathrm{Weyl}(\mathbb{U})$. The Weyl
group, which for suitable choice of $N$ is  a subgroup of the symmetric group
$\mathcal{S}_N$  admits a natural ordering in terms of
the minimal number $\ell_T$ of reflections  with respect to simple
roots $\alpha_i$ necessary to reproduce any considered element $w$. The number
$\ell_T(w)$, named the height of $w \in \mathrm{Weyl}(\mathrm{U})$, is
the same as the number of transpositions of the corresponding
permutation when $\mathrm{Weyl}(\mathrm{U})$ is embedded in the
symmetric group. Time flows goes always in the direction of
increasing $\ell_T$ which, therefore, plays the role of entropy.
$>>$
\end{principio}
\section{The paint group and the Tits Satake projection}
In \cite{noipaintgroup} first and then more systematically in
\cite{contoine} it was observed that the  Tits-Satake theory of
non-compact cosets, which is a classical chapter of modern
differential geometry, provides a natural frame to discuss the
structure of the $\mathrm{U/H}$ cosets appearing in supergravity with particular reference to
their role in billiard dynamics. In \cite{noipaintgroup} a new
concept was introduced, that of \textbf{paint group}, which plays a
fundamental role in classifying the relevant $\mathrm{U/H}$ manifolds
and grouping them into universality classes with respect to the Tits
Satake projection. The systematics of these universality classes was
developed in \cite{contoine}.
\par
In the present paper we will clarify and illustrate by means of explicit
examples the  meaning of these universality classes showing
that the essential features of billiard dynamics are just a property
of the class, independently from the choice of the representative,
namely independently from the choice of the paint group.
 In particular the Weyl group and the asymptotic states
are common to the whole class. On the other hand the notion of the paint
group enters in the precise definition of the parameter space for the
general integral. Let us therefore recall the essential notions
relevant to our subsequent discussion.
\subsection{The solvable algebra}
Following the discussion of \cite{noipaintgroup} let us recall that in the
case the scalar manifold of supergravity is a \textit{non maximally
non-compact manifold} $\mathrm{U/H}$ the Lie algebra $\mathbb{U}$ of the
numerator group is some appropriate real form
\begin{equation}
  \mathbb{U} \, = \, \mathbb{U}_R
\label{realformaGR}
\end{equation}
of a complex Lie algebra $\mathbb{U}(\mathbb{C})$ of rank $r
=\mbox{rank}(\mathbb{U})$. The Lie algebra $\mathbb{H}$ of the
denominator $\mathrm{H}$ is the maximal compact subalgebra $\mathbb{H}
\subset \mathbb{U}_R$, which has typically rank $r_{\rm c} < r$. Denoting,
as usual, by $\mathbb{K}$ the orthogonal complement of $\mathbb{H}$ in
$\mathbb{U}_R$
\begin{equation}
  \mathbb{U}_R = \mathbb{H} \, \oplus \,\mathbb{K}
\label{Grdecompo}
\end{equation}
and defining as non-compact rank, or rank of the coset $\mathrm{U/H}$, the
dimension of the noncompact Cartan subalgebra
\begin{equation}
  r_{\rm nc}\, = \, \mbox{rank} \left( \mathrm{U/H}\right)  \, \equiv \, \mbox{dim} \,
  \mathcal{H}^{\rm nc} \quad ; \quad \mathcal{H}^{\rm nc} \, \equiv \,
  \mbox{CSA}_{\mathbb{U}(\mathbb{C})} \, \bigcap \, \mathbb{K} ~,
\label{rncdefi}
\end{equation}
we obtain that $r_{\rm nc} < r$.
\par
The manifold $\mathrm{U_R/H}$ is always metrically equivalent to a
solvable group manifold $\mathcal{M}_{Solv} \equiv \exp
[Solv(\mathrm{U_R/H})]$ although the form of the solvable Lie algebra
$Solv(\mathrm{U_R/H})$, whose structure constants define the Nomizu
connection,  is  more complicated when $r_{\rm nc} \, < \, r$ than
in the  \textit{maximally split case} $r_{\rm nc} \, = \, r$.
For the details on the construction of the solvable Lie algebra
we refer to the literature \cite{primisolvi,myparis}.
The important thing in our present context is that it exists.
Furthermore, using a general theorem proven in such textbooks like
\cite{Helgason} we know that every linear representation of a
solvable Lie algebra can be written in a basis where all of its
elements are given by upper triangular matrices. Hence for any of the
$\mathrm{U/H}$ cosets of supergravity we can choose a coset
representative $\mathbb{L}(\phi)$ given by the matrix
exponential of an upper triangular matrix. This is the so named
solvable parametrization of the coset manifold which plays a
fundamental role in our subsequent discussion of the general
integral.
\subsection{The paint group and its Lie algebra}
Naming $\mathcal{M}=\mathrm{U/H}$ the considered coset manifold
and $Solv_\mathcal{M} \, \subset \, \mathbb{U}$ the corresponding
solvable algebra, there exists a \textit{compact algebra} $\mathbb{G}_{\rm paint} $ which
acts as  an algebra of outer automorphisms ({\it i.e.} outer derivatives)
of the  solvable algebra $Solv_{\mathcal{M}}$
\begin{equation}
 \mathrm{Aut} \, \left[ Solv_\mathcal{M}\right] \,  \quad = \quad \left\{  X \, \in \, \mathbb{U} \,  \mid \, \forall \, \Psi \,
 \in \, Solv_\mathcal{M}\ :\ \left[ X\, ,\, \Psi \right] \, \in \,
 Solv_\mathcal{M} \, \right\}.
\label{automorfismi}
\end{equation}
By its own definition the algebra $\mathrm{Aut} \, \left[
Solv_\mathcal{M}\right] $ contains $Solv_\mathcal{M}$ as an ideal.
Hence  we can define the algebra of external automorphisms as the
quotient
\begin{equation}
  \mathrm{Aut}_{\mathrm{Ext}} \, \left[ Solv_\mathcal{M}\right] \,
  \equiv \, \frac{\mathrm{Aut} \, \left[
  Solv_\mathcal{M}\right]}{Solv_\mathcal{M}},
\label{outerauto}
\end{equation}
and we identify $\mathbb{G}_{\mathrm{paint}}$ as the maximal compact
subalgebra of $\mathrm{Aut}_{\mathrm{Ext}} \, \left[
Solv_\mathcal{M}\right]$. Actually we immediately see that
\begin{equation}
  \mathbb{G}_{\mathrm{paint}} \, = \, \mathrm{Aut}_{\mathrm{Ext}} \, \left[
Solv_\mathcal{M}\right]. \label{pittureFuori}
\end{equation}
Indeed, as a consequence of its own definition the algebra
$\mathrm{Aut}_{\mathrm{Ext}} \, \left[ Solv_\mathcal{M}\right]$ is
composed of isometries which belong to the stabilizer subalgebra
$\mathbb{H} $ of any
point of the manifold, since $Solv_\mathcal{M}$ acts transitively. In
virtue of the Riemannian  structure of $\mathcal{M}$ we have $\mathbb{H}
\subset \so(n) $ where $n = \mbox{dim} \left(Solv_{\mathcal{M}} \right)$
and hence also $\mathrm{Aut}_{\mathrm{Ext}} \, \left[
Solv_\mathcal{M}\right] \, \subset \, \so(n)$ is a compact Lie algebra.
\par
The paint group is now defined by exponentiation of the paint
algebra
\begin{equation}
  \mathrm{G}_{\mathrm{paint}} \, \equiv \, \exp \,
  \left[ \, \mathbb{G}_{\mathrm{paint}} \, \right]~.
\label{pittapitta}
\end{equation}
The notion of maximally
split algebras  can be formulated in terms of the paint algebra by stating that
\begin{equation}
\mathbb{U} = \mbox{maximally split} \, \Leftrightarrow \, \mathrm{Aut}_{\mathrm{Ext}} \, \left[
Solv_{\mathrm{U/H}}\right] = \emptyset~. \label{maxsplittus}
\end{equation}
Namely $\mathbb{U}$ is maximally split if and only if the paint group is just the trivial
identity group.
\subsection{The subpaint group and the Tits Satake subalgebra}
Making a long story short, once the paint algebra has been defined,
the solvable Lie algebra falls into a linear representation of
$\mathbb{G}_{\mathrm{paint}}$ and one can define  its little group,
generated by the stability subalgebra of a generic element $X \, \in \,
Solv_{\mathcal{M}}$. In other words, viewed as
a representation of $\mathbb{G}_{\mathrm{paint}}$, under the subalgebra
\begin{equation}
  \mathbb{G}_{\mathrm{subpaint}} \, \subset \, \mathbb{G}_{\mathrm{paint}}
\label{subpainta}
\end{equation}
the solvable Lie algebra decomposes into a singlet subalgebra
$Solv_{TS}$ plus a bunch of non trivial irreducible representations
of $\mathbb{G}_{\mathrm{subpaint}}$.
 We name such a Lie subalgebra the
\textbf{subpaint algebra}. Then
the Tits Satake subalgebra of the original algebra $\mathbb{U}$ is defined as
the set of all elements which are invariant with respect to
$\mathbb{G}_{\mathrm{subpaint}}$:
\begin{equation}
  X \, \in \, \mathbb{U}_{TS} \, \subset \, \mathbb{U} \, \Leftrightarrow \, \forall \Psi \, \in \, \mathbb{G}_{\mathrm{subpaint}} \quad : \quad  \left
  [ X \, ,\, \Psi \right]  \, = \, 0~.
\label{GTS}
\end{equation}
By construction the Tits Satake subalgebra $\mathbb{U}_{\mathrm{TS}}$ is maximally
split and the Tits Satake projection is defined as the following
mapping of coset manifolds:
\begin{equation}
  \Pi_{\mathrm{TS}} \quad : \quad \frac{\mathrm{U}}{\mathrm{H}} \,\rightarrow
  \, \frac{\mathrm{U}_{\mathrm{TS}}}{\mathrm{H}_{\mathrm{TS}}}~.
\label{Tsproj1}
\end{equation}
In terms of root systems the Tits Satake projection has a natural and
simple intepretation. The root system $\Delta_\mathrm{U}$ of the
original algebra is composed by a set of vectors in $r$-dimension
where $r$ is the rank of $\mathbb{U}$. This system of vectors can be
projected onto the $r_{nc}$-dimensional subspace dual to the
non-compact Cartan subalgebra. Somewhat surprisingly, with just one
exception, the projected set of vectors is a new root system in
rank $r_{\rm nc}$, which we name $\Delta_{\mathrm{TS}}$.  Indeed the
corresponding Lie algebra is precisely the Tits Satake subalgebra
$\mathbb{U}_{\mathrm{TS}}\, \subset \, \mathbb{U}$ of the original algebra.
\section{Triangular embedding in $\mathrm{SL(N,\mathbb{R})/SO(N)}$ and integrability}
As a consequence of all the algebraic structures we have described we
can conclude with the following statement.
\par
\begin{statement}\label{statamento}
$<<$
Let $N$ be the real dimension of the fundamental representation of
$\mathbb{U}$. Then there is a canonical embedding
\begin{eqnarray}
 \mathbb{ U} & \hookrightarrow & \slal(N,\mathbb{R})~,\nonumber\\
\mathbb{ U} \, \supset \, \mathbb{H} & \hookrightarrow & \so(N) \,
\subset \, \slal(N,\mathbb{R})~.
\label{triaembed}
\end{eqnarray}
This embedding is determined by the choice of the basis where $Solv\left(\mathrm{U/H}
\right)$ is made by upper triangular matrices. In the same basis the
elements of $\mathbb{K}$ are symmetric matrices while those of $\mathbb{H}$ are
antisymmetric ones.
$>>$
\end{statement}
The embedding (\ref{triaembed}) defines also a canonical embedding of
the relevant Weyl group $Weyl(\mathbb{U})$ of $\mathbb{U}$ into that of
$\slal(N,\mathbb{R})$ namely into the symmetric group
$\mathcal{S}_N$.
\par
The existence of (\ref{triaembed}) is the key-point in order to
extend the integration algorithm of supergravity billiards presented
in \cite{sahaedio} from the case of maximally-split cosets to the
generic case. Indeed that algorithm is defined for $\mathrm{SL(N,\mathbb{R})/SO(N)}$
and it has the property that if initial data are defined in a
submanifold $\mathrm{U/H}$ where $\mathrm{U} \, \subset \, \mathrm{SL(N,\mathbb{R})}$
and $\mathrm{H} \, \subset \, \mathrm{SO(N)}$, then the entire time
flow occurs in the same submanifold. Hence the embedding
(\ref{triaembed}) suffices to define explicit integration formulae for
all supergravity billiards.
\par
Let us review the steps of the procedure.
\begin{enumerate}
  \item First one defines a coset representative for $\mathrm{U/H}$ in the
  solvable parametrization as follows:
\begin{equation}
  \mathbb{L}\left( \phi\right)  \, = \, \prod_{I=m}^{I=1} \,
  \exp \left [\varphi_I \,
  E^{\alpha_I}\right] \, \exp \left[h_i \mathcal{H}^i\right]
\label{cosettorepresentat}
\end{equation}
where the roots pertaining to the solvable Lie algebra are ordered in
ascending order of height ($\alpha_I \le \alpha_J$ if $I < J$), $\mathcal{H}^i$ denote the non compact Cartan generators
and the product of matrix exponentials appearing in (\ref{cosettorepresentat}) goes from the highest
on the left, to lowest root on the right. In this
way the parameters $\left \{\phi\right\} \, \equiv \, \left \{\varphi_I \, , \, h_i \right\} $ have a precise and uniquely
defined correspondence with the fields of supergravity by means of dimensional oxidation \cite{noiconsasha,Weylnashpaper}.
  \item Restricting all the fields $\phi$ of supergravity to pure
  time dependence $\phi \, = \, \phi(t)$, the coset representative becomes also a function of
  time $\mathbb{L}\left(\phi(t) \right) = \mathbb{L}(t)$ and we define the Lax operator
  $L(t)$ and the connection $W(t)$ as follows:
\begin{eqnarray}
  L(t) & = & \sum_{i} \, \mbox{Tr} \left(\mathbb{L}^{-1}
  \frac{d}{dt}\mathbb{L} \, \mathrm{K}_i\right) \mathrm{K}_i~,
  \nonumber\\
    W(t) & = & \sum_{\ell} \, \mbox{Tr} \left(\mathbb{L}^{-1}
  \frac{d}{dt}\mathbb{L} \, \mathrm{H}_\ell\right)\,
  \mathrm{H}_\ell
\label{postayanna}
\end{eqnarray}
where $\mathrm{K}_i$ and $\mathrm{H}_\ell$ denote an orthonormal
basis of generators for $\mathbb{K}$ and $\mathbb{H}$, respectively.
\item With these definitions the field equations of supergravity,
which are just the geodesic equations for the manifold $\mathrm{U/H}$ in
the solvable parametrization, reduce to the single matrix valued Lax
equation \cite{sahaedio}
\begin{eqnarray}
\label{Lax}
\frac{d}{dt} L=\left [W \, , \, L\right ]~.
\label{newLax}
\end{eqnarray}
\item If we are able to write the general integral of the Lax
equation, depending on $p=\mbox{dim}(\mathrm{U/H})$ integration
constants, then comparison of the definition of the Lax operator
(\ref{postayanna},\ref{cosettorepresentat}) with its explicit form
in the integration reduces the differential equations of supergravity
to quadratures
\begin{equation}
  \frac{d}{dt} \phi(t) \, = \, F(t) \quad = \quad \mbox{known function of time.}
\label{quadrature}
\end{equation}
\end{enumerate}
\subsection{The integration algorithm for the Lax Equation}
Let us assume that we have explicitly constructed the embedding
(\ref{triaembed}). In this case, in the decomposition
\begin{equation}
  \mathbb{U} = \mathbb{K} \oplus \mathbb{H}
\label{UKH}
\end{equation}
of the relevant Lie algebra $\mathbb{U}$, the matrices representing
the elements of $\mathbb{K}$ are all symmetric while those
representing the elements of $\mathbb{H}$ are all antisymmetric as we have already pointed out.
Furthermore the matrices representing  the solvable Lie algebra
$Solv(\mathrm{U}/\mathrm{H})$ are all upper triangular. These are the
necessary and sufficient conditions to apply to the relevant Lax equation (\ref{Lax}) the integration
algorithm originally described in \cite{kodama} and reviewed in
\cite{sahaedio}.
The key point is that the connection $W(t)$
appearing in eq.(\ref{newLax}) is related to the Lax operator by
means of an algebraic projection operator as follows:
\begin{eqnarray}
 W=\Pi (L):=L_{>0}-L_{<0},
 \label{Lprojection}
\end{eqnarray}
$L_{>0~(<0)}$ denoting the strictly upper (lower) triangular
part of the $N \times N$ matrix $L$.
The relation (\ref{Lprojection}) is nothing else but the statement that
the coset representative $\mathbb{L}(\phi)$ from which the Lax operator is extracted is
taken in the solvable parametrization.
\par
This established, we can proceed to apply the integration algorithm.  Actually this is nothing else
but an instance of the inverse scattering method. Indeed
equation \p{Lax} represents the compatibility condition for the
following linear system exhibiting the iso-spectral property of $L$:
\begin{eqnarray}
\label{LaxIs}
L\Psi=\Psi \Lambda,\nonumber\\
\frac{d}{dt} \Psi=P\Psi
\end{eqnarray}
where $\Psi(t)$ is the eigenmatrix, namely the matrix whose $i$-th row is the
eigenvector $\varphi(t,\lambda_i)$ corresponding to the eigenvalue $\lambda_i$ of the Lax operator
$L(t)$ at time $t$ and $\Lambda$ is the diagonal
matrix of eigenvalues, which are constant throughout the whole time flow
\begin{eqnarray}
\Psi&=&\left [\varphi(\lambda_1),\dots,\varphi(\lambda_n)]\equiv[\varphi_i(\lambda_j)\right ]_{1\leq
i,j\leq n},\nonumber\\
\Psi^{-1}&=&\left [\psi(\lambda_1),\dots,\psi(\lambda_n)]^T\equiv[\psi_j(\lambda_i)\right ]_{1\leq
i,j\leq n},\nonumber\\
 \Lambda&=& \mathrm{diag}\left (\lambda_1,\dots, \lambda_n \right ).
 \end{eqnarray}
 The solution of \p{LaxIs} for the Lax operator  is given by  the following explicit  form
 of the matrix elements:
\begin{eqnarray}
\label{sol}\left[ L(t) \right]_{ij}=\sum_{k=1}^n \lambda_k
\varphi_i(\lambda_k,t) \psi_j(\lambda_k,t)~.
\end{eqnarray}
The eigenvectors of the Lax operator at each instant of time, which define the eigenmatrix $\Psi(t)$, and the
columns of its inverse $\Psi^{-1}(t)$, are expressed in closed form in terms of
the initial data  at some conventional instant of time, say at
$t=0$.
\par
Explicitly we have
\begin{eqnarray}
\varphi_i(\lambda_j,t)&=&\frac{e^{-\lambda_j
t}}{\sqrt{D_i(t)D_{i-1}(t)}} \, \mathrm{Det} \, \left ( \begin{array}{cccc}
c_{11}&\dots &c_{1,i-1}& \varphi_1^0(\lambda_j)\\
\vdots&\ddots&\vdots&\vdots\\
c_{i1}&\dots &c_{i,i-1}& \varphi_i^0(\lambda_j)\\
\end{array}\right
),\nn\\
\psi_j(\lambda_i,t)&=&\frac{e^{-\lambda_i t}}{\sqrt{D_j(t)D_{j-1}(t)}}
\, \mathrm{Det} \, \left ( \begin{array}{ccc}
c_{11}&\dots &c_{1,j}\\
\vdots&\ddots&\vdots\\
c_{j-1,1}&\dots &c_{j-1,j}\\
 \psi_1^0(\lambda_i)&\dots &\psi_j^0(\lambda_i)
\end{array}\right)
\label{Psi-1}
\end{eqnarray}
where the time dependent matrix $c_{ij}(t)$ is defined below
\begin{eqnarray}
c_{ij}(t)=\sum_{k=1}^N e^{-2\lambda_k t}
\varphi_i^0(\lambda_k)\psi_j^0(\lambda_k)
\end{eqnarray}
and
\begin{eqnarray}
  \varphi_i^0(\lambda_k) & := & \varphi_i(\lambda_k,0)~, \nonumber\\
  \psi_i^0(\lambda_k) & := &\psi_i(\lambda_k,0)
\label{eigenvecat0}
\end{eqnarray}
are the eigenvectors and their adjoints calculated at $t=0$.
These constant vectors as well as eigenvalues $\lambda_k$ constitute the initial data of the
problem and provide the integration constants. Finally $D_k(t)$ denotes the determinant of the $k\times k$ matrix
with entries $c_{ij}(t)$
 \begin{eqnarray}
 D_k(t)=\mathrm{Det} \Biggr [ \Bigr ( c_{ij}(t) \Bigr )_{1\leq i,j \leq k} \Biggr
 ].
 \label{Ddefi}
\end{eqnarray}
Note that $c_{ij}(0)=\delta_{ij}$ and $D_k(0)=1$.
\section{Properties of the general
integral and the parameter space}
The algorithm we have described in the previous section realizes a map
\begin{equation}
  \mathcal{I}_K \quad : \quad L_0 \, \mapsto \, L\left(t,L_0\right)
\label{Imap}
\end{equation}
which, starting from the initial data, i.e. the Lax
operator $ L(0) \, = \, L_0  \, \in \, \mathbb{K}$  at some conventional time $t=0$,
produces a flow, namely a map of the infinite time line into the
subspace $\mathbb{K} \, \subset \, \mathbb{U}$
\begin{equation}
  L\left(t,L_0\right) \quad : \quad\underbrace{ \mathbb{R}}_{-\infty \, \le \, t \, \le +\infty} \,
  \mapsto\, \mathbb{K}~.
\label{flowmap}
\end{equation}
It is of the outmost interest to enumerate the  properties
of the maps (\ref{Imap},\ref{flowmap}). A first set of four fundamental properties
are listed below:
\begin{enumerate}
  \item The flow $L\left(t,L_0\right)$ is iso-spectral. This means the
  following. The Lax operator is a symmetric matrix and therefore can
  be diagonalized at every instant of time. Calling $\lambda _1 \,
  \dots \lambda_N$ the set of its $N$ eigenvalues, we have that this
  set is time--independent, namely the numerical values of the
  eigenvalues remain the same throughout the entire motion.
  \item If the Lax operator $L(t)$ is diagonal at any finite time $t
  \, \ne \, \pm \infty$, then it is actually constant $L(t) = L_0$
  \item The asymptotic limits of the Lax operator for $t \, \mapsto
  \, \pm \infty$ are diagonal matrices $L_{\pm \, \infty}$.
  \item If $L_0 \, \in \, \mathbb{K}_\mathbb{U} $ belongs to the
  symmetric part of a proper Lie subalgebra $\mathbb{U} \, \subset \,
  \slal\left(N, \mathbb{R}\right)$, then
  the entire motion remains in that subalgebra, namely $\forall \, t
  \, , \, L(t) \, \in \, \mathbb{K}_\mathbb{U}$.
\end{enumerate}
Relying on this first set of properties we can refine our formulation
of the initial conditions and of the asymptotic limits in terms of the generalized Weyl group and of its Tits Satake projection. This leads to
state  further properties of the map (\ref{Imap}) which are even more striking.
\par
\subsection{Discussion of the generalized Weyl group}
Diagonal matrices are just elements of the non-compact Cartan subalgebra
$\mathcal{C} \,  \subset \, \mathbb{K} \, \subset \, \mathbb{U}$.
The Lax operator at $t=0$ can be diagonalized by means of an
orthogonal matrix $\mathcal{O} \, \in \, \mathrm{SO(N)}$ which
actually lies in the subgroup $\mathrm{H} \, \subset \, \mathrm{SO(N)}$.
Hence by writing
\begin{equation}
  L_0 \, = \, \mathcal{O}^T \, \mathcal{C}_0 \, \mathcal{O}
\label{iniziadato}
\end{equation}
initial data can be given as a pair
\begin{equation}
  \mathcal{C}_0 \, \in \, \mbox{CSA} \, \bigcap \, \mathbb{K} \quad ; \quad
  \mathcal{O} \, \in \, \mathrm{H}~.
\label{inidata}
\end{equation}
Let us now introduce the notion of generalized Weyl group
$\mathcal{W}(\mathbb{U})$. To understand its definition let us review
the definition of the standard Weyl group. This latter is an intrinsic
attribute of a complex Lie algebra.
For a  complex Lie algebra $\mathbb{U}_\mathbb{C}$, the Weyl group $\mbox{Weyl}(\mathbb{U}_\mathbb{C})$ is the
finite group generated by the reflections
$\sigma_\alpha$ with respect to all the roots $\alpha$. Actually as generators of $\mbox{Weyl}(\mathbb{U})$ it suffices
to consider the reflections with respect to the simple roots
$\sigma_{\alpha_i}$. It turns out that if we consider the maximally split
real section $\mathbb{U}_{split}$ of the complex Lie algebra
$\mathbb{U}_\mathbb{C}$ then the Weyl group $\mbox{Weyl}(\mathbb{U}_\mathbb{C})$
is realized as a subgroup of the maximal
compact subgroup $\mathrm{H}_{split}\, \subset \,\mathrm{U}_{split}$. This isomorphism is realized as follows.
Consider the integer valued elements of $\mathrm{H}$ defined below
\begin{equation}
\mathrm{H} \, \ni \,   \gamma_\alpha \, \equiv \, \exp\left[\frac{\pi}{2} \, \left( E^{\alpha} - E^{-\alpha}\right)
  \right] \, \quad , \quad \alpha \, > \, 0
\label{alfarut}
\end{equation}
and take them as generators. These generators produce a finite subgroup $\mathcal{W}(\mathbb{U})$
which we name \textit{generalized Weyl group}. It contains a normal subgroup $\mathrm{N}(\mathbb{U})\,  \subset
\, \mathcal{W}(\mathbb{U})$ whose adjoint action on any Cartan Lie
algebra element is just the identity. The factor group
$\mathcal{W}(\mathbb{U})/\mathrm{N}(\mathbb{U}) \, \sim \, \mbox{Weyl}(\mathbb{U}_\mathbb{C})$ is isomorphic to
the abstract Weyl group of the complex Lie algebra.
\par
Imitating such a construction also in the non maximally split cases we can
introduce the following
\begin{definizione}
\label{genWeyl}
$<<$
Let $\mathbb{U}$ be a not necessarily  maximally split real section of the complex
Lie algebra $\mathbb{U}_\mathbb{C}$ and $\mathbb{H} \, \subset \, \mathbb{U}$ its maximal
compact subalgebra. Let $\left\{\alpha_{[K]}\right\}$ be the set of positive roots
which are not in the kernel of the Tits Satake projection and which
therefore participate in the construction of the solvable Lie algebra
of $\mathrm{U/R}$. The generalized Weyl group $\mathcal{W}(\mathbb{U})$ is the finite subgroup
of $\mathrm{H}$ generated by the following generators:
\begin{equation}
\mathrm{H} \, \ni \,   \gamma_{\alpha_{[K]}} \, \equiv \, \exp\left[\frac{\pi}{2} \,
\left( E^{\alpha_{[K]}} - E^{-\alpha_{[K]}}\right)
  \right] \, \quad , \quad \alpha_{[K]} \, > \, 0
\label{alfarutK}
\end{equation}
whose number is $\mbox{dim}(\mathrm{U/H}) \, - \,
\mbox{rank}(\mathrm{U/H})$.
$>>$
\end{definizione}
As we already noted, the generalized Weyl
group is typically bigger and has more elements than the ordinary Weyl group.
\par
By construction the adjoint action of the generalized Weyl group maps the non-compact Cartan
subalgebra into itself
\begin{equation}
  \forall \, \mathcal{O}_w \, \in \, \mbox{W}(\mathbb{U}) \,\,
  \mbox{and} \,\,
  \forall \, C \, \in \, \mbox{CSA} \, \bigcap \, \mathbb{K} \quad : \quad \mathcal{O}_w^T \, C \, \mathcal{O}_w \,
  \in \, \mbox{CSA}\, \bigcap \, \mathbb{K}~.
\label{weylandcsa}
\end{equation}
This  can be verified by means of the same calculation which shows
that the ordinary Weyl group, as defined in eq.(\ref{alfarut}), maps
the Cartan subalgebra into itself for the maximally split case.
\par
This observation shows that giving the initial data as we did in
eq.(\ref{inidata}) actually corresponds to an over-counting. Indeed the
generalized Weyl group should be modded out since it amounts to a
redefinition of the Cartan subalgebra data $\mathcal{C}_0$. So we are
led to guess that for each choice of the eigenvalues of the Lax
operator, namely at fixed $\mathcal{C}_0$, the parameter space of the Lax
equation is $\mathcal{P} \, = \, \mathrm{H}/\mathcal{W}(\mathbb{U})$. This however is not
yet the complete truth. Indeed there is also a continuous group,
whose adjoint action on the non-compact Cartan subalgebra is the
identity map. This is the paint group $\mathrm{G}_{\mathrm{paint}}$.
Hence the true parameter space of the Lax equation is the orbifold
with respect to the generalized Weyl group, not of a group, rather of
a compact coset manifold. Indeed we can write
\begin{equation}
  \mathcal{P} \, = \, \frac{\mathrm{H}}{\mathrm{G}_{\mathrm{paint}}}
  \, / \, \mathcal{W}(\mathbb{U})~.
\label{bunduparametru}
\end{equation}
Furthermore we can consider a normal
subgroup $\mathrm{N}_\mathcal{W}(\mathbb{U}) \, \subset \, \mathcal{W}(\mathbb{U})$ of the generalized Weyl
group defined by the following condition:
\begin{equation}
  \gamma \, \in \, \mathrm{N}_\mathcal{W}(\mathbb{U}) \, \subset \,
  \mathcal{W}(\mathbb{U}) \quad  \mbox{iff} \quad \forall C_0 \, \in
  \, \mbox{CSA}\, \bigcap \, \mathbb{K} \, \quad \quad \gamma^T \, \mathcal{C}_0 \, \gamma
  = \mathcal{C}_0
\label{Normalsugruppo}
\end{equation}
and we can state the proposition which is true for all non-compact
cosets $\mathrm{U/H}$:
\begin{statement}\label{fattoregruppo}
$<<$
The factor group of the generalized Weyl group with respect to its
normal subgroup stabilizing all elements of the non-compact Cartan
subalgebra is just isomorphic to the ordinary Weyl group of the Tits Satake
subalgebra:
\begin{equation}
  \frac{\mathcal{W}(\mathbb{U})}{\mathrm{N}_\mathcal{W}(\mathbb{U}) }
  \, \simeq \, \mbox{$\mathrm{Weyl}$}\left( \mathbb{U}_{TS}\right)~.
\label{riguardoso}
\end{equation}
$>>$
\end{statement}
This shows that the only relevant Weyl group is just the Weyl group
of the Tits-Satake subalgebra $\mathbb{U}_{\mathrm{TS}}$.
\par
In view of the iso-spectral property and of the asymptotic property
of the Lax operator which becomes diagonal at $t=\pm \infty$ we
conclude that,  once $\mathcal{C}_0$ is chosen, the available end-points of the
flows at the remote past and at the remote future are in one-to-one
correspondence with the elements of the Weyl group
$\mathrm{Weyl}(\mathbb{U}_{TS})$.
Indeed diagonal matrix means an element of the non-compact Cartan
subalgebra and, since the eigenvalues are numerically fixed by the
original choice of $\mathcal{C}_0$, the only thing which can happen is
a permutation. The available permutations are on the other hand
dictated by the embedding of the Weyl group into the symmetric group:
\begin{equation}
  \mathrm{Weyl}\left(\mathbb{U}_{\mathrm{TS}}\right) \, \hookrightarrow \,
  \mathcal{S}_\mathrm{N} \, \simeq \, \mathrm{Weyl}\left(A_{N-1}\right)
\label{symmetricembed}
\end{equation}
which is induced by the embedding (\ref{triaembed}) of the Lie
algebra $\mathbb{U}$ into $\slal(\mathrm{N},\mathbb{R})$. The latter,
as we already stressed, follows by the choice of the upper triangular
basis for the solvable Lie algebra in the fundamental representation
of $\mathbb{U}$.
\subsection{The arrow of time, trapped and critical surfaces}
\label{genprper}
In view of the above discussion we conclude that the integration
algorithm (\ref{Imap}) realizes a map of the following type:
\begin{equation}
  \mathcal{T}_K \quad : \quad
  \frac{\mathrm{H}/\mathrm{G}_{\mathrm{paint}}}{\mathcal{W}(\mathbb{U})}
  \, \Longrightarrow \,
  \mathrm{Weyl}\left(\mathbb{U}_{\mathrm{TS}}\right)_- \, \otimes \,
  \mathrm{Weyl}\left(\mathbb{U}_{\mathrm{TS}}\right)_+
\label{wwmappa}
\end{equation}
where $\mp$ refer to the choice of a Weyl group element at $\mp
\infty$ realized by the asymptotic limits of the Lax operator.
\par
It is of the outmost interest to explore the general properties of
the map $\mathcal{T}_K$.
\par
Let $\overrightarrow{\mathbf{w}}^I$, ($I=1,\dots , \mathrm{N})$ be the weights of $\mathbb{U}$
in its fundamental $\mathrm{N}-dimensional$ representation $\mathcal{R}_N$
and let
\begin{equation}
  \overrightarrow{h} \, = \,\left\{\underbrace{ h_1,..,h_{r_{nc}}}_{\overrightarrow{h}_{TS}},\underbrace{0,0,\dots,0}_{r-r_{nc}}\right\}
\label{hvector}
\end{equation}
be the $r$-vector of parameters identifying the $\mathcal{C}_0$
element in the non compact Cartan subalgebra
\begin{equation}
\mbox{CSA}\,\bigcap \, \mathbb{K} \, \ni \,   \mathcal{C}_0 \, = \,
\sum_{i=1}^{r_{nc}} \, h_i \, \mathcal{H}^{i}~.
\label{Cnotto}
\end{equation}
The $r -r_{nc}$ zeros in eq.(\ref{hvector}) correspond to the
statement that all components of $\mathcal{C}_0$ in the compact directions of
the Cartan subalgebra vanish. The sub-vector
$\overrightarrow{h}_{TS}$ is the only non vanishing one and it is the
same as we would have in the Tits Satake projected case.
With these notations the $\mathrm{N}$ eigenvalues of the Lax operator $\lambda ^1,\lambda
^2,\dots , \lambda ^\mathrm{N}$ are represented as follows:
\begin{equation}
  \lambda ^I \, = \, \overrightarrow{\mathbf{w}}^I\left(\mathcal{C}_0\right) \,
  = \, \overrightarrow{\mathbf{w}}^I \, \cdot \, \overrightarrow{h} \, .
\label{lambdafot}
\end{equation}
Consider now the branching of the fundamental representation of $\mathbb{U}$
with respect to the Tits Satake subalgebra times the $\mathbb{G}_{\mathrm{subpaint}}$ algebra:
\begin{equation}
  \mathcal{R}_\mathrm{N} \,
  \stackrel{\mathbb{U}_{\mathrm{TS}} \, \times \, \mathbb{G}_{\mathrm{subpaint}} }{\Longrightarrow}\,
  \left( \mathcal{R}_{\mathrm{N_{TS}}},\mathbf{1}\right)  \, \oplus \, \left(
  \mathbf{p},\mathbf{q}\right)~.
\label{brancolando}
\end{equation}
By definition $\mathcal{R}_{\mathrm{N_{TS}}}$ is the fundamental
representation of the Tits Satake subalgebra of dimension $\mathrm{N_{TS}} \, < \,
\mathrm{N}$ which is a singlet under the subpaint algebra
$\mathbb{G}_{\mathrm{subpaint}}$, while the remaining representation
$\left( \mathbf{p},\mathbf{q}\right)$ is non trivial both with respect to
the Tits-Satake and with respect to the subpaint Lie algebra. Obviously we have
$p\times q = \mathrm{N}-\mathrm{N_{TS}}$. Correspondingly the
eigenvalues of the Lax operator organize in the way we are going to describe. Let
$\mathbf{w}^i_{-|TS} < 0$ ($i=1,\dots,m$) be the negative weights of the representation
$\mathcal{R}_{\mathrm{N_{TS}}}$, let $\mathbf{w}^0_{TS} = 0$ be the
null-weight of the same representation (if it exists) and let $\mathbf{w}^i_{+|TS} > 0$ ($i=1,\dots,m$)
be the positive weights. If there is a null-weight we have $\mathrm{N_{TS}} =
2m+1$, otherwise $\mathrm{N_{TS}} = 2m$. A conventional order  for the $\mathrm{N}$ eigenvalues is given by the following vector :
\begin{equation}
\overrightarrow{ \lambda}_{[1]} \, = \, \left(\begin{array}{ccc}
    \lambda^1 & = & \overrightarrow{\mathbf{w}}^1_{-|TS}\, \cdot \, \overrightarrow{h}_{TS} \\
    \lambda^2 & = & \overrightarrow{\mathbf{w}}^2_{-|TS}\, \cdot \, \overrightarrow{h}_{TS} \\
    \dots & \dots & \dots \\
    \lambda^{m} & = & \overrightarrow{\mathbf{w}}^m_{-|TS}\, \cdot \, \overrightarrow{h}_{TS} \\
    \hline
    \lambda^{m+1} & = & 0 \\
    \lambda_{m+2} & = & 0\\
    \dots& \dots &\dots\\
     \lambda^{m+pq+1} & = & 0\\
     \hline
     \lambda^{m+pq+2} & = & \overrightarrow{\mathbf{w}}^1_{+|TS}\, \cdot \, \overrightarrow{h}_{TS} \\
    \lambda^{m+pq+3} & = & \overrightarrow{\mathbf{w}}^2_{+|TS}\, \cdot \, \overrightarrow{h}_{TS} \\
    \dots & \dots & \dots \\
    \lambda^{N} & = & \overrightarrow{\mathbf{w}}^m_{+|TS}\, \cdot \, \overrightarrow{h}_{TS} \
  \end{array}\right)
\label{organizzo}
\end{equation}
where the weights are organized from the lowest to the highest. The vector $\overrightarrow{ \lambda}_{[1]}$ corresponds to the
diagonal entries of the matrix $\mathcal{C}_0$ defined in eq.(\ref{inidata}). All
the other possible orders of the same eigenvalues are obtained from the action
of the Weyl group
$\mathrm{Weyl}\left(\mathbb{U}_{\mathrm{TS}}\right)$ on the weights
of $\mathcal{R}_{\mathrm{N_{TS}}}$.
By construction, such an action  permutes the positions of the
non-vanishing eigenvalues while all the zeros stay at their place. In
this way starting from $\overrightarrow{ \lambda}_{[1]}$ we obtain $n=|\mathrm{Weyl}\left(\mathbb{U}_{\mathrm{TS}}\right)|$
such vectors $\overrightarrow{ \lambda}_{[x]}$ in one-to-one correspondence with
the elements of the Tits Satake Weyl group. Schematically, naming
$\Omega_{x}$ the elements of
$\mathrm{Weyl}\left(\mathbb{U}_{\mathrm{TS}}\right)$, we obtain
\begin{equation}
  \overrightarrow{ \lambda}_{[x]} \, = \, \Omega_{x} \, \overrightarrow{ \lambda}_{[1]}
\label{schematical}
\end{equation}
with the understanding that $\Omega_1$ is the identity element of the Weyl
group. Each element $\Omega_x$ is represented by a permutation  of the
non vanishing eigenvalues, hence  by an element
$P(\Omega_x) \, \in \, \mathcal{S}_{\mathrm{N_{TS}}} \, \subset \,
\mathcal{S}_{\mathrm{N}}$. Among the vectors $\overrightarrow{
\lambda}_{[x]}$ there will be one $\overrightarrow{ \lambda}_{[min]}$
where the eigenvalues are organized in decreasing order
\begin{equation}
  \lambda_{[min]}^1 \, \ge \, \lambda_{[min]}^2 \, \ge \, \dots \, \ge \,
  \lambda_{[min]}^{N-1}\, \ge \, \lambda_{[min]}^N
\label{decreasorder}
\end{equation}
and there will be another one $\overrightarrow{ \lambda}_{[max]}$
where the eigenvalues are instead organized in increasing order
\begin{equation}
  \lambda_{[max]}^1 \, \le \, \lambda_{[max]}^2 \, \le \, \dots \, \le \,
  \lambda_{[max]}^{N-1}\, \le \, \lambda_{[max]}^N~.
\label{increasorder}
\end{equation}
Name $\Omega_{min/max}$ the corresponding Weyl elements. It follows
that equation (\ref{schematical}) can be rewritten as
\begin{equation}
  \overrightarrow{ \lambda}_{[x]} \, = \, \Omega_{x} \,
  \Omega_{min}^{-1} \, \overrightarrow{\lambda}_{[min]}~.
\label{fertilidea}
\end{equation}
The symmetric group admits a partial ordering of its elements given
by the number $\ell_T(P)$ of elementary transpositions necessary to obtain a
given permutation $P$ starting from the fundamental one $P_0$. The
embedding of the Weyl group into the symmetric group allows to
transfer this partial ordering to the Weyl group as well. We define
the length of a Weyl element as follows. Taking the permutation of
$\lambda_{[min]}$ as fundamental we set
\begin{equation}
\forall \, \Omega_x \, \in \, \mathrm{Weyl}
\left (\mathbb{U}_{TS}\right) \quad : \quad \ell_T \left(\Omega_{x}\right) \,
\equiv \, \ell_T\left[ P\left(
  \Omega_x \, \Omega_{min}^{-1}\right) \right] \,~.
\label{lungadiOmega}
\end{equation}
With this definition the Weyl element $\Omega_{min}$ has length
$\ell_T=0$ while the Weyl element $\Omega_{max}$ has the maximal length
$\ell_T = \ft 12 \left( \mathrm{N_{TS}} -1 \right) \mathrm{N_{TS}}$
and an element $\Omega_x$ is higher than an element
$\Omega_y$ if $\ell_T(\Omega_x) \, > \, \ell_T(\Omega_y)$. We
can observe that the partial ordering induced by the immersion in the
symmetric group is, up to some rearrangement, the intrinsic ordering
of the Weyl group provided by counting the minimal number of reflections with
respect to  simple roots necessary to construct the considered
element. In our context formalising   the precise correspondence between the
two ordering procedures is not necessary since the relevant one is
that with respect to permutations and this is well and uniquely
defined.
\par
Having introduced the above ordering of Weyl elements we can now
state the main and most significant property of the map
(\ref{wwmappa}) and of the Toda flows realized by the integration
algorithm (\ref{Imap}).
\begin{principio}\label{timearrowprinc}
$<<$
In any flow the arrow of time is so directed that the state at
$t=-\infty$ corresponds to the lowest accessible Weyl element and the
state at $t=+\infty$ corresponds to the highest accessible one.
$>>$
\end{principio}
To make the principle \ref{timearrowprinc} precise we need to
define the notion of accessible Weyl elements. This latter relies on another
remarkable and striking property of the Toda flows (\ref{Imap}) which
we have numerically verified in a large variety of cases never finding
any counterexample.
\begin{property}\label{kkminorprinc} $<<$
At any instant of time the Lax operator $L(t)$ can be diagonalized by
a time dependent orthogonal matrix $\mathcal{O}(t) \, \in \, H$,
writing $L(t) = \mathcal{O}^T(t) \, \mathcal{C}_0 \, \mathcal{O}(t)$.
Consider now the $\mathrm{N}^2-1$ minors of $\mathcal{O}(t)$ obtained by
intersecting the first $k$ columns  with any set of
$k$-rows, for $k=1,\dots,\mathrm{N}-1$. If any of these minors
vanishes at any finite time $t \,\ne \, \pm \infty$ then it is constant and
vanishes at all times.
$>>$
\end{property}
The remarkable conservation law stated in property \ref{kkminorprinc}
which has the mathematical status of a \textbf{conjecture}
implies that there are generic initial data, namely points of the
parameter space $\mathcal{P}$ defined in eq.(\ref{bunduparametru}) and $N^2 -1$
\textbf{trapped hypersurfaces} $\Sigma_i \, \subset \, \mathcal{P}$
defined by the vanishing of one of the minors. These
trapped surfaces can also be intersected creating trapped
sub-varieties of equal or lower dimensions. If the initial data are generic,
then principle \ref{timearrowprinc} implies that the flow will
necessarily be from $\Omega_{min}$ to $\Omega_{max}$. On the other
hand if we are on a trapped surface we have to see which elements
of the generalized Weyl group $\mathcal{W}(\mathbb{U})$ belong to
that surface. As we know from (\ref{riguardoso}) each element of
$\mathcal{W}(\mathbb{U})$ is equivalent to an element of
$\mathrm{Weyl}(\mathbb{U}_{\mathrm{TS}})$ modulo an element in the
normal subgroup. Hence we can introduce the following definition:
\begin{definizione}\label{accessibili}
$<<$
A Weyl group element $\gamma \, \in \, \mathrm{Weyl}\left(\mathbb{U}_{TS}\right )$ is accessible to a
trapped surface $\Sigma$ if there exists a representative $\mu \, \in \, \mathcal{W}(\mathbb{U})$ of its equivalence class
in the generalized Weyl group which belongs to $\Sigma$.
$>>$
\end{definizione}
The set of Weyl elements $\mathcal{A}_\Sigma $ accessible to a trapped surface $\Sigma$ inherits an
ordering from the general ordering of $\mathrm{Weyl}\left(\mathbb{U}_{TS}\right
)$, namely we can write:
\begin{equation}
  \mathcal{A}_\Sigma \, = \, \left\{ \Omega_{x_1} \,, \Omega_{x_2}\,, \dots \,,
  \Omega_{x_\sigma}\right\}
\label{Asetti}
\end{equation}
where $\sigma$ is the cardinality of the set $\sigma = \mbox{card}
\,\mathcal{A}_\Sigma$ and $\Omega_{x_i} \, \le \, \Omega_{x_j}$ if
$i<j$. Then the flow is always from the lowest Weyl element of
$\mathcal{A}_\Sigma$ at $t=-\infty$ (i.e., from $\Omega_{x_1}$) to
the highest one at $t=\infty$ (i.e., to $\Omega_{x_\sigma}$)
as stated in principle \ref{timearrowprinc}.
\par
If we consider lower dimensional trapped surfaces obtained by intersection,
then the set of Weyl elements accessible to the intersection is
simply given by the intersection of the accessible sets:
\begin{equation}
  \mathcal{A}_{\Sigma \, \bigcap \, \Pi} \, = \,
\mathcal{A}_{\Sigma} \, \bigcap \, \mathcal{A}_{\Pi}
\label{intersettone}
\end{equation}
and the flow is from the lowest element of $\mathcal{A}_{\Sigma \, \bigcap \,
\Pi}$ to its highest one.
\par
We can now introduce a further
\begin{definizione}\label{criticasurfacia}
$<<$
A trapped surface $\Sigma$ is named \textbf{critical} if the set of
 Weyl elements $\mathcal{A}_{\Sigma}$ accessible to the surface is a proper subset of
 the Weyl group, in other words if
\begin{equation}
  \mathrm{card} \,\mathcal{A}_{\Sigma} \, < \, |\mathrm{Weyl}\left
  (\mathbb{U}_{TS}\right)|~.
\label{critical}
\end{equation}
$>>$
\end{definizione}
Note that the property of criticality does not necessarily imply a
variance of asymptotics from the generic case $\Omega_{min} \,
\rightarrow \, \Omega_{max}$. Indeed, although the cardinality of
$\mathcal{A}_{\Sigma}$ is lower than the order of the Weyl group so
that some elements are missing, yet it suffices that both
$\Omega_{min} \, \in \, \mathcal{A}_{\Sigma}$ and $\Omega_{max} \, \in \, \mathcal{A}_{\Sigma}$
to guarantee that the infinite past and infinite future states of the Universe will be the
same as in the generic case. This observation motivates the further
definition:
\begin{definizione}\label{supcriticasurfacia}
$<<$
A trapped surface $\Sigma$ is named \textbf{super-critical} if it is
critical and moreover either the maximal or the minimal Weyl elements are
missing from $\mathcal{A}_{\Sigma}$:
\begin{equation}
\Omega_{min} \, \notin \, \mathcal{A}_{\Sigma} \quad \mbox{and/or}
\quad \Omega_{max} \, \notin \, \mathcal{A}_{\Sigma}~.
\label{supercritical}
\end{equation}
$>>$
\end{definizione}
This discussion shows that the truly relevant concept is that of
trapped surface which streams from the remarkable conservation law
given by property \ref{kkminorprinc}, criticality or super-criticality
being, from the mathematical point of view,  just
accessory features although of the highest physical relevance.
If we just focus our attention on the initial and final states the
intermediate concept of critical surface seems to be unmotivated.
The reason why it is useful  is that critical surfaces as defined in
\ref{criticasurfacia} can be computed in an intrinsic way  taking a
dual point of view. Rather than computing accessible Weyl elements
one can define forbidden ones by using the embedding of the
Weyl group into the symmetric group mentioned in
eq.(\ref{symmetricembed}). It follows from this that to each element
$\Omega_x \, \in \, \mathrm{Weyl}\left
  (\mathbb{U}_{TS}\right)$ we can associate a permutation $P_x \, \in
  \, \mathcal{S}_{\mathrm{N}}$, where $\mathrm{N}$ is the dimension
  of the fundamental representation of $\mathbb{U}$ and hence of the
  orthogonal matrix $\mathcal{O}$ we are discussing. This fact allows
  to associate to $\Omega_x$ a set of $\mathrm{N-1}$ minors defined as
  follows:
\begin{eqnarray}
\mathrm{Weyl}\left
  (\mathbb{U}_{TS}\right) \, \ni \,\Omega_x & \rightarrow & \left\{\mbox{min}^{(1)}_x\,\left[\mathcal{O}\right] \, , \,
  \mbox{min}^{(2)}_x\,\left[\mathcal{O}\right]\,, \dots \, , \, \mbox{min}^{(N-1)}_x\,\left[\mathcal{O}\right]\right\} \\
   \mbox{min}^{(k)}_x\,\left[\mathcal{O}\right]
 & = & \mbox{Det} \, \left( \mathcal{O}\left[\left(P_x(1),\dots,P_x(k)\right),\left(1,\dots,k\right)\right]\right)
\label{minorati}
\end{eqnarray}
where
$M\left[\left(a_1,\dots,a_k\right),\left(1,\dots,k\right)\right]$
denotes the minor of the matrix $M$ obtained by intersecting the $k$-rows
$a_1,\dots,a_k$ with the first $k$-columns.
Using a dual view-point it was shown in \cite{leitepoli} that  in any
flow, in order for a permutation $P$ of the eigenvalues to be a candidate for asymptotics (i.e. to be available),
its associated minors should all be non zero. Hence relying on the embedding of the Weyl group into the
symmetric group we conclude that if any
minor $\mbox{min}^{(k)}_x\,\left[\mathcal{O}\right]$ vanishes then $\Omega_x$ is excluded from
the set $\mathcal{A}_{\Sigma_{x|k}}$ of Weyl elements accessible to
the surface $\Sigma_{x|k}$ defined by the vanishing of the minor
$\mbox{min}^{(k)}_x\,\left[\mathcal{O}\right]$. We can write:
\begin{equation}
  \Sigma_{x|k} \, \equiv \, \left\{ \mathcal{O} \, \in \, \mathrm{H}
  \, \backslash \, \mbox{min}^{(k)}_x\,\left[\mathcal{O}\right] \, =
  \, 0 \right\} \quad \Rightarrow \quad \Omega_x \, \notin \, \mathcal{A}_{\Sigma_{x|k}}~.
\label{Xsurface}
\end{equation}
The same minor is produced by more than one element and hence
identifying all the $\Omega_x$ for which
$\mbox{min}^{(k)}_x\,\left[\mathcal{O}\right] \, = \,
\mbox{min}^{(k)}_{[0]}$ we immediately calculate the set of Weyl
elements excluded from $\mathcal{A}_{\Sigma_{0}}$ and by complement
we also know the set $\mathcal{A}_{\Sigma_{0}}$.
\par
If all the possible minors considered in property (\ref{kkminorprinc}) could be
produced by Weyl elements, then what we have just described would
be a quick and efficient way to obtain all trapped surfaces.
In that case all trapped surfaces would also be critical. The fact is
that not all minors can be obtained from Weyl elements and this
implies that there are \textit{trapped surfaces which are not critical}.
The reason of this difference is  evident from our
discussion. It is due to the fact that the Weyl group is in general
only a proper subgroup of the symmetric group
$\mathcal{S}_\mathrm{N}$. Therefore there are permutations and
therefore minors which do not correspond to any Weyl element and for
that reason they define non-critical trapped surfaces. In the case of
$\mathrm{SL(N,\mathbb{R})}/\mathrm{SO(N)}$ flows the Weyl group is
just the full $\mathcal{S}_\mathrm{N}$  and all trapped surfaces are
critical. In conclusion trapped but not critical surfaces are
critical surfaces of the embedding $\mathrm{SL(N)}$ where the missing
Weyl elements are in the kernel of the projection $\mathcal{S}_\mathrm{N} \, \mapsto \,
\mathrm{Weyl}(\mathbb{U})$.
\par
This concludes the general presentation of our results.
By means of some case studies the next part illustrates the principles formulated
in this part.
\part{Examples illustrating the principles}
\section{Choice of the examples}
In this part we make three case studies:
\begin{description}
  \item[1] We survey the flows on the simplest example
  $\mathrm{SL(3,\mathbb{R})}/\mathrm{SO(3)}$ of maximally split
 coset manifolds in order to demonstrate the relation between the
 Weyl group and the arrow of time by calculating explicitly all the critical
 surfaces which are two-dimensional and can be visualized. The
 parameter space is a three dimensional cube with some vertices identified and can also be
 visualized.
  \item[2] Next we make a detailed study of the flows on
  $\mathrm{Sp(4,\mathbb{R})}/\mathrm{U(2)}$. This manifold is the
  Tits Satake projection of an entire universality class of
  manifolds, $\mathrm{SO(2,2+2s)}/\mathrm{SO(2)}\times
  \mathrm{SO(2+2s)}$ which, on the other hand, is for $r=2$ of the type
  $\mathrm{SO(r,r+2s)}/\mathrm{SO(r)}\times
  \mathrm{SO(r+2s)}$. For the latter we make the general construction
  of the triangular basis of the solvable algebra illustrating the
  embedding:
\begin{equation}
  \so(r,r+2s) \, \hookrightarrow \, \slal(2r+2s)
\label{embeddatone}
\end{equation}
which is crucial in order to establish the integration algorithm. In
the case of $\sym(4) \, \sim \, \so(2,3)$, which is maximally split of
rank two, the parameter space is a $4$-dimensional hypercube also
with vertices identified.
  \item[3] Finally we study the case of $\mathrm{SO(2,4)}$ in comparison with
  that of $\mathrm{SO(2,3)}$ in order to illustrate the properties of the Tits
  Satake projection and the meaning of Tits Satake universality
  classes.
\end{description}
\section{The simplest maximally split case: $\mathrm{SL(3,\mathbb{R})}/\mathrm{SO(3)}$}
\begin{figure}
\caption{\it The Dynkin diagram of the  $A_2$ Lie algebra.
\label{A2dynk}}
\centering
\begin{picture}(10,100)
\put (-70,65){$A_2$}
\put (-20,70){\circle {10}}
\put (-23,55){$\alpha_1$}
\put (-15,70){\line (1,0){20}}
\put (10,70){\circle {10}}
\put (7,55){$\alpha_2$}
\end{picture}
\end{figure}
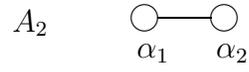

\begin{figure}[!hbt]
\begin{center}
\iffigs
 \includegraphics[height=60mm]{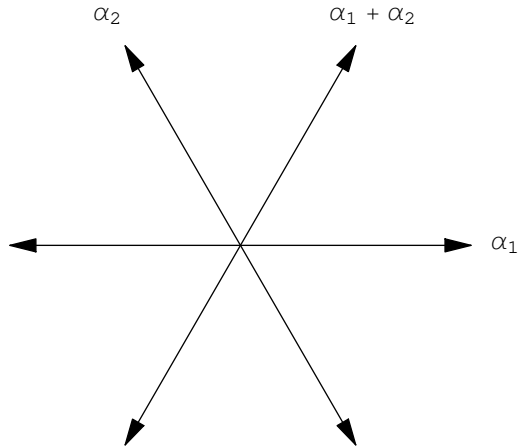}
\else
\end{center}
 \fi
\caption{\it The $A_2$ root system.}
\label{A2picroot}
 \iffigs
 \hskip 1.5cm \unitlength=1.1mm
 \end{center}
  \fi
\end{figure}

In order to illustrate the general ideas discussed in the previous
part and as a preparation to  the study of more general cases, we begin with a
detailed analysis of the time flows  in the simplest instance of maximally
split coset manifolds, namely for
\begin{equation}
  \mathcal{M}_5 \, = \,
  \frac{\mathrm{SL(3,\mathbb{R})}}{\mathrm{SO(3)}}~.
\label{SL3defi}
\end{equation}
The $\slal(3,\mathbb{R})$ Lie algebra is the maximally split real
section of the $A_2$ Lie algebra, encoded in the Dynkin diagram of
fig.\ref{A2dynk}.
The root system has rank two and it is composed by the six vectors
displayed below and pictured in fig.\ref{A2picroot}:
\begin{equation}
  \Delta_{A_2} \, = \, \left \{ \begin{array}{rcl}
    \alpha_1 & = & \left( \sqrt{2} , 0 \right)~, \\
    \alpha_2 & = & \left(-\, \frac{1}{\sqrt{2}} ,
    \sqrt{\frac{3}{2}} \right)~, \\
    \alpha_1\, +\, \alpha_2 & = & \left( \frac{1}{\sqrt{2}} \, ,
    \, \sqrt{\frac{3}{2}} \right)~, \\
    - \,\alpha_1 & = & \left(- \sqrt{2} , 0 \right)~,  \\
   - \,  \alpha_2 & = & \left( \frac{1}{\sqrt{2}} ,
   - \, \sqrt{\frac{3}{2}} \right)~, \\
    - \, \alpha_1\, - \, \alpha_2 & = &\left( -\frac{1}{\sqrt{2}}
     ,
  \, -\, \sqrt{\frac{3}{2}} \right) ~.\
  \end{array}\right.
\label{A2roots}
\end{equation}
The simple roots are $\alpha_1$ and $\alpha_2$.
\par
A complete set of
generators for the Lie algebra is provided by the following $3 \times
3$ matrices:
\begin{eqnarray}
 && \begin{array}{ccccccc}
    H_1 & = & \left(
\begin{array}{lll}
 \frac{1}{\sqrt{2}} & 0 & 0 \\
 0 & -\frac{1}{\sqrt{2}} & 0 \\
 0 & 0 & 0
\end{array}
\right) & ; & H_2 & = & \left(
\begin{array}{lll}
 -\frac{1}{\sqrt{6}} & 0 & 0 \\
 0 & -\frac{1}{\sqrt{6}} & 0 \\
 0 & 0 & \sqrt{\frac{2}{3}}
\end{array}
\right)~, \\
    E^{\alpha_1} & = & \left(
\begin{array}{lll}
 0 & 1 & 0 \\
 0 & 0 & 0 \\
 0 & 0 & 0
\end{array}
\right) & ; & E^{\alpha_2} & = &  \left(
\begin{array}{lll}
 0 & 0 & 0 \\
 0 & 0 & 1 \\
 0 & 0 & 0
\end{array}
\right)~, \\
     E^{\alpha_1+\alpha_2} & = & \left(
\begin{array}{lll}
 0 & 0 & 1 \\
 0 & 0 & 0 \\
 0 & 0 & 0
\end{array}
\right), & \null & \null & \null & \null \
  \end{array}\nonumber\\
  && E^{- \alpha_1} \, = \, \left( E^{\alpha_1}\right) ^T \quad ;
  \quad E^{- \alpha_2} \, = \, \left( E^{\alpha_2}\right) ^T \quad ;
  \quad E^{- \alpha_1- \alpha_2} \, = \, \left( E^{\alpha_1+ \alpha_2}\right) ^T
\label{a2Liealg}
\end{eqnarray}
where $H_{1,2}$ are the two Cartan generators and $E^{\alpha}$ are
the step operators associated to the corresponding roots.
The solvable Lie algebra generating the coset (\ref{SL3defi}) is
composed by the following five operators:
\begin{equation}
  Solv\left(\frac{\mathrm{SL(3,\mathbb{R})}}{\mathrm{SO(3)}} \right) =
  \mbox{span}\, \left\{ H_1\, , \, H_2\, , \, E^{\alpha_1} \, , \,
  E^{\alpha_2}\, , \, E^{\alpha_1 + \alpha_2} \right\}
\label{solvA2}
\end{equation}
and it is clearly represented by upper triangular matrices. The
orthogonal decomposition
\begin{equation}
  \mathbb{G} \, = \, \mathbb{H} \, \oplus \, \mathbb{K}
\label{orthodecompo2}
\end{equation}
of the Lie algebra with respect to its maximal compact subalgebra:
\begin{equation}
  \so(3) \, \equiv \, \mathbb{H} \, \subset \, \mathbb{G} \, \equiv
  \, \slal(3,\mathbb{R})
\label{orthodecompo1}
\end{equation}
is performed by defining the following generators:
\begin{eqnarray}
  \mathbb{K} &=& \mbox{span}\left\{ K_1,\dots \,, K_5\right\}
  \nonumber\\
  &\equiv &
  \left\{ H_1\, , \,  H_2\, , \, \ft{1}{\sqrt{2}} \,  \, \left( E^{\alpha_1}
  +  E^{- \alpha_1} \right) \, \,, \ft{1}{\sqrt{2}} \,  \,\left( E^{\alpha_2}
  +  E^{- \alpha_2} \right) \, , \,\ft{1}{\sqrt{2}} \,\left( E^{\alpha_1 \, + \, \alpha_2}
  +  E^{-\,\alpha_1 \,- \alpha_2} \right)\right\}~,\nonumber\\
\label{Kgeneri}
\end{eqnarray}
\begin{eqnarray}
  \mathbb{H} &=& \mbox{span}\left\{ J_1,\dots \,, J_3\right\}
  \nonumber\\
  &\equiv &
  \left\{ \ft{1}{\sqrt{2}} \, \, \left( E^{\alpha_1}
  -  E^{- \alpha_1} \right) \,, \, \ft{1}{\sqrt{2}} \, \,\left( E^{\alpha_2}
  -  E^{- \alpha_2} \right) \, , \,\ft{1}{\sqrt{2}} \,\left( E^{\alpha_1 \, + \, \alpha_2}
  -  E^{-\,\alpha_1 \,- \alpha_2} \right)\right\}~.
\label{Hgeneri}
\end{eqnarray}
By definition the Lax operator $L(t)$ is a symmetric $3 \times 3$ matrix
which can be decomposed along the generators of the subspace
$\mathbb{K}$:
\begin{equation}
  L(t) \, = \, \sum_{i=1}^{5} k^i(t) \, K_i
\label{Lt}
\end{equation}
and once the functions $k^i(t)$ have been determined, by means of  an
oxidation procedure which was fully described in \cite{noiconsasha}, the fields of supergravity
can be extracted by simple quadratures.
\par
As we explained in \cite{sahaedio} and we recalled in the
introduction, the initial data for the integration of the Lax
equation are provided by the choice of an element of the Cartan
subalgebra, namely by a diagonal matrix of the form:
\begin{equation}
 \mbox{CSA} \, \ni \, \mathcal{C}\left(\left\{ \lambda_1,\lambda_2\right\} \right) \, = \, \left( \begin{array}{ccc}
   \lambda_1 & 0 & 0 \\
   0 & \lambda_2 & 0 \\
   0 & 0 & -\lambda_1-\lambda_2 \
 \end{array}\right)
\label{Lam123}
\end{equation}
and by a finite element $\mathcal{O}\, \in \,
\mathrm{SO(3)}$ of the compact subgroup  which together with $\mathcal{C}$ determines the value of the Lax
operator at time $t=0$:
\begin{equation}
  L_0 \, = \, \mathcal{O}^T \,
  \mathcal{C}\left(\left\{ \lambda_1,\lambda_2\right\}\right )\, \mathcal{O}~.
\label{Lzero}
\end{equation}
We stressed that the choice of the group element $\mathcal{O}$ is
actually defined modulo multiplication on the left by any element $w
\in \mathcal{W}\, \subset \, \mathrm{H} \, \simeq \, \exp \mathbb{H}$ of the discrete Weyl subgroup.
By definition the Weyl group maps the Cartan subalgebra into itself, so that we
have:
\begin{equation}
  \forall \, w \, \in \, \mathcal{W} \, \subset \, \mathrm{SO(3})
  \quad : \quad w^T \, \mathcal{C}\left(\left\{ \lambda_1,\lambda_2\right\}\right) \,
  w \, = \, \mathcal{C}\left(w\left\{
  \lambda_1,\lambda_2\right\}\right) \,  \in
  \mbox{CSA}
\label{weylaction}
\end{equation}
where $\mathcal{C}\left(w\left\{\lambda_1,\lambda_2\right\}\right)$ denotes the diagonal matrix of
type (\ref{Lam123}) with eigenvalues
$\lambda_1^\prime,\lambda_2^\prime,-\lambda_1^\prime-\lambda_2^\prime$
obtained from the action of the Weyl group on the original ones. So
the actual moduli space of the Lax equation is not $\mathrm{H}$ but the
quotient $\mathrm{H}/\mathcal{W}$.
\par
In the case of the Lie algebras
$A_n$ the Weyl group is the symmetric group $\mathcal{S}_{n+1}$ and
its action on the eigenvalues $\lambda_1,\lambda_2, \dots\,,
\lambda_n,\lambda_{n+1}\,=\, -\sum_{i=1}^{n}\lambda_i$ is just that of permutations on
these $n+1$-eigenvalues. For $A_2$ we have $\mathcal{S}_3$ whose order is six.
The six group elements can be enumerated in the following way:
\begin{equation}
  \begin{array}{ccccccc}
    w_1 & = & \left(
\begin{array}{lll}
 1 & 0 & 0 \\
 0 & 1 & 0 \\
 0 & 0 & 1
\end{array}
\right) & ; & \left( \lambda_1,\lambda_2,\lambda_3\right) &
    \mapsto & \left( \lambda_1,\lambda_2,\lambda_3\right)~, \\
    w_2 & = & \left(
\begin{array}{lll}
 0 & 1 & 0 \\
 1 & 0 & 0 \\
 0 & 0 & 1
\end{array}
\right) &; & \left( \lambda_1,\lambda_2,\lambda_3\right) &
    \mapsto & \left( \lambda_2,\lambda_1,\lambda_3\right)~, \\
    w_3 & = & \left(
\begin{array}{lll}
 0 & 0 & 1 \\
 0 & 1 & 0 \\
 1 & 0 & 0
\end{array}
\right) & ; & \left( \lambda_1,\lambda_2,\lambda_3\right) &
    \mapsto & \left( \lambda_3,\lambda_2,\lambda_1\right)~, \
    \end{array}
    \end{equation}
    \begin{equation}
     \begin{array}{ccccccc}
    w_4 & = & \left(
\begin{array}{lll}
 1 & 0 & 0 \\
 0 & 0 & 1 \\
 0 & 1 & 0
\end{array}
\right) & ; & \left( \lambda_1,\lambda_2,\lambda_3\right) &
    \mapsto & \left( \lambda_1,\lambda_3,\lambda_2\right)~,\\
    w_5 & = & \left(
\begin{array}{lll}
 0 & 0 & 1 \\
 1 & 0 & 0 \\
 0 & 1 & 0
\end{array}
\right) & ; & \left( \lambda_1,\lambda_2,\lambda_3\right) &
    \mapsto & \left( \lambda_2,\lambda_3,\lambda_1\right)~, \\
  w_6 & = & \left(
\begin{array}{lll}
 0 & 1 & 0 \\
 0 & 0 & 1 \\
 1 & 0 & 0
\end{array}
\right) & ; & \left( \lambda_1,\lambda_2,\lambda_3\right) &
    \mapsto & \left( \lambda_3,\lambda_1,\lambda_2\right)~. \
    \end{array}
\label{weylus}
\end{equation}
Let us now choose as eigenvalues $\lambda_1,\lambda_2,\lambda_3$
defined at the central time $t=0$ the conventional set
\begin{equation}
\lambda _1 \, = \, 1 \quad ; \quad \lambda_2 = 2 \quad ; \quad
\lambda_3 \, = \, -3~.
\label{convenset}
\end{equation}
In this case the decreasing sorting to be expected at past infinity
is given by: $2,1,-3$ which, according to eq.(\ref{weylus}),
corresponds to the Weyl element $w_2$. Hence we can use $w_2$ as the
fundamental permutation and rate all the other Weyl group elements according
to the number of transpositions $\ell_T$ needed to bring their
corresponding permutation to that of $w_2$.
\par
In this way we obtain
a partial ordering of the Weyl group where the highest element is the
unique $w_6$ corresponding to the increasing sorting of eigenvalues
$-3,1,2$. Indeed we have the result shown in table \ref{SL3Weylordo}
\begin{table}
  \centering
  $$
  \begin{array}{|c|c|}
  \hline
    \ell_T & \mbox{Weyl group}\\
    \null & \mbox{of $\mathrm{SL(3,\mathbb{R})}$} \\
    \hline
    0 & w_2 \\
    1 & w_1 \\
    1 & w_5 \\
    2 & w_3 \\
    2 & w_4 \\
    3 & w_6 \\
    \hline
  \end{array}
  $$
  \caption{Partial ordering of the Weyl group of $\mathrm{SL(3,\mathbb{R}).}$}
  \label{SL3Weylordo}
\end{table}
and if all the Weyl elements are accessible there is a unique
predetermined process: the state of the universe at past infinity is
the Kasner era $w_2$, while the state of the Universe at future
infinity is the Kasner era $w_6$. If not all the Weyl elements are
accessible, then we can have different situations. In order to
discuss them we have to study the structure of the orbifold
$\mathrm{SO(3)}/\mathcal{W}$.
\par
To parametrize the $\mathrm{SO(3)}$ compact group we introduce three
Euler angles $\theta_i \,(i=1,2,3)$ and we write
\begin{eqnarray}
\mathcal{O}(\theta_i) & \equiv & \exp\left[\theta_1 \,J_1 \right] \,\exp\left[ \theta_2 \, J_2 \right] \,
\exp\left[\theta_3 \,J_3  \right] \,  =  \,\left(
\begin{array}{ccc}
  O_{11} & O_{12} & O_{13} \\
  O_{21} & O_{22} & O_{23} \\
  O_{31} & O_{32} & O_{33}
\end{array}
\right) \nonumber\\
\label{Pspec}
\end{eqnarray}
where:
\begin{equation}
  \begin{array}{cclc}
    O_{11} & = & \cos \left(\theta _1\right)
   \cos \left(\theta
   _3\right)-\sin \left(\theta
   _1\right) \sin \left(\theta
   _2\right) \sin \left(\theta
   _3\right) & ; \\ O_{12} & = & \cos \left(\theta _2\right)
   \sin \left(\theta _1\right) & ; \\
    O_{13} & = & \cos \left(\theta _3\right)
   \sin \left(\theta _1\right)
   \sin \left(\theta
   _2\right)+\cos \left(\theta
   _1\right) \sin \left(\theta
   _3\right) & ; \\ O_{21} & = & -\cos \left(\theta _3\right)
   \sin \left(\theta
   _1\right)-\cos \left(\theta
   _1\right) \sin \left(\theta
   _2\right) \sin \left(\theta
   _3\right) & ; \\
    O_{22} & = & \cos \left(\theta _1\right)
   \cos \left(\theta _2\right) & ; \\ O_{23} & = & \cos
   \left(\theta _1\right)
   \cos \left(\theta _3\right)
   \sin \left(\theta
   _2\right)-\sin \left(\theta
   _1\right) \sin \left(\theta
   _3\right) & ; \\
    O_{31} & = & -\cos \left(\theta _2\right)
   \sin \left(\theta _3\right) & ; \\ O_{32} & = & -\sin \left(\theta _2\right) & ; \\
    O_{33} & = & \cos \left(\theta _2\right)
   \cos \left(\theta _3\right) & . \
   \end{array}
\label{Pmatraelelmenta}
\end{equation}
In this parametrization, if we introduce the notation
\begin{equation}
  \mathcal{O}_{xyz} \, = \, \mathcal{O}\left( x \frac{\pi}{2} \, , \, y
  \frac{\pi}{2} \, , \, z \frac{\pi}{2}\right )
\label{notazia}
\end{equation}
we obtain:
\begin{equation}
  \begin{array}{ccccccc}
    \mathcal{O}_{000} & = & \left(
\begin{array}{lll}
 1 & 0 & 0 \\
 0 & 1 & 0 \\
 0 & 0 & 1
\end{array}
\right) & ; &  \mathcal{O}_{100} & = & \left(
\begin{array}{lll}
 0 & 1 & 0 \\
 -1 & 0 & 0 \\
 0 & 0 & 1
\end{array}
\right)~, \\
   \mathcal{O}_{010} & = & \left(
\begin{array}{lll}
 1 & 0 & 0 \\
 0 & 0 & 1 \\
 0 & -1 & 0
\end{array}
\right) & ; &  \mathcal{O}_{001} & = & \left(
\begin{array}{lll}
 0 & 0 & 1 \\
 0 & 1 & 0 \\
 -1 & 0 & 0
\end{array}
\right)~, \
\end{array}
\label{weylini1}
\end{equation}
\begin{equation}
  \begin{array}{ccccccc}
    \mathcal{O}_{110} & = & \left(
\begin{array}{lll}
 0 & 0 & 1 \\
 -1 & 0 & 0 \\
 0 & -1 & 0
\end{array}
\right) & ; &  \mathcal{O}_{101} & = & \left(
\begin{array}{lll}
 0 & 1 & 0 \\
 0 & 0 & -1 \\
 -1 & 0 & 0
\end{array}
\right)~, \\
    \mathcal{O}_{011} & = & \left(
\begin{array}{lll}
 0 & 0 & 1 \\
 -1 & 0 & 0 \\
 0 & -1 & 0
\end{array}
\right) & ; &  \mathcal{O}_{111} & = & \left(
\begin{array}{lll}
 -1 & 0 & 0 \\
 0 & 0 & -1 \\
 0 & -1 & 0
\end{array}
\right)~. \
  \end{array}
\label{weylini2}
\end{equation}
\subsection{Discussion of the generalized Weyl group}
\begin{figure}[!hbt]
\begin{center}
\iffigs
 \includegraphics[height=70mm]{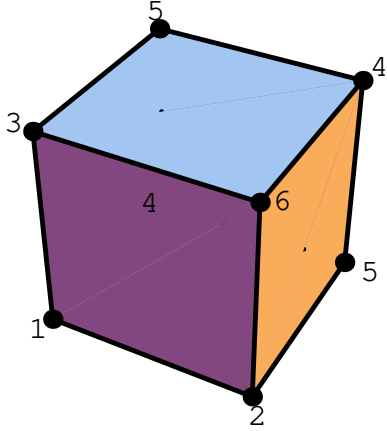}
\else
\end{center}
 \fi
\caption{\it
A three-dimensional cube $0\le \theta_i \le \frac{\pi}{2}$ ($i=1,2,3$)
whose eight vertices
are identified with the six Weyl group elements as shown in the picture. The 12 edges of the cube represent
one parameter submanifolds of $\mathrm{SO(3)}$ where just one angle varies while the other two are at fixed values,
either $0$ or $\pi/2$.}
\label{cuboide}
 \iffigs
 \hskip 1.5cm \unitlength=1.1mm
 \end{center}
  \fi
\end{figure}

Let us now construct the generalized Weyl group, according to the
definition \ref{genWeyl}. This case is maximally split and all
roots participate in the construction. Hence as generators we take
the three matrices
\begin{equation}
  \mbox{generators} =
  \left\{ \mathcal{O}_{100},\mathcal{O}_{010},\mathcal{O}_{001}\right\}
\label{generaSL3W}
\end{equation}
as defined above in eq.(\ref{weylini1}). Closing the shell of
products we find a group $\mathcal{W}(\slal(3))$ containing $24$
elements organized in $6$ equivalence classes with respect to a
normal subgroup $\mathrm{N}\left (\slal(3)\right) \, \sim \,\mathbb{Z}_2 \times \mathbb{Z}_2$. The
four elements of $\mathrm{N}\left (\slal(3)\right)$ are the following four matrices:
\begin{equation}
  \begin{array}{ccccccc}
    n_1 & = & \left(
\begin{array}{lll}
 1 & 0 & 0 \\
 0 & 1 & 0 \\
 0 & 0 & 1
\end{array}
\right) & ; & n_2 & = & \left(
\begin{array}{lll}
 1 & 0 & 0 \\
 0 & -1 & 0 \\
 0 & 0 & -1
\end{array}
\right)~, \\
    n_3 & = & \left(
\begin{array}{lll}
 -1 & 0 & 0 \\
 0 & 1 & 0 \\
 0 & 0 & -1
\end{array}
\right) & ; & n_4 & = & \left(
\begin{array}{lll}
 -1 & 0 & 0 \\
 0 & -1 & 0 \\
 0 & 0 & 1
\end{array}
\right)~. \
  \end{array}
\label{Nsl3}
\end{equation}
The factor group is isomorphic to the Weyl group of $\slal(3)$
\begin{equation}
  \frac{\mathcal{W}\left(\slal(3)\right)}{\mathrm{N}\left(\slal(3)\right)}
  \, \sim \, \mathrm{Weyl}\left( \slal(3)\right) \, \equiv \, \mathcal{S}_3
\label{identificosl3}
\end{equation}
and a representative of the six equivalence classes is listed below
\begin{equation}
  \begin{array}{ccccccc}
    w_1 & \sim & \left(
\begin{array}{lll}
 1 & 0 & 0 \\
 0 & 1 & 0 \\
 0 & 0 & 1
\end{array}
\right) \, \mathrm{N} \left (\slal(3)\right) & ; & w_2 & \sim & \left(
\begin{array}{lll}
 0 & 1 & 0 \\
 1 & 0 & 0 \\
 0 & 0 & -1
\end{array}
\right) \, \mathrm{N} \left (\slal(3)\right)~, \\
    w_3 & \sim & \left(
\begin{array}{lll}
 0 & 0 & 1 \\
 0 & 1 & 0 \\
 -1 & 0 & 0
\end{array}
\right) \, \mathrm{N} \left (\slal(3)\right) & ; & w_4 & \sim & \left(
\begin{array}{lll}
 1 & 0 & 0 \\
 0 & 0 & 1 \\
 0 & -1 & 0
\end{array}
\right) \, \mathrm{N} \left (\slal(3)\right)~, \\
    w_5 & \sim & \left(
\begin{array}{lll}
 0 & 0 & 1 \\
 1 & 0 & 0 \\
 0 & 1 & 0
\end{array}
\right) \, \mathrm{N} \left (\slal(3)\right) & ; & w_6 & \sim & \left(
\begin{array}{lll}
 0 & 1 & 0 \\
 0 & 0 & 1 \\
 1 & 0 & 0
\end{array}
\right) \, \mathrm{N} \left (\slal(3)\right)~. \
  \end{array}
\label{eqclasSL3}
\end{equation}
Hence modulo the normal subgroup  the eight matrices
listed in eq.s(\ref{weylini1},\ref{weylini2}) can be identified with
the six elements of the Weyl group in the following way:
\begin{equation}
  \begin{array}{ccccccccccc}
    \mathcal{O}_{000} & \sim & w_1 & ; & \mathcal{O}_{100} & \sim & w_2 & ; & \mathcal{O}_{010} & \sim & w_4 \\
    \mathcal{O}_{001} & \sim & w_3 & ; & \mathcal{O}_{110} & \sim & w_5 & ; & \mathcal{O}_{101} & \sim & w_5 \\
     \mathcal{O}_{011} & \sim & w_6 & ; & \mathcal{O}_{111} & \sim & w_4 & . & \null & \null & \null \
  \end{array}
\label{weylidentificazie}
\end{equation}
Let us now consider the general form of the $\mathrm{SO(3)}$ matrix
as given in eq.(\ref{Pspec}) and the modding by the generalized Weyl
group. Precisely, with our conventions this means the following\footnote{Modding
is done by left multiplication because, if sitting on the left of
$\mathcal{O}$ the generalized Weyl group element
$\gamma$ will act on the Cartan element $\mathcal{C}$ by conjugation
$\gamma^T \, \mathcal{C} \, \gamma$ (see
eq.(\ref{Lzero})).}:
\begin{equation}
  \forall \, \gamma \, \in \, \mathcal{W}(\slal(3))
  \quad \mbox{and}\quad \forall \, \mathcal{O} \,\in \,
  \mathrm{SO(3)}\quad \, :
  \quad \,\,\,\gamma \, \mathcal{O} \,\sim \, \mathcal{O}~.
\label{identifia}
\end{equation}
In terms of the matrix entries $O_{ij}$ the operation
(\ref{identifia}) is quite simple, it just implies that all
orthogonal matrices which differ by an arbitrary permutation of the
rows accompanied by overall changes of signs rows by rows are to be
identified. On the other hand transferring the multiplication by
$\gamma$ on the theta angles is a highly non trivial and
complicated operation. In other words the map
\begin{equation}
  \theta_i \, \rightarrow \, f_\gamma^i\left( \theta\right)
\label{fmap}
\end{equation}
defined by:
\begin{equation}
  \mathcal{O}\left(f_\gamma^i\left( \theta\right) \right) \, = \,
  \gamma \, \mathcal{O}\left(\theta \right)
\label{fmap2}
\end{equation}
is quite involved and not handy. This implies that displaying a
fundamental cell in $\theta$-space is not an easy task and does not
lead to any illuminating picture. This is no serious problem, since
it is just a coordinate artifact. Furthermore precisely since we are
finally interested in equivalence classes with respect to the
algebraic Weyl group, $i.e.$ in sets of $4$-matrices of the form
\begin{equation}
  \mathrm{N}(\slal(3)) \, \mathcal{O}
\label{Nequiclassi}
\end{equation}
then it just suffices to identify a minimal  neighborhood of $\mathbb{R}^3$
in the open chart of the group manifold $\mathrm{SO(3)}$ defined by the Euler angle
parameterization (\ref{Pspec}) such that it contains at least one
copy of each Weyl group element $w_i \, \in \, \mathrm{Weyl}(\slal(3))$.
An example of such a minimal submanifold is provided by the  cube
$0\le \theta_i \le \frac{\pi}{2}$ shown in fig.\ref{cuboide} whose
vertices are just the required representatives of the Weyl group
elements. In all cases we can focus our attention on the hypercube in
Euler angle spaces defined by the vertices which correspond to the
nearest copies of all the Weyl group elements. We stress that these
hypercubes are not fundamental cells for the equivalence classes
$\mathrm{H}/\mathcal{W}(\mathbb{U})$ but are just sufficient for our
purposes, in particular in order to study the flow diagram produced
by links, namely flows on one dimensional trapped surfaces.
\subsection{The flow diagram and the critical surfaces for $\mathrm{SL(3,\mathbb{R})}$}
We can now explore the behaviour of Lax equation on the vertices, the
edges and the interior of the parameter space we have described in
the previous section.
\paragraph{Vertices}
As we know from the general properties of the
integral discussed in section 5 if the Lax operator lies
in the Cartan subalgebra at the initial point $t=0$, namely it is
diagonal it will remain constant all the time from $-\infty$ to
$+\infty$. Hence on each vertex of the cube, which corresponds to a Weyl
group element, we have constant Lax operators, corresponding to as
many Kasner epochs.
\par
\begin{figure}[!hbt]
\begin{center}
\iffigs
 \includegraphics[height=60mm]{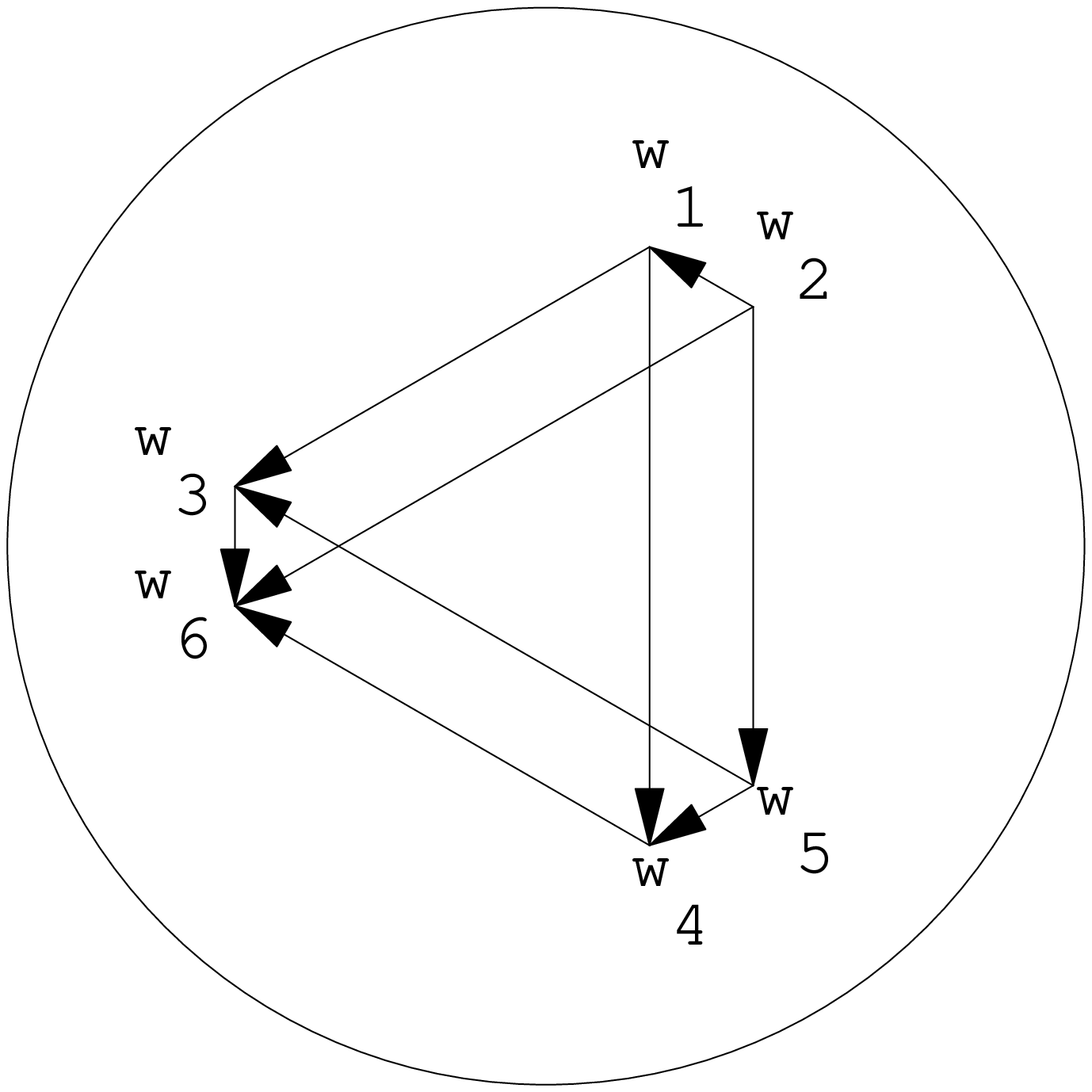}
 \includegraphics[height=65mm]{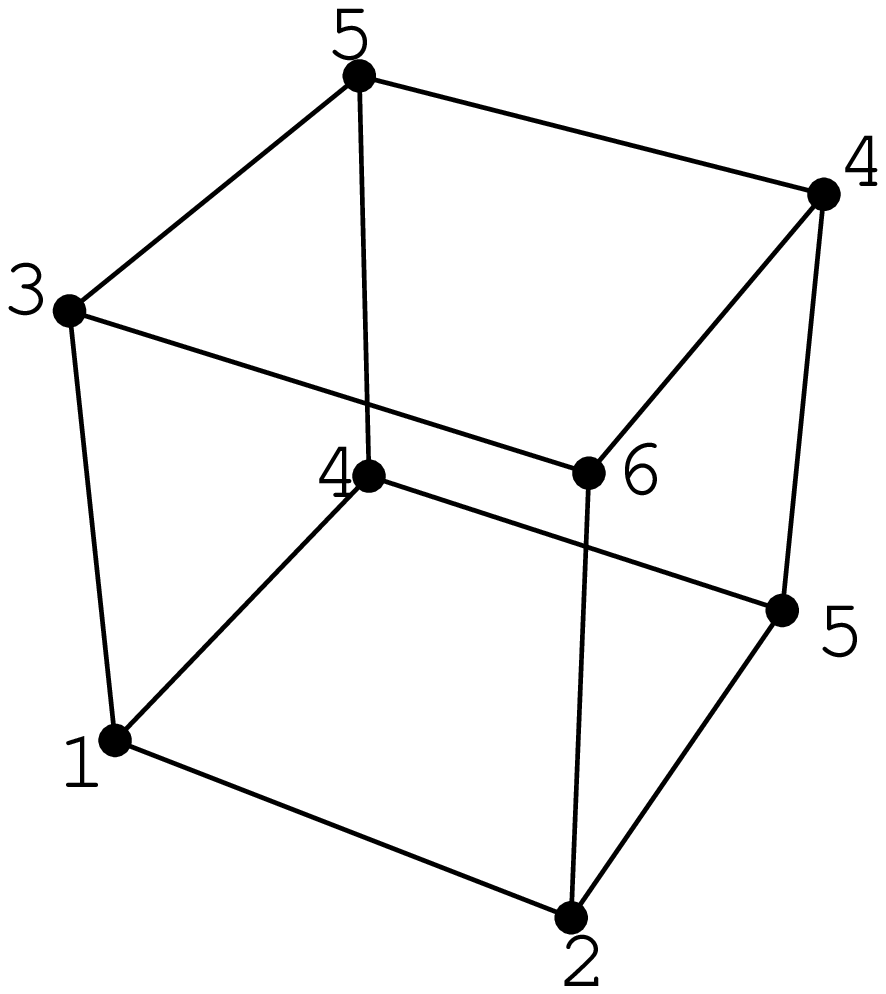}
\else
\end{center}
 \fi
\caption{\it The oriented diagram of the $\mathrm{SL(3,\mathbb{R})}/\mathrm{SO(3)}$ flows.
The Lie algebra $\mathrm{sl(3,\mathbb{R})}$ is the
maximally split real section of the complex Lie algebra $\mathrm{A}_2$.
Its Weyl group is $\mathrm{S}_3$ and has six elements identified by their action on
the eigenvalues $\lambda_1,\lambda_2,\lambda_3$ of the Lax operator. Six are
therefore the possible asymptotic states of the universe at
$\pm\infty$ and each possible motion is an oriented flow from one
\textbf{lower} Weyl element to another \textbf{higher} one. The lines of the graph on the right represent possible oriented
flows along one-dimensional submanifolds of the parameter space located on the edges of the cube defined
by restricting the range of the three Euler angles $\left\{\theta_1,\theta_2,\theta_3 \right\}$
to the closed interval $\left[ 0,\ft{\pi}{2}\right] $. On the vertices of the cube we
find $\mathrm{SO(3)}$ group elements lying in the Weyl group (modulo the center $\mathbb{Z}^3_2$),
just as shown in the three-dimensional picture on the
right. In the two-dimensional picture on the left, by choosing as fundamental
eigenvalues $\lambda_1=1,\lambda_2=2,\lambda_3=-3$ the Weyl group element
$w_i\,  \in \, \mathcal{W} $ is identified by the point in the plane that has coordinates
equal to the projections of $w_i\,(\lambda _1, \lambda _2, \lambda _3)$
along an orthonormal basis of vectors spanning the plane orthogonal to $(1,1,1).$ }
\label{sl3graphus}
 \iffigs
 \hskip 1.5cm \unitlength=1.1mm
 \end{center}
  \fi
\end{figure}

\paragraph{Edges}
It is interesting to see what happens on the
twelve edges of the cube. Let us display the form of the matrix
$\mathcal{O}$ on each of these edges.
\begin{equation}
  \begin{array}{ccccc}
    1) & (000) & \leftrightarrow & (100) & \mathcal{O}\, = \,\left(
\begin{array}{lll}
 \cos \left(\theta _1\right) & \sin \left(\theta _1\right) & 0
   \\
 -\sin \left(\theta _1\right) & \cos \left(\theta _1\right) & 0
   \\
 0 & 0 & 1
\end{array}
\right)~, \\
    2) & (000) & \leftrightarrow & (010) & \mathcal{O}\, = \,\left(
\begin{array}{lll}
 1 & 0 & 0 \\
 0 & \cos \left(\theta
   _2\right) & \sin
   \left(\theta _2\right) \\
 0 & -\sin \left(\theta
   _2\right) & \cos
   \left(\theta _2\right)
\end{array}
\right)~, \\
    3) & (000) & \leftrightarrow & (001) & \mathcal{O}\, = \,\left(
\begin{array}{lll}
 \cos \left(\theta _3\right) &
   0 & \sin \left(\theta
   _3\right) \\
 0 & 1 & 0 \\
 -\sin \left(\theta _3\right) &
   0 & \cos \left(\theta
   _3\right)
\end{array}
\right)~, \
\end{array}
\label{linkaggi1}
\end{equation}
\begin{equation}
  \begin{array}{ccccc}
    4) & (100) & \leftrightarrow & (110) & \mathcal{O}\, = \,\left(
\begin{array}{lll}
 0 & \cos \left(\theta
   _2\right) & \sin
   \left(\theta _2\right) \\
 -1 & 0 & 0 \\
 0 & -\sin \left(\theta
   _2\right) & \cos
   \left(\theta _2\right)
\end{array}
\right)~, \\
    5) & (100) & \leftrightarrow & (101) & \mathcal{O}\, = \,\left(
\begin{array}{lll}
 0 & 1 & 0 \\
 -\cos \left(\theta _3\right) &
   0 & -\sin \left(\theta
   _3\right) \\
 -\sin \left(\theta _3\right) &
   0 & \cos \left(\theta
   _3\right)
\end{array}
\right)~, \\
    6) & (010) & \leftrightarrow & (011) & \mathcal{O}\, = \,\left(
\begin{array}{lll}
 \cos \left(\theta _3\right) &
   0 & \sin \left(\theta
   _3\right) \\
 -\sin \left(\theta _3\right) &
   0 & \cos \left(\theta
   _3\right) \\
 0 & -1 & 0
\end{array}
\right)~, \\
\end{array}
\label{linkaggi2}
\end{equation}
\begin{equation}
  \begin{array}{ccccc}
    7) & (010) & \leftrightarrow & (110) & \mathcal{O}\,=\, \left(
\begin{array}{lll}
 \cos \left(\theta _1\right) &
   0 & \sin \left(\theta
   _1\right) \\
 -\sin \left(\theta _1\right) &
   0 & \cos \left(\theta
   _1\right) \\
 0 & -1 & 0
\end{array}
\right)~, \\
    8) & (001) & \leftrightarrow & (101) & \mathcal{O}\,=\,\left(
\begin{array}{lll}
 0 & \sin \left(\theta
   _1\right) & \cos
   \left(\theta _1\right) \\
 0 & \cos \left(\theta
   _1\right) & -\sin
   \left(\theta _1\right) \\
 -1 & 0 & 0
\end{array}
\right)~, \\
    9) & (001) & \leftrightarrow & (011) & \mathcal{O}\,=\,\left(
\begin{array}{lll}
 0 & 0 & 1 \\
 -\sin \left(\theta _2\right) &
   \cos \left(\theta _2\right)
   & 0 \\
 -\cos \left(\theta _2\right) &
   -\sin \left(\theta _2\right)
   & 0
\end{array}
\right)~, \
\end{array}
\label{linkaggi3}
\end{equation}
\begin{equation}
  \begin{array}{ccccc}
    10) & (110) & \leftrightarrow & (111) & \mathcal{O}\,=\,\left(
\begin{array}{lll}
 -\sin \left(\theta _3\right) &
   0 & \cos \left(\theta
   _3\right) \\
 -\cos \left(\theta _3\right) &
   0 & -\sin \left(\theta
   _3\right) \\
 0 & -1 & 0
\end{array}
\right)~, \\
    11) & (011) & \leftrightarrow & (111) & \mathcal{O}\,=\,\left(
\begin{array}{lll}
 -\sin \left(\theta _1\right) &
   0 & \cos \left(\theta
   _1\right) \\
 -\cos \left(\theta _1\right) &
   0 & -\sin \left(\theta
   _1\right) \\
 0 & -1 & 0
\end{array}
\right)~, \\
    12) & (101) & \leftrightarrow & (111) & \mathcal{O}\,=\, \left(
\begin{array}{lll}
 -\sin \left(\theta _2\right) &
   \cos \left(\theta _2\right)
   & 0 \\
 0 & 0 & -1 \\
 -\cos \left(\theta _2\right) &
   -\sin \left(\theta _2\right)
   & 0
\end{array}
\right). \
  \end{array}
\label{linkaggi4}
\end{equation}
On each link we have one of the three one-parameter subgroups
respectively generated by $J_{1,2,3}$ multiplied on the left or on
the right by a Weyl group element.
By means of a computer programme we can then easily evaluate the
general integral on each of these links.
For instance on the link number $1$ we obtain:
\begin{eqnarray}
  L(t)& = & \left( \begin{array}{ccc}
    L_{11}(t) & L_{12}(t) & 0 \\
    L_{12}(t) & L_{22}(t) & 0 \\
    0 & 0 & -\lambda_1\, - \, \lambda_2 \
  \end{array}\right)~,\nonumber\\
L_{11}(t)&=& \frac{e^{2 t \lambda _2} \lambda _1 \cos ^2\left(\theta _1\right)+e^{2 t \lambda
   _1} \sin ^2\left(\theta _1\right) \lambda _2}{e^{2 t \lambda _2} \cos
   ^2\left(\theta _1\right)+e^{2 t \lambda _1} \sin ^2\left(\theta
   _1\right)}~,\nonumber\\
 L_{22}(t)&=&  \frac{e^{2 t \lambda _2} \lambda _2 \cos ^2\left(\theta _1\right)+e^{2 t \lambda
   _1} \sin ^2\left(\theta _1\right) \lambda _1}{e^{2 t \lambda _2} \cos
   ^2\left(\theta _1\right)+e^{2 t \lambda _1} \sin ^2\left(\theta
   _1\right)}~,\nonumber\\
 L_{12}(t) &=& \frac{e^{t \left(\lambda _1+\lambda _2\right)} \sin \left(2 \theta _1\right)
   \left(\lambda _1-\lambda _2\right)}{\left(-e^{2 t \lambda _1}+e^{2 t \lambda
   _2}\right) \cos \left(2 \theta _1\right)+e^{2 t \lambda _1}+e^{2 t \lambda _2}}~.
\label{pensaunpo}
\end{eqnarray}
We can also calculate the asymptotic limits of the Lax operator at
$\pm\infty$ for each of these flows. As it follows from the
properties of the general integral, at remotely early or at remotely late
times the Lax operator is always diagonal and its eigenvalues are
organized in one of the possible six ways corresponding to the six Weyl
group elements acting on their reference order $
(\lambda_1,\lambda_2,-\lambda_1 \, - \, \lambda_3)$. If we associate
an arrow to each of these twelve links and we take into account the
identification of vertices as displayed in eq.(\ref{weylidentificazie})
we obtain the flow diagram shown in fig.\ref{sl3graphus}.
As it is clear from the quoted picture, the flows on the edges of the
cube relate states of the Universe where there is no complete sorting
of the eigenvalues at past and future infinities. Indeed if we use
as reference set the eigenvalues of eq.(\ref{convenset})
then complete sorting would require $w_2\, \in \, \mathcal{W}$ at
$-\infty$, corresponding to the decreasing ordering $2,1,-3$ and
$w_6$ at $+\infty$ corresponding to the increasing ordering $-3,1,2$
as we already observed. As it is evident by inspection, the matrices
located on the twelve edges define one-dimensional
critical surfaces. It is a a fundamental property of the Lax equation
that flows touching upon a critical surface are completely constrained on it.
Hence flows touching one link just lie on that link at all instants of
time and the asymptotic states correspond to the vertices located at the
endpoints of that link. The orientation of the link is also decided a
priori. Past infinity is the lower of the two end point Weyl elements
while future infinity is the higher one. This is just evident by
comparing the graph in fig.(\ref{sl3graphus}) with the ordering of
Weyl group elements as displayed in table \ref{SL3Weylordo}.
\par
\begin{figure}[!hbt]
\begin{center}
\iffigs
\includegraphics[height=70mm]{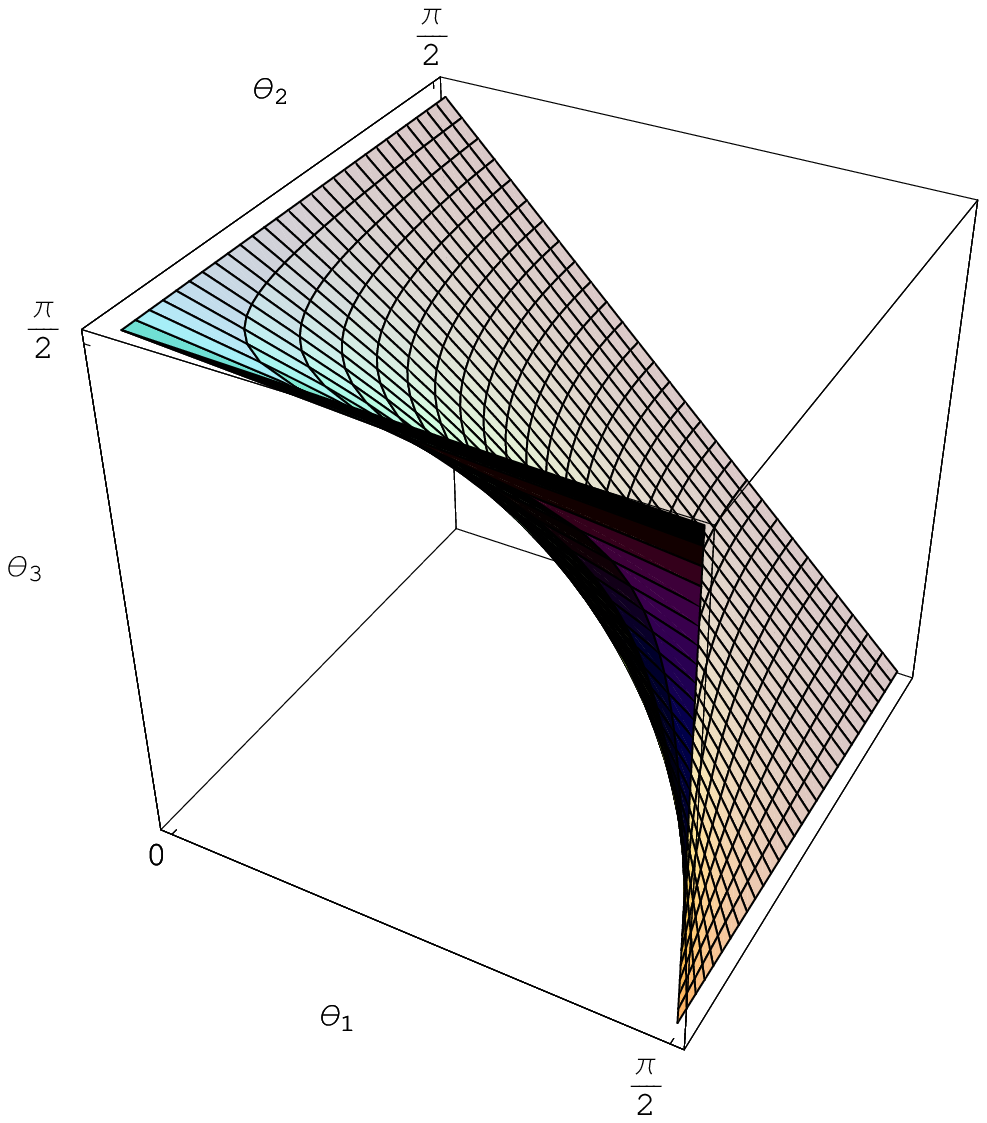}
 \includegraphics[height=73mm]{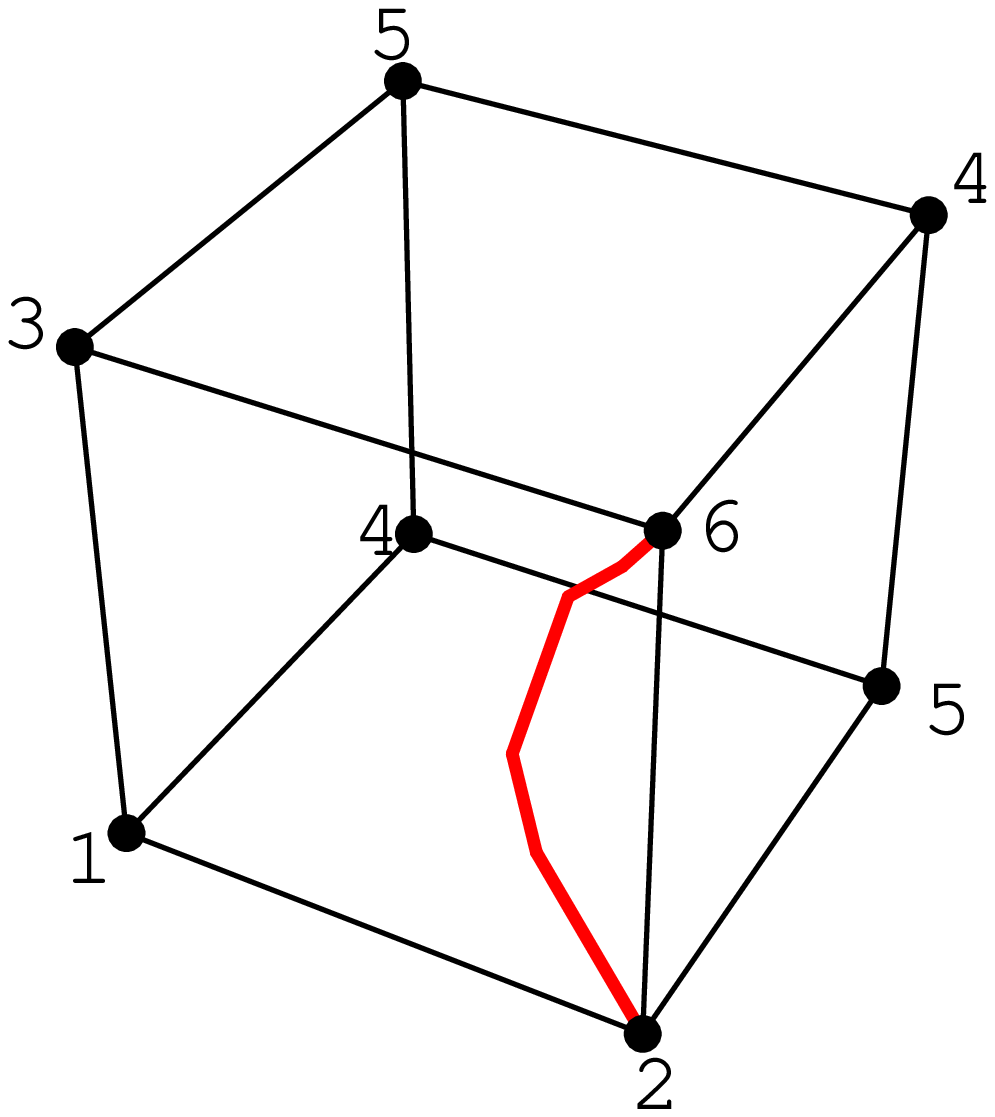}
\else
\end{center}
 \fi
\caption{\it The picture on the left shows the  critical surfaces $\Sigma_1$
defined by the equation $O_{1,1}=0$ imposed on
the  $\mathrm{SO(3)}$ group
element. The picture on the right reminds
the reader of the Weyl group elements located
at the vertices of the parameter space. The vertices belonging
the surface are (in increasing order) $w_2,\, w_5, \, w_3, \, w_6$ so
that the flow is $w_2 \, \mapsto \, w_6$.}
\label{sl3ritical1}
 \iffigs
 \hskip 1.5cm \unitlength=1.1mm
 \end{center}
  \fi
\end{figure}
\begin{figure}[!hbt]
\begin{center}
\iffigs
\includegraphics[height=70mm]{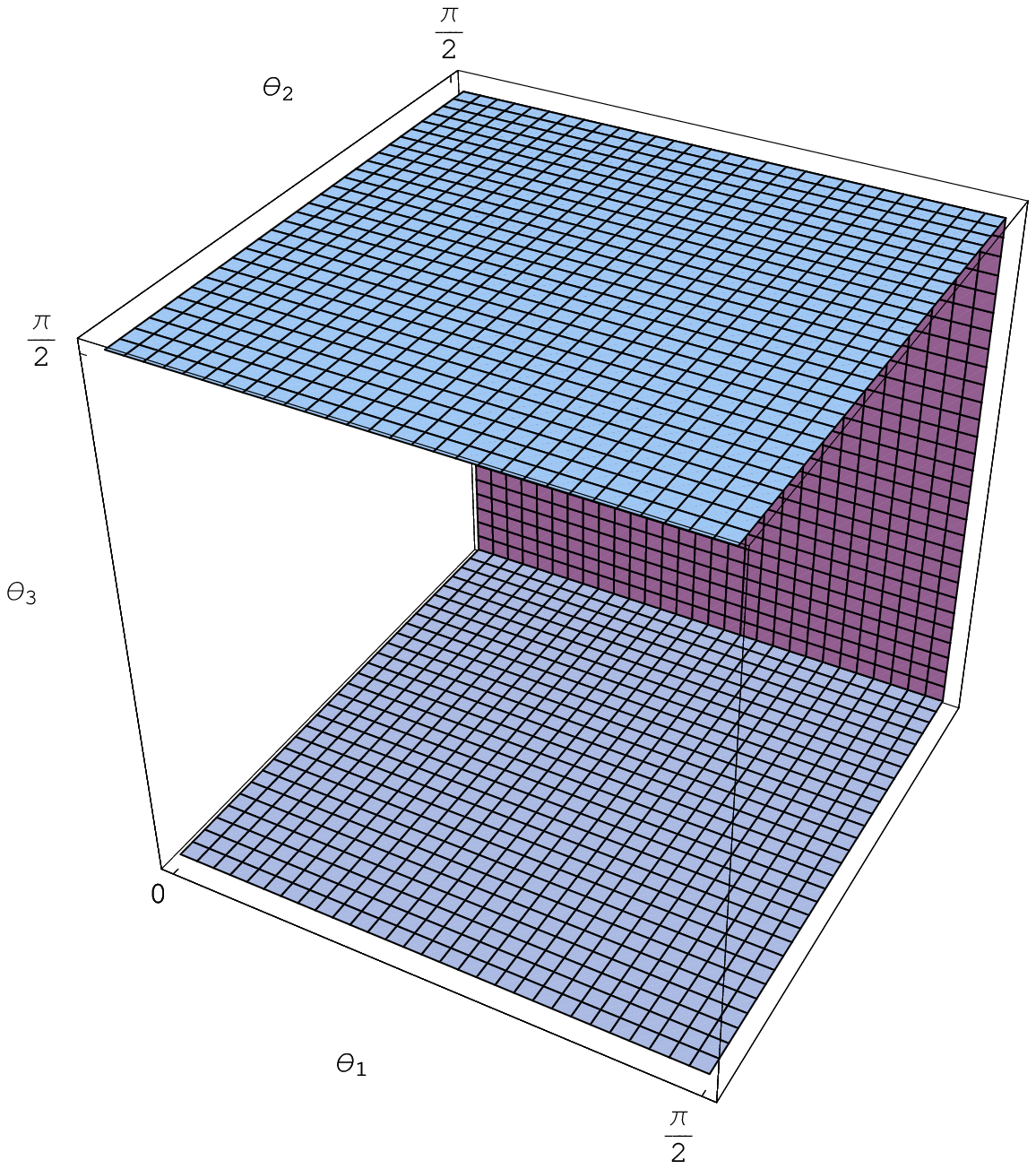}
 \includegraphics[height=73mm]{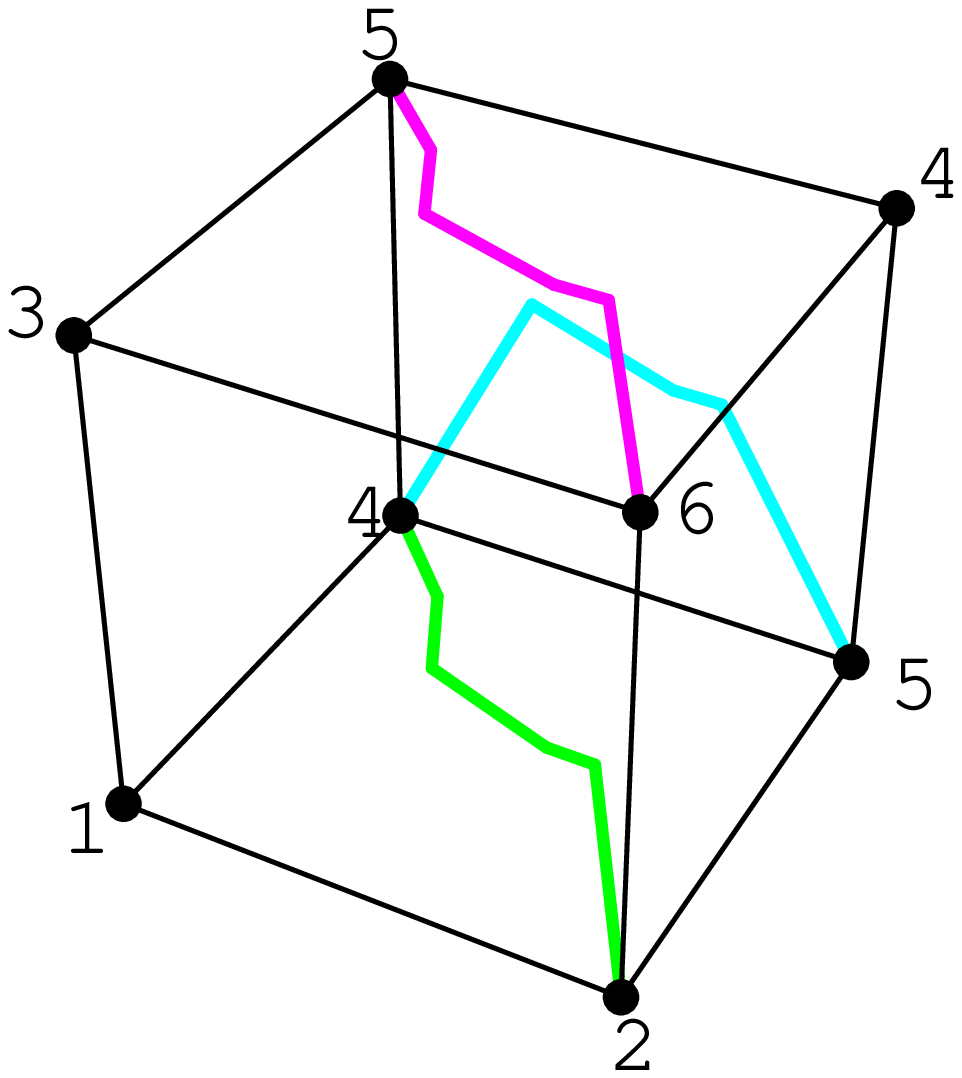}
\else
\end{center}
 \fi
\caption{\it The picture on the left shows the union of
the critical surfaces $\Sigma_3$ and $\Sigma_4$,
respectively
defined by the equations $O_{3,1}=0$ and
$O_{1,1} O_{2,2}-O_{1,2} O_{2,1}=0$ imposed on
the minors of the $\mathrm{SO(3)}$ group
element. The picture on the right reminds
the reader of the Weyl group elements located
at the vertices of the parameter space
and shows the possible oriented flows
on the critical surfaces. The vertices belonging to $\Sigma_3$
are $w_2,\, w_1, \, w_5, \, w_4$  and the flow on this surface
is $w_2 \, \mapsto \, w_4$. The vertices belonging to $\Sigma_4$
are instead $w_5,\, w_3, \, w_4, \, w_6$  and the flow on this surface
is $w_5 \, \mapsto \, w_6$. The plaquette $\theta_2 = \frac{\pi}{2}$ is actually the
intersection $\Sigma_3 \, \bigcap \, \Sigma_4$ and on this surfaces the flow
goes from the lowest to the highest of the elements in the set
of the vertices accessible to both surfaces, namely we
have $w_5\, \mapsto \, w_4$.}
\label{sl3ritical34}
 \iffigs
 \hskip 1.5cm \unitlength=1.1mm
 \end{center}
  \fi
\end{figure}
\begin{figure}[!hbt]
\begin{center}
\iffigs
\includegraphics[height=70mm]{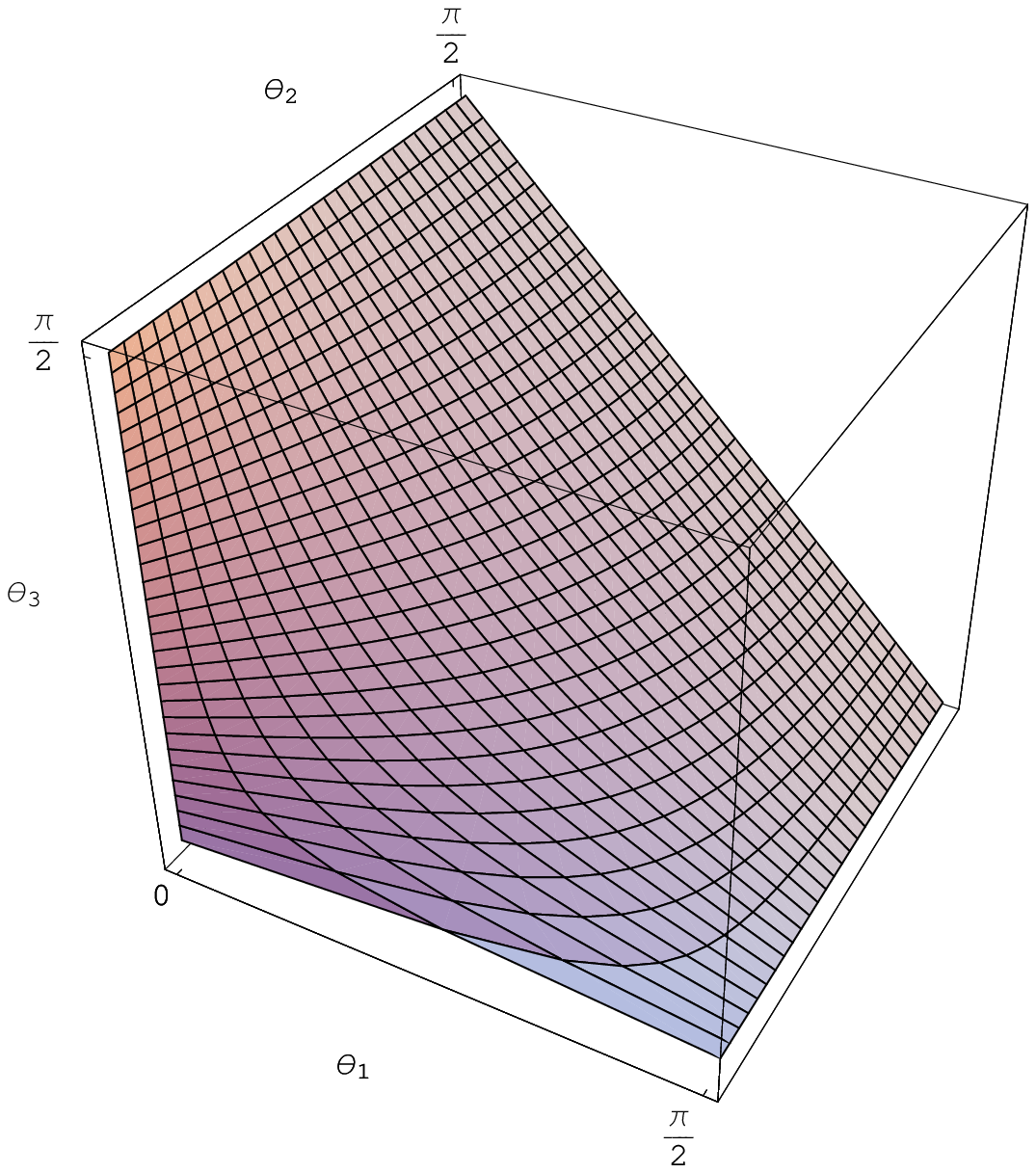}
 \includegraphics[height=73mm]{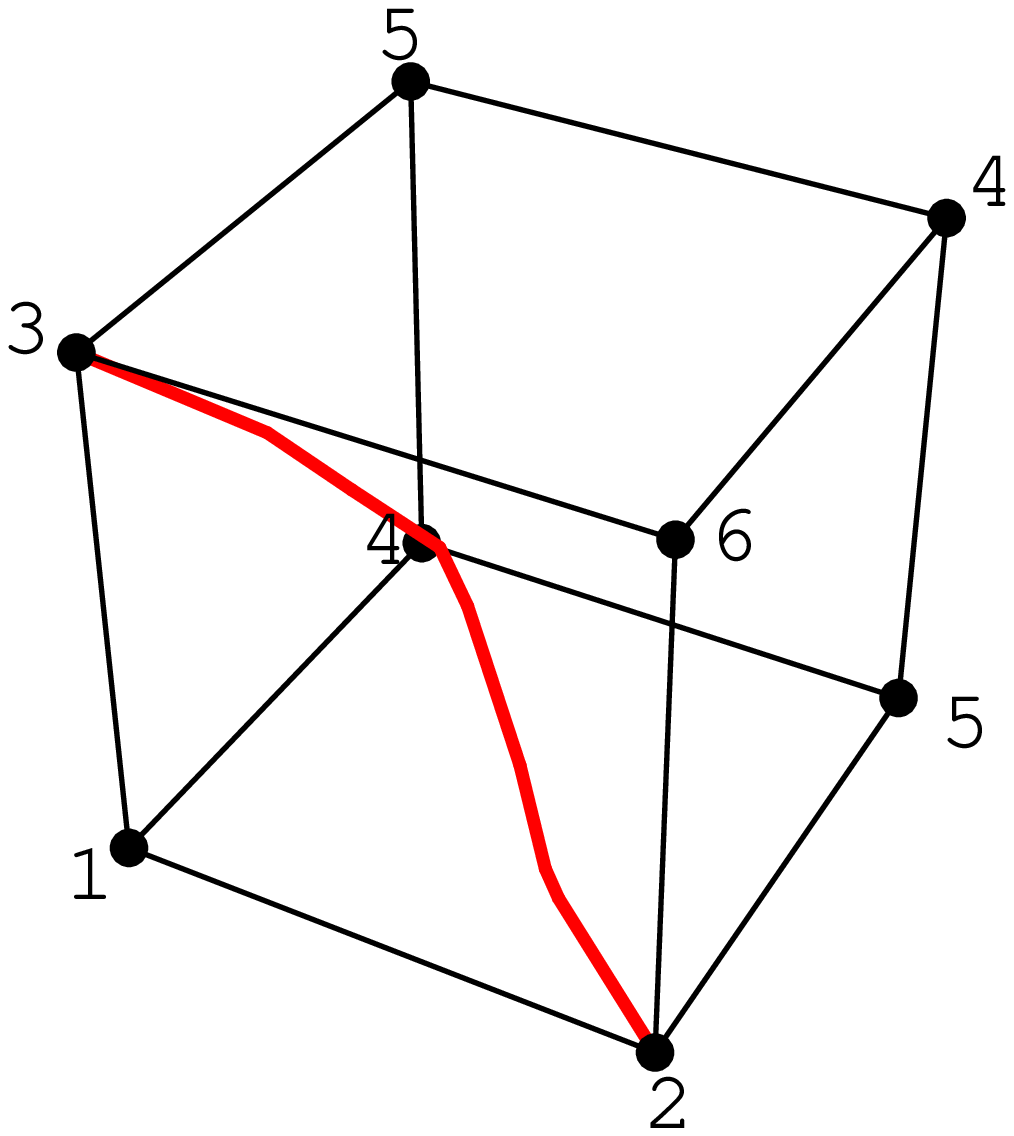}
\else
\end{center}
 \fi
\caption{\it The picture on the left shows the  critical surfaces $\Sigma_5$
defined by the equation $O_{1,1} O_{3,2}-O_{1,2} O_{3,1}=0$ imposed on
the  $\mathrm{SO(3)}$ group
element. The identification of the cube vertices with Weyl group
elements is shown on the right. Here the accessible vertices are
$w_2,\, w_1, \, w_5, \, w_3$  and the flow on this surface
is $w_2 \, \mapsto \, w_3$.}
\label{sl3ritical5}
 \iffigs
 \hskip 1.5cm \unitlength=1.1mm
 \end{center}
  \fi
\end{figure}

Besides one-dimensional critical surfaces (the links) there are also
two-dimensional ones (the faces) and these are obtained by studying the minors of
$\mathcal{O}$.
\paragraph{Faces}
Let us consider the orthogonal matrix $\mathcal{O} \, \in \, \mathrm{SO(3)}$
and let us name and parameterize its entries as in
eq.s(\ref{Pspec},\ref{Pmatraelelmenta}). Then there are in general
exactly $6$ minors that correspond to the conditions involved in the
definition of trapped surfaces in Section 5.2. Since we are dealing
with $\slal(3)$ all trapped surfaces are also critical. Three of the
relevant minors are $1 \times 1$
minors and three of them are $2\times 2$ minors. Imposing their
vanishing one obtains  equations on the three parameters
$\theta_1,\theta_2,\theta_3$ which would define as many critical surfaces,
namely six. Let us enumerate these candidate trapped and critical surfaces
\begin{equation}
  \begin{array}{ccrcccl}
    \Sigma_1 & : & O_{1,1} & = & 0 & = & \cos \left(\theta _1\right) \cos
   \left(\theta _3\right)-\sin
   \left(\theta _1\right) \sin
   \left(\theta _2\right) \sin
   \left(\theta _3\right)~,   \\
    \Sigma_2 & : & O_{2,1} & = & 0 & = & -\cos \left(\theta _3\right) \sin
   \left(\theta _1\right)-\cos
   \left(\theta _1\right) \sin
   \left(\theta _2\right) \sin
   \left(\theta _3\right)~, \\
    \Sigma_3 & : & O_{3,1} & = & 0 & = & -\cos \left(\theta _2\right) \sin
   \left(\theta _3\right)~,  \\
    \Sigma_4 & : & O_{1,1} O_{2,2}-O_{1,2} O_{2,1} & = & 0 & = & \cos \left(\theta _2\right) \cos
   \left(\theta _3\right)~, \\
    \Sigma_5 & : & O_{1,1} O_{3,2}-O_{1,2} O_{3,1} & = & 0 & = & \sin \left(\theta _1\right) \sin
   \left(\theta _3\right)-\cos
   \left(\theta _1\right) \cos
   \left(\theta _3\right) \sin
   \left(\theta _2\right)~, \\
   \Sigma_6 & :  & O_{2,1} O_{3,2}-O_{2,2} O_{3,1} & =  & 0 & = & \cos \left(\theta _3\right) \sin
   \left(\theta _1\right) \sin
   \left(\theta _2\right)+\cos
   \left(\theta _1\right) \sin
   \left(\theta _3\right)~.  \
  \end{array}
\label{criticalfacce}
\end{equation}
It is now fairly simple to verify that, while the equations for
$\Sigma_1,\Sigma_3,\Sigma_4,$ and $\Sigma_5$ can be solved inside the
cube $0\le \theta_i \le \frac{\pi}{2}$, all solutions of
the equations for $\Sigma_2$ and $\Sigma_6$ are located outside this range.
The existing inside the cube critical surfaces are shown
in fig.s \ref{sl3ritical1}, \ref{sl3ritical34}, \ref{sl3ritical5}.
By means of a computer programme we can evaluate the flows on all
these critical surfaces and we find the following results for the
asymptotic values of the Lax operator:
\begin{equation}
  \begin{array}{ccrcl}
    \Sigma_1 & : & w_2 & \rightarrow & w_6
    \mbox{ (surface equation $\theta_1 = \arccos\left(\frac{\sin \left(\theta _2\right) \sin \left(\theta
   _3\right)}{\sqrt{\cos ^2\left(\theta _3\right)+\sin ^2\left(\theta
   _2\right) \sin ^2\left(\theta _3\right)}}\right) $)}~,\\
    \Sigma_3 &:& w_2 & \rightarrow & w_4 \mbox{ (for $\theta_3 = 0$)}~, \\
    \Sigma_4 &:& w_5 & \rightarrow & w_6 \mbox{ (for $\theta_3 = \ft{\pi}{2}$)}~, \\
    \Sigma_3 \, \bigcap \, \Sigma_4 &:& w_5 & \rightarrow & w_4 \mbox{ (for $\theta_2 = \ft{\pi}{2}$)}~, \\
    \Sigma_5 &:& w_2 & \rightarrow & w_3 \mbox{ (surface equation $\theta_3 =
    \arccos\left(\frac{\sin \left(\theta _1\right)}{\sqrt{\sin ^2\left(\theta _1\right)
    +\cos ^2\left(\theta _1\right)
   \sin ^2\left(\theta _2\right)}}\right) $)}~. \
  \end{array}
\label{universedestiny}
\end{equation}
The fourth case listed in eq.(\ref{universedestiny}) needs a comment.
When we set $\theta_2 = \ft{\pi}{2}$ it happens that both $O_{3,1}
=0$ and $O_{1,1}\,O_{2,2}\, - \, O_{1,2}\, O_{2,1} \, = \,0$. Hence
this plaquette of the hypercube is actually the intersection of two
critical surfaces. Altogether the result displayed in eq.(\ref{universedestiny})
could be predicted
a priori relying on  the notion of accessible vertices. Given the equation
of a critical surface, the accessible vertices are defined as those
Weyl elements which have at least one representative satisfying the
defining condition and therefore belong to the  surface.
Once the accessible set is defined, the flow is easily singled out. It
goes from the lowest Weyl member of the set to the highest one.
This task is easily carried through in the present case.
For the six surfaces defined in equation (\ref{criticalfacce}) the
corresponding accessible sets are rapidly calculated and we find the
result displayed in table \ref{sl3accessible}.
\begin{table}
  \centering
  $$
\begin{array}{|c|c|c|c|}
\hline
\mbox{Surf.}  & \mbox{Equation} & \mbox{Accessible Vertex} & \mbox{Flow}\\
\hline
  \Sigma_1 & O_{1,1}\,=\, 0 & \left( \begin{array}{cc}
    0 & w_2 \\
    1 & w_5 \\
    2 & w_3 \\
    3 & w_6 \
  \end{array}\right)  & w_2 \, \mapsto \, w_6 \\
  \hline
  \Sigma_2  & O_{2,1}\,=\, 0 & \left( \begin{array}{cc}
    1 & w_1 \\
    2 & w_3 \\
    2 & w_4 \\
    3 & w_6 \
  \end{array}\right) & w_1 \, \mapsto \, w_6 \\
  \hline
  \Sigma_3  & O_{3,1}\,=\, 0 & \left( \begin{array}{cc}
    0 & w_2 \\
    1 & w_1 \\
    1 & w_5 \\
    2 & w_4 \
  \end{array}\right) & w_2 \, \mapsto \, w_4 \\
  \hline
  \Sigma_4  &  -O_{1,2}\, O_{2,1} \, + \, O_{1,1}\, O_{2,2} \,=\, 0 & \left( \begin{array}{cc}
    1 & w_5 \\
    2 & w_3 \\
    2 & w_4 \\
    3 & w_6 \
  \end{array}\right) & w_5 \, \mapsto \, w_6 \\
  \hline
  \Sigma_5  &  -O_{1,2}\, O_{3,1} \, + \, O_{1,1}\, O_{3,2} \,=\, 0 & \left( \begin{array}{cc}
    0 & w_2 \\
    1 & w_1 \\
    1 & w_5 \\
    2 & w_3 \
  \end{array}\right) & w_2 \, \mapsto \, w_3 \\
  \hline
  \Sigma_6  &  -O_{2,2}\, O_{3,1} \, + \, O_{2,1}\, O_{3,2} \,=\, 0 & \left( \begin{array}{cc}
    0 & w_2 \\
    1 & w_1 \\
    2 & w_4 \\
    3 & w_6 \
  \end{array}\right) & w_2 \, \mapsto \, w_6 \\
  \hline
\end{array}
  $$
  \caption{The accessible vertices for each of the six two-dimensional critical
  surfaces in the case $\mathrm{SL(3,\mathbb{R})/\mathrm{SO(3)}}$.}\label{sl3accessible}
\end{table}
Expunging surfaces $\Sigma_2$ and $\Sigma_6$ which fall outside
the cube, we find that the available flows on critical
two dimensional surfaces inside the cube are just only four,
namely the following ones:
\begin{eqnarray}
w_2 \, \mapsto \, w_6~, && \nonumber\\
w_2 \, \mapsto \, w_4~, && \nonumber\\
w_5 \, \mapsto \, w_6~, && \nonumber\\
w_2 \, \mapsto \, w_3~. &&
\label{possibleSl3flows}
\end{eqnarray}
The only non vanishing intersection of these surfaces is the afore-mentioned
plaquette $\Sigma_{3}\, \bigcap \, \Sigma_4$. We can easily calculate the
intersection of vertices accessible to both $\Sigma_3$ and
$\Sigma_4$. We find
\begin{equation}
  \left\{ w_2,w_1,w_5,w_4\right\} \,\bigcap \,\left\{w_5,w_3,w_4,w_6 \right\}
  \, = \, \left\{w_5 , w_4 \right\}
\label{intersetto}
\end{equation}
where all sets are written in ascending order. It follows that on
the surface $\Sigma_{3}\, \bigcap \, \Sigma_4$ the oriented flow is
\begin{equation}
  w_5 \, \mapsto \, w_4
\label{w5tow4}
\end{equation}
as indeed it is verified by numerical calculation on the computer.
\par
This concludes our discussion of the $\mathrm{SL(3,\mathbb{R})}$
which has been instrumental to illustrate the involved mathematical
structures.
\par
We have seen that the topology of the parameter space $\mathrm{H}/\mathcal{W}(\mathbb{U})$
is indeed complicated and cannot be easily displayed as an hypercube.
Yet it is completely defined by the trapped hypersurfaces which admit
a clear definition in terms of algebraic equations. These surfaces
split the parameter space  $\mathrm{H}/\mathcal{W}(\mathbb{U})$ into convex
hulls which are separated from each other. Indeed the walls are impenetrable
according to Toda evolution. Moreover
we can relate the initial and final states of the flows to these
critical surfaces defined by the vanishing of the relevant minors
in the orthogonal matrix $\mathcal{O}$.
\par
Our next section is devoted to another maximal split case of rank
$r=2$ which will correspond to an entire Tits Satake universality
class of cases.
\section{The maximally split case  $\mathrm{Sp(4,\mathbb{R})}/\mathrm{U(2)}$}
\label{sectsp4}
As we explained in the introduction our goal is the illustration of
the Lax integration formula in the non-maximally split case
$\mathrm{SO(r,r+2s)}/\mathrm{SO(r)}\times \mathrm{SO(r+2s)}$. The
Tits Satake projection of these manifolds is provided by the
maximally split coset
\begin{equation}
  \mathcal{M}^{TS}_r \, \equiv \,
  \frac{\mathrm{SO(r,r+1)}}{\mathrm{SO(r)} \times \mathrm{SO(r+1)}}~.
\label{MTSr}
\end{equation}
In the case of rank $r=2$ we have
\begin{equation}
  \mathcal{M}^{TS}_2 \, \equiv \, \frac{\mathrm{SO(2,3)}}{\mathrm{SO(2)} \times
  \mathrm{SO(3)}}\,
  =\,\frac{\mathrm{Sp(4,\mathbb{R})}}{\mathrm{U(2)}}
\label{sp4u2}
\end{equation}
due to the accidental isomorphism between the $B_2$ and $C_2$ Lie
algebras whose Dynkin diagram is displayed in fig. \ref{B2dynk}.
\begin{figure}
\caption{\it The Dynkin diagram of the  $B_2\sim C_2$ Lie algebra.
\label{B2dynk}}
\centering
\begin{picture}(10,100)
\put (-70,65){$C_2$}
\put (-25,70){\circle {10}}
\put (-33,55){$\alpha_1$}
\put (-20,69){\line (1,0){30}}
\put (-10,69){\line (1,-1){10}}
\put (-10,72){\line (1,1){10}}
\put (-20,72){\line (1,0){30}}
\put (15,70){\circle {10}}
\put (14,55){$\alpha_2$}
\put (-70,35){$B_2$}
\put (-25,40){\circle {10}}
\put (-33,25){$\alpha_1$}
\put (-20,39){\line (1,0){30}}
\put (-5,39){\line (-1,-1){10}}
\put (-5,42){\line (-1,1){10}}
\put (-20,42){\line (1,0){30}}
\put (15,40){\circle {10}}
\put (14,25){$\alpha_2$}
\end{picture}
\end{figure}
For this reason we can rely on either formulation in terms of $4
\times 4$ symplectic matrices or $5\times 5$ pseudo-orthogonal matrices
to obtain the same result.
\par
\begin{figure}[!hbt]
\begin{center}
\iffigs
 \includegraphics[height=70mm]{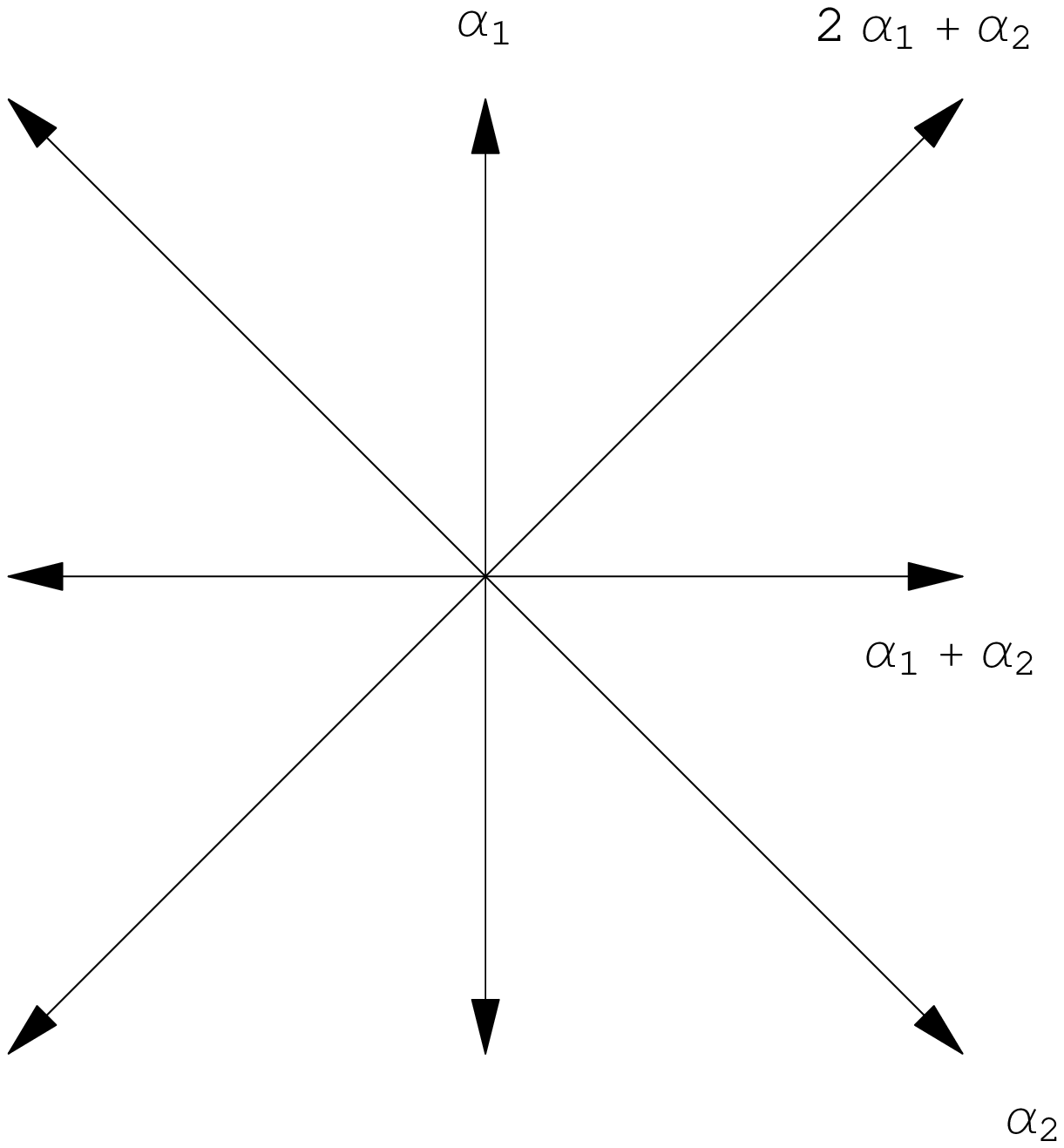}
 \includegraphics[height=70mm]{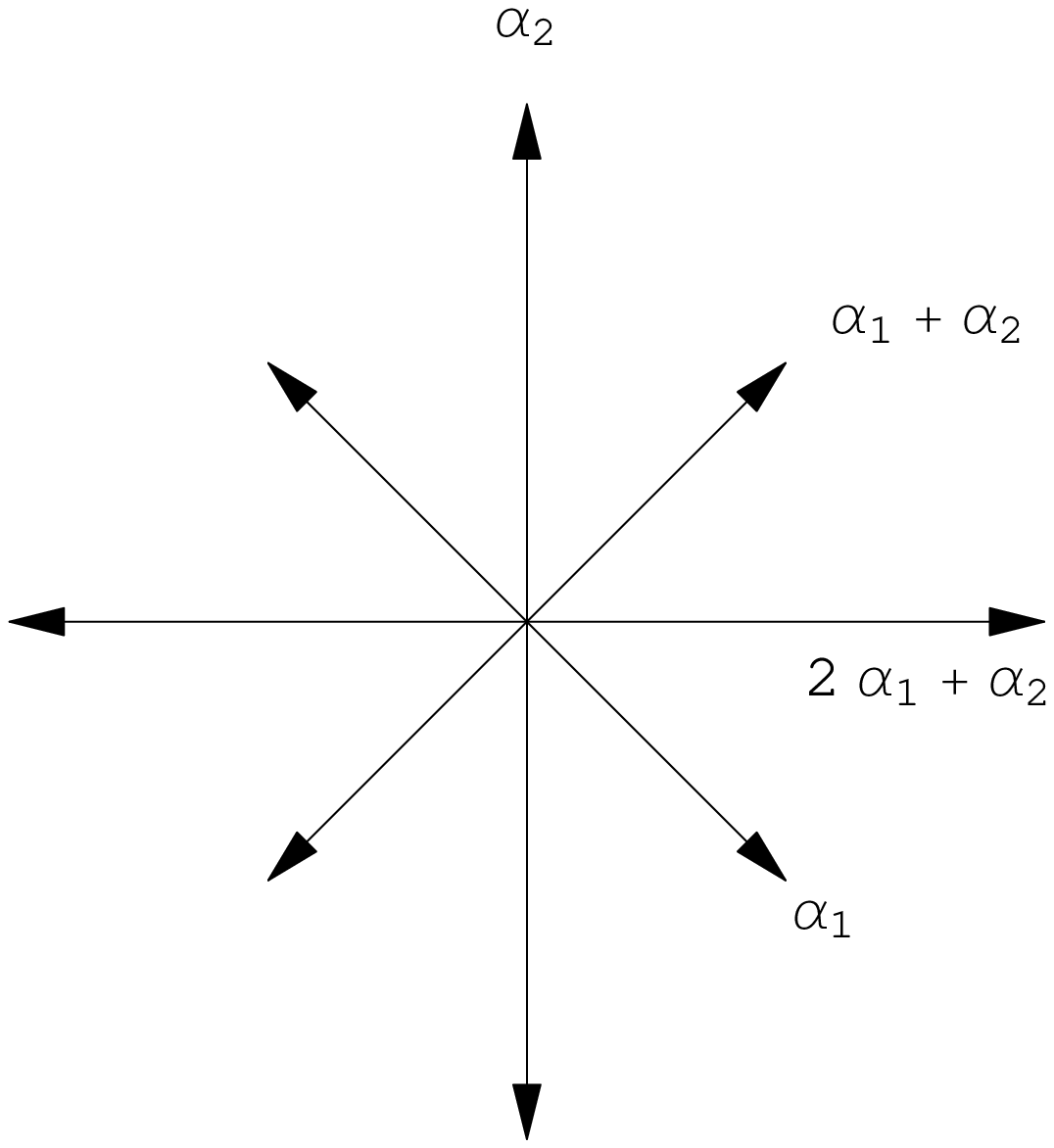}
\else
\end{center}
 \fi
\caption{\it The $C_2$ and $B_2$ root systems. They are related by the exchange
of the long with the short roots and viceversa.}
\label{c2b2rutte}
 \iffigs
 \hskip 1.5cm \unitlength=1.1mm
 \end{center}
  \fi
\end{figure}

In the symplectic $\sym(4)$ interpretation, the $C_2$ root system can be
realized by the following eight two-dimensional vectors:
\begin{equation}
  \Delta_{C_2} = \left\{  \pm \epsilon^{i} \, \pm \, \epsilon^j \, ,
  \, \pm \epsilon^i \right\}
\label{Deltac2}
\end{equation}
where $\epsilon^i \, (i=1,2)$ denotes a basis of orthonormal unit
vectors. In the pseudo-orthogonal $\so(2,3)$ interpretation of the
same algebra the $B_2$ root system is instead realized by the
following eight vectors:
\begin{equation}
  \Delta_{B_2} = \left\{  \pm \epsilon^{i} \, \pm \, \epsilon^j \, ,
  \, \pm 2 \, \epsilon^i \right\}~.
\label{Deltab2}
\end{equation}
The two root systems are displayed in fig. \ref{c2b2rutte}.
\subsection{The Weyl group and the generalized Weyl group of $\sym(4,\mathbb{R})$}
Abstractly the Weyl group $\mathrm{Weyl}(C_2)$ of the Lie algebra $\sym(4,\mathbb{R})$ is given by $(\mathbb{Z}_2 \times \mathbb{Z}_2) \, \ltimes \,
\mathrm{S}_2$ and its eight elements $w_i \in \mathrm{Weyl}(C_2)$ can be described by
their action on the two Cartan fields $h_1,h_2$
\begin{equation}
\begin{array}{ccccl}
w_1 & : & (h_1,h_2) & \rightarrow & (h_1,h_2)~, \\
w_2 & : & (h_1,h_2) & \rightarrow & (-h_1,-h_2)~, \\
w_3 & : & (h_1,h_2) & \rightarrow & (-h_1,h_2)~, \\
w_4 & : & (h_1,h_2) & \rightarrow & (h_1,-h_2)~, \\
w_5 & : & (h_1,h_2) & \rightarrow & (h_2,h_1)~, \\
w_6 & : & (h_1,h_2) & \rightarrow & (h_2,-h_1)~, \\
w_7 & : & (h_1,h_2) & \rightarrow & (-h_2,h_1)~, \\
w_8 & : & (h_1,h_2) & \rightarrow & (-h_2,-h_1)~. \\
\end{array}
\label{weylsp4intri}
\end{equation}
Just as we did in the previous case study we can introduce a partial
ordering of the Weyl group elements which will govern the orientation
of all dynamical flows. The key point is the embedding
\begin{equation}
  \mathrm{Weyl}\left(\sym(4)\right) \, \hookrightarrow \, \mathcal{S}_4
  \, \simeq \, \mathrm{Weyl}\left( \slal(4)\right)
\label{embeddusweylus}
\end{equation}
of the Weyl group into the symmetric group
$\mathcal{S}_4$ induced by the $4 \times 4$ representation of the Lie
algebra $\sym(4)$ in which the solvable Lie algebra is made of upper
triangular matrices. The explicit construction of such a representation
is performed in the next section.  For the purpose of the
considered issue, namely discovering the structure of the generalized Weyl group
and ordering of Weyl group elements, we anticipate some results.
The matrices corresponding to Cartan subalgebra elements are of the following
form:
\begin{equation}
  \mbox{CSA} \, \ni \, \left(\begin{array}{cccc}
    h_1 & 0 & 0 & 0 \\
    0 & h_2 & 0 & 0 \\
    0 & 0 & -h_2 & 0 \\
    0 & 0 & 0 & -h_1 \
  \end{array} \right)~.
\label{oppaCSA}
\end{equation}
In this way, the action of each Weyl group element as defined in
eq.(\ref{weylsp4intri}) can be reinterpreted as a particular
permutation of the set $\left(h_1,h_2,-h_2,-h_1 \right) $ and this
interpretation provides the embedding
(\ref{embeddusweylus}). To make it precise let
us derive the structure of the generalized Weyl group.
Following the definition \ref{genWeyl} we introduce as generators the
 operator defined in eq.(\ref{Omatra44}) for the following four
 choices of the $\theta$-angles:
\begin{equation}
  \left\{ \theta_1 \, , \,  \theta_2 \, , \, \theta_3 \, , \,
  \theta_4 \right\} \, = \, \left\{ \begin{array}{cccc}
    \frac{\pi}{2} & 0 & 0 & 0 \\
    0 &  \frac{\pi}{2} & 0 & 0 \\
    0 & 0 &  \frac{\pi}{2} & 0 \\
    0 & 0 & 0 &  \frac{\pi}{2} \
  \end{array}\right.
\label{urcagenera}
\end{equation}
which just corresponds to the $4$ roots of $\sym(4)$. By closing the
shell of products we obtain a group with $32$-elements,
$\mathcal{W}(\sym(4))$. This group has an order four normal subgroup $\mathrm{N}(\sym(4))$ with the
structure of $ \mathbb{Z}_2 \, \times \, \mathbb{Z}_2$ whose adjoint
action on any of the Cartan matrices (\ref{oppaCSA}) is the identity.
Explicitly $\mathrm{N}(\sym(4))$ is made by the following four
symplectic matrices:
\begin{equation}
  \begin{array}{rclcrcl}
    N_1 & = & \left(
\begin{array}{llll}
 -1 & 0 & 0 & 0 \\
 0 & -1 & 0 & 0 \\
 0 & 0 & -1 & 0 \\
 0 & 0 & 0 & -1
\end{array}
\right)   & ; & N_2 & \simeq &
\left(
\begin{array}{llll}
 -1 & 0 & 0 & 0 \\
 0 & 1 & 0 & 0 \\
 0 & 0 & 1 & 0 \\
 0 & 0 & 0 & -1
\end{array}
\right)~, \
\end{array}
\label{NN12}
\end{equation}
\begin{equation}
  \begin{array}{rclcrcl}
    N_3 & \simeq &  \left(
\begin{array}{llll}
 1 & 0 & 0 & 0 \\
 0 & -1 & 0 & 0 \\
 0 & 0 & -1 & 0 \\
 0 & 0 & 0 & 1
\end{array}
\right)  & ; & N_4 & \simeq & \left(
\begin{array}{llll}
 1 & 0 & 0 & 0 \\
 0 & 1 & 0 & 0 \\
 0 & 0 & 1 & 0 \\
 0 & 0 & 0 & 1
\end{array}
\right)~. \\
 \end{array}
\label{NN34}
\end{equation}
As expected the factor group
$\mathcal{W}(\sym(4))/\mathrm{N}(\sym(4))$ is isomorphic to the Weyl
group $\mathrm{Weyl}\left(\sym(4)\right)$ since the adjoint action of
each equivalence class produces the same transformation on the eigenvalues
$h_1,h_2$ as the abstract Weyl elements listed in eq.(\ref{weylsp4intri}).
Explicitly the $8$-equivalence classes of $4$-elements each are
displayed below
\begin{equation}
  \begin{array}{rclcrcl}
    \Omega_1 & \simeq & \left( \begin{array}{llll}
 1 & 0 & 0 & 0 \\
 0 & 1 & 0 & 0 \\
 0 & 0 & 1 & 0 \\
 0 & 0 & 0 & 1
\end{array}\right) \, \mathrm{N}(\sym(4))  & ; & \Omega_2 & \simeq &
\left(
\begin{array}{llll}
 0 & 0 & 0 & 1 \\
 0 & 0 & 1 & 0 \\
 0 & -1 & 0 & 0 \\
 -1 & 0 & 0 & 0
\end{array}
\right) \, \mathrm{N}(\sym(4))~, \
\end{array}
\label{Omega12}
\end{equation}
\begin{equation}
  \begin{array}{rclcrcl}
    \Omega_3 & \simeq &  \left(
\begin{array}{llll}
 0 & 0 & 0 & 1 \\
 0 & 1 & 0 & 0 \\
 0 & 0 & 1 & 0 \\
 -1 & 0 & 0 & 0
\end{array}
\right) \, \mathrm{N}(\sym(4))  & ; & \Omega_4 & \simeq & \left(
\begin{array}{llll}
 1 & 0 & 0 & 0 \\
 0 & 0 & 1 & 0 \\
 0 & -1 & 0 & 0 \\
 0 & 0 & 0 & 1
\end{array}
\right) \, \mathrm{N}(\sym(4))  \\
 \end{array}~,
\label{Omega34}
\end{equation}
\begin{equation}
  \begin{array}{rclcrcl}
    \Omega_5 & \simeq & \left(
\begin{array}{llll}
 0 & 1 & 0 & 0 \\
 1 & 0 & 0 & 0 \\
 0 & 0 & 0 & 1 \\
 0 & 0 & 1 & 0
\end{array}
\right) \, \mathrm{N}(\sym(4))  & ; & \Omega_6 & \simeq & \left(
\begin{array}{llll}
 0 & 0 & 1 & 0 \\
 1 & 0 & 0 & 0 \\
 0 & 0 & 0 & 1 \\
 0 & -1 & 0 & 0
\end{array}
\right) \, \mathrm{N}(\sym(4))~,  \\
 \end{array}
\label{Omega56}
\end{equation}
\begin{equation}
\begin{array}{rclcrcl}
    \Omega_7 & \simeq & \left(
\begin{array}{llll}
 0 & 1 & 0 & 0 \\
 0 & 0 & 0 & 1 \\
 -1 & 0 & 0 & 0 \\
 0 & 0 & 1 & 0
\end{array}
\right) \, \mathrm{N}(\sym(4))  & ; & \Omega_8 & \simeq & \left(
\begin{array}{llll}
 0 & 0 & 1 & 0 \\
 0 & 0 & 0 & 1 \\
 -1 & 0 & 0 & 0 \\
 0 & -1 & 0 & 0
\end{array}
\right) \, \mathrm{N}(\sym(4))~.  \
  \end{array}
\label{Omega12345678}
\end{equation}
Considering now a Cartan Lie algebra element in the fundamental
representation of $\sym(4,\mathbb{R})$ as given in eq.(\ref{oppaCSA})
we have
\begin{equation}
  \forall w_i \, \in \, Weyl(C_2) \quad : \quad \Omega_i^T \,
  \mathcal{C}\left(\{h_1,h_2\}\right) \, \Omega_i \, =
  \, \mathcal{C}\left(w_i\{h_1,h_2\}\right)~.
\label{adjoinrepre}
\end{equation}
This being established let us choose as  conventional reference set of
eigenvalues the following one:
\begin{equation}
  h_1\, = \, 1 \quad ; \quad h_2 \, = \, 2~,
\label{h12values}
\end{equation}
then the decreasing sorting to be expected at past infinity is
$2,1,-1,-2$ and corresponds to the Weyl element $\Omega_5$. If we
take this as the fundamental permutation, all the other eight
permutations belonging to the Weyl group can be ranked with the
number of transpositions needed to bring them to the fundamental one.
This procedure provides the partial ordering of the Weyl group
displayed in table \ref{Sp4Weylordo}.
\begin{table}
  \centering
  $$
  \begin{array}{|c|c|}
  \hline
    \ell_T & \mbox{Weyl group}\\
    \null & \mbox{of $\sym(4,\mathbb{R})$} \\
    \hline
    0 & w_5 \\
    1 & w_6 \\
    2 & w_1 \\
    3 & w_3 \\
    3 & w_4 \\
    4 & w_2 \\
    5 & w_7 \\
    6 & w_8 \\
    \hline
  \end{array}
  $$
  \caption{Partial ordering of the Weyl group of $\sym(4,\mathbb{R}).$}
  \label{Sp4Weylordo}
\end{table}
\par
We can now study the general features of the flows
associated with the maximally split coset manifold (\ref{sp4u2}) and see
how they follow the general principles and connect past and future Kasner epochs
ordered according to table \ref{Sp4Weylordo}. To realize this study the first
essential step is the construction of the $\sym(4,\mathbb{R})$ Lie
algebra in a basis which fulfils the condition that the solvable Lie
algebra generating the coset is represented by upper triangular
matrices. The form of the Cartan subalgebra in such a basis was
already anticipated in eq.(\ref{oppaCSA}), the full construction is
presented in the next section.
\subsection{Construction of the $\sym(4,\mathbb{R})$ Lie algebra}
The most compact way of presenting our basis is the
following. Let us begin with the solvable Lie algebra
$Solv(\mathrm{Sp(4,\mathbb{R})}/\mathrm{U(2)})$. Abstractly the most
general element of this algebra is given by
\begin{equation}
\mathcal{T}\, =\, h_1 \, \mathcal{H}_1 \, + \, h_2 \, \mathcal{H}_2 \, + \, e_1 \,
E^{\alpha_1} \, + \, e_2 \, E^{\alpha_2} \,  + \, e_3 \, E^{\alpha_1 + \alpha_2}
\, + \, e_4 \, E^{2\alpha_1 + \alpha_2}~.
\label{Tdefi}
\end{equation}
If we write the explicit form of $\mathcal{T}$ as a $4 \times 4$,
upper triangular symplectic matrix
\begin{eqnarray}
\mathcal{T}_{sym}& = &  \left(
\begin{array}{llll}
 h_1 & e_1 & e_3 & -\sqrt{2} e_4 \\
 0 & h_2 & \sqrt{2} e_2 & e_3 \\
 0 & 0 & -h_2 & -e_1 \\
 0 & 0 & 0 & -h_1
\end{array}
\right) \, \in \, \sym(4,\mathbb{R})
\label{sp4fundrep}
\end{eqnarray}
which satisfies the condition
\begin{equation}
  \mathcal{T}_{sym}^T \, \left( \begin{array}{cc}
    \mathbf{0}_{2} & \mathbf{1}_{2} \\
    -\mathbf{1}_{2} & \mathbf{0}_{2} \
  \end{array}\right) \, + \, \left( \begin{array}{cc}
    \mathbf{0}_{2} & \mathbf{1}_{2} \\
    -\mathbf{1}_{2} & \mathbf{0}_{2} \
  \end{array}\right) \, \mathcal{T}_{sym} \, = \, 0
\label{symleconda}
\end{equation}
all the generators of the solvable algebra are defined in the four dimensional symplectic
representation.
\par
By writing the same Lie algebra element (\ref{Tdefi}) as a $5 \times
5$ matrix
\begin{eqnarray}
\mathcal{T}_{so}& = & \left(
\begin{array}{lllll}
 h_1+h_2 & -\sqrt{2} e_2 & -\sqrt{2} e_3 & -\sqrt{2} e_4 & 0 \\
 0 & h_1-h_2 & -\sqrt{2} e_1 & 0 & \sqrt{2} e_4 \\
 0 & 0 & 0 & \sqrt{2} e_1 & \sqrt{2} e_3 \\
 0 & 0 & 0 & h_2-h_1 & \sqrt{2} e_2 \\
 0 & 0 & 0 & 0 & -h_1-h_2
\end{array}
\right) \, \in \, \so(2,3)
\label{so23fundrep}
\end{eqnarray}
which satisfies the condition
\begin{equation}
  \mathcal{T}_{so}^T \, \left( \begin{array}{cc|c|cc}
    0 & 0 & 0 & 0 & 1 \\
    0 & 0 & 0 & 1 & 0 \\
    \hline
    0 & 0 & 1 & 0 & 0 \\
    \hline
    0 & 1 & 0 & 0 & 0  \\
    1 & 0 & 0 & 0 & 0 \
  \end{array}\right) \, + \, \left( \begin{array}{cc|c|cc}
    0 & 0 & 0 & 0 & 1 \\
    0 & 0 & 0 & 1 & 0 \\
    \hline
    0 & 0 & 1 & 0 & 0 \\
    \hline
    0 & 1 & 0 & 0 & 0  \\
    1 & 0 & 0 & 0 & 0 \
  \end{array}\right)\,  \mathcal{T}_{so} \, = \, 0
\label{so23conda}
\end{equation}
we define the same generators also in the five dimensional
pseudo-orthogonal representation. The choice of the invariant metric
displayed in eq.(\ref{so23conda}) is that which guarantees the upper triangular structure of the solvable Lie algebra generators.
We shall come back on this point in later sections.
\par
Once the generators of the solvable Lie algebra are given the full
Lie algebra can be completed by defining the orthonormal generators
of the $\mathbb{K}$ subspace as follows:
\begin{eqnarray}
\mathrm{K}_1 & = & \mathcal{H}_1~, \nonumber\\
\mathrm{K}_2 & = & \mathcal{H}_2~, \nonumber\\
\mathrm{K}_3 & = & \ft{1}{\sqrt{2}}\left( E^{\alpha_1} \, + \, (E^{\alpha_1})^T\right)~,  \nonumber\\
\mathrm{K}_4 & = & \ft{1}{\sqrt{2}}\left( E^{\alpha_2} \, + \, (E^{\alpha_2})^T\right)~, \nonumber\\
\mathrm{K}_5 & = & \ft{1}{\sqrt{2}}\left( E^{\alpha_1+\alpha_2} \, + \, (E^{\alpha_1+\alpha_2})^T\right)~, \nonumber\\
\mathrm{K}_6 & = & \ft{1}{\sqrt{2}}\left( E^{\alpha_1+2\alpha_2} \, + \, (E^{\alpha_1+2\alpha_2})^T\right),
\label{Kgenerisp4}
\end{eqnarray}
and those of the maximal compact subalgebra $\mathbb{H}=\uu(2)$ as
follows:
\begin{eqnarray}
\mathrm{J}_1 & = & \ft{1}{\sqrt{2}}\left( E^{\alpha_1} \, - \, (E^{\alpha_1})^T\right)~,  \nonumber\\
\mathrm{J}_2 & = & \ft{1}{\sqrt{2}}\left( E^{\alpha_2} \, - \, (E^{\alpha_2})^T\right)~, \nonumber\\
\mathrm{J}_3 & = & \ft{1}{\sqrt{2}}\left( E^{\alpha_1+\alpha_2} \, - \, (E^{\alpha_1+\alpha_2})^T\right)~, \nonumber\\
\mathrm{J}_4 & = & \ft{1}{\sqrt{2}}\left( E^{\alpha_1+2\alpha_2} \, - \, (E^{\alpha_1+2\alpha_2})^T\right)~.
\label{HHgenerisp4}
\end{eqnarray}
In this way we have constructed all the relevant generators in both
representations. The flows are clearly an intrinsic property of the
algebra and will not depend on the representation chosen.
\par
\subsection{Parameterization of the compact group $\mathrm{U(2)}$ and critical
submanifolds}
In a way completely analogous to the previous case-study we can now
parameterize the compact subgroup by writing
\begin{equation}
  \mathcal{O} \, = \, \exp \left[\sqrt{2} \,\theta_1 \, \mathrm{J_1} \right] \, \exp \left[\theta_2 \, \mathrm{J_2}  \right] \,
  \exp \left[ \sqrt{2} \,\theta_3 \, \mathrm{J_3} \right] \, \exp \left[\theta_4 \, \mathrm{J_4}  \right]~.
\label{Omatra44}
\end{equation}
The square root of two factors have been introduced in equation
(\ref{Omatra44}) in such a way as to normalize the theta angles so
that the  group element $\mathcal{O}$ becomes an integer valued matrix at
$\theta_i = \frac{\pi}{2}$. Obviously we have two instances of
$\mathcal{O}$: the $4 \times 4 $ symplectic $\mathcal{O}_{sp}$ and the $5
\times 5$ pseudo-orthogonal $\mathcal{O}_{so}$. Both of them become
integer valued for the same choice of the angles and when   acting by similarity
transformation  on the Cartan subalgebra they correspond to
Weyl group elements.
\par
\begin{figure}[!hbt]
\begin{center}
\iffigs
 \includegraphics[height=60mm]{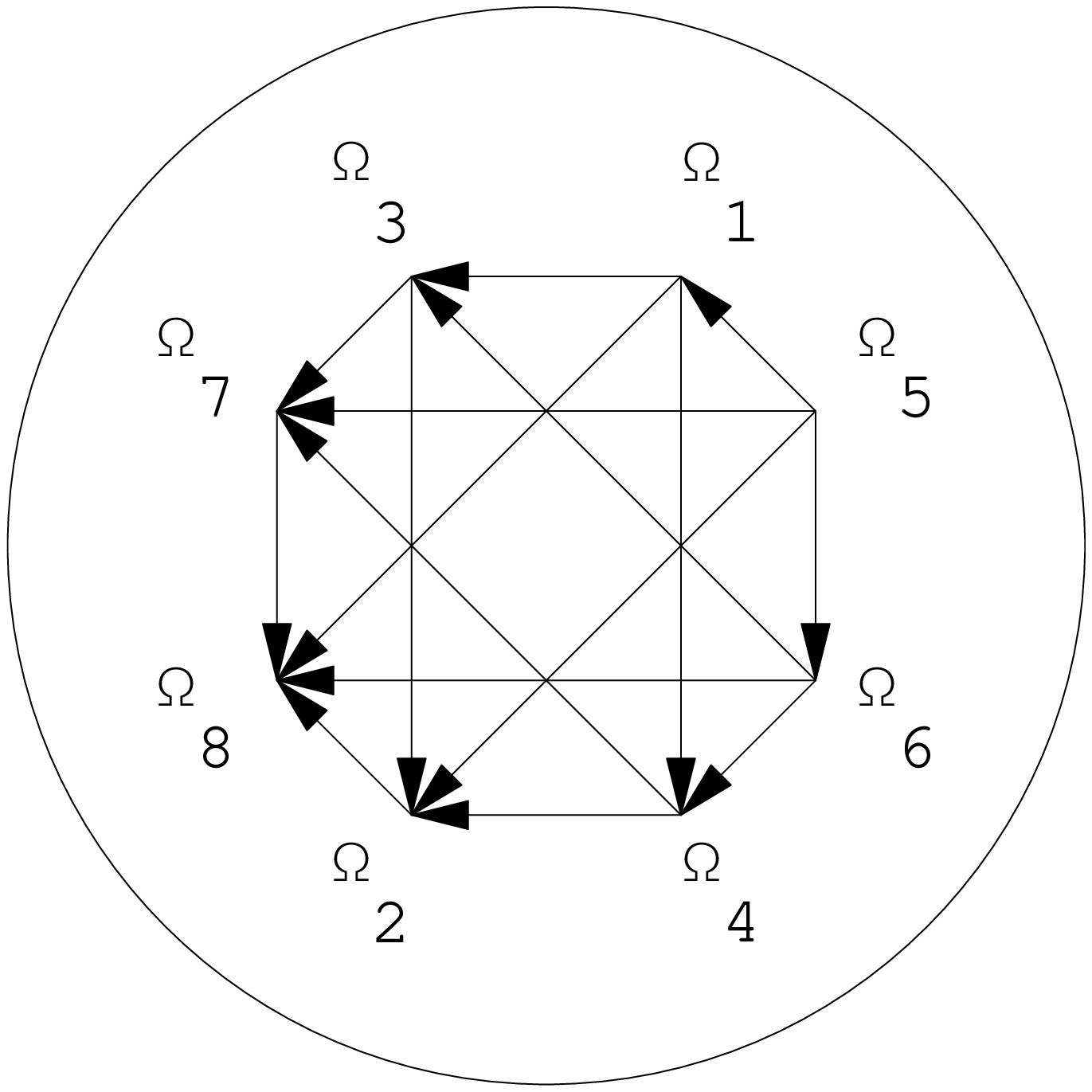}
{\scriptsize
 $$
\begin{array}{l|llllllll}
\null&1 & 2 & 3 & 4 & \mathbf{5} & 6 & 7 & 8\\
\hline
1& 1 & 2 & 3 & 4 & 5 & 6 & 7 & 8 \\
\mathbf{2}& 2 & 1 & 4 & 3 & \mathbf{8} & 7 & 6 & 5 \\
3& 3 & 4 & 1 & 2 & 6 & 5 & 8 & 7 \\
4& 4 & 3 & 2 & 1 & 7 & 8 & 5 & 6 \\
5& 5 & 8 & 7 & 6 & 1 & 4 & 3 & 2 \\
6& 6 & 7 & 8 & 5 & 3 & 2 & 1 & 4 \\
7& 7 & 6 & 5 & 8 & 4 & 1 & 2 & 3 \\
8& 8 & 5 & 6 & 7 & 2 & 3 & 4 & 1
\end{array}$$}
\else
\end{center}
 \fi
\caption{\it The oriented phase diagram of the $\mathrm{Sp(4,\mathbb{R})}/\mathrm{U(2)}$ flows.
The Lie algebra $\sym(4,\mathbb{R})$ is the
maximally split real section of the complex Lie algebra $\mathrm{C}_2 \sim
\mathrm{B}_2$. Its Weyl group is $(\mathbb{Z}_2 \times \mathbb{Z}_2) \, \ltimes \,
\mathrm{S}_2$ and has eight elements identified by their action on
the eigenvalues $h_1,h_2$ of the Lax operator. Eight are
therefore the possible asymptotic states of the universe at
$t= \pm\infty$ and each possible motion is an oriented flow from one
Weyl element to another one. The orientation follows the ordering of
Weyl group elements: it is always from a lower to a higher one.
In this picture, choosing as fundamental eigenvalues $h_1=1,h_2=2$
the Weyl group element
$\Omega_i\,  \in \, Weyl(C_2) $ is identified by the point in the plane that has
coordinates $\Omega_i(h_1,h_2)$. Each link is therefore associated with a Weyl group element which
multiplying on the left the past infinity element produces the future infinity one.
By comparison we display below the graph the multiplication table
of the Weyl group. Note that in each vertex of the diagram there meet just four lines.
In vertex $\Omega_5$ there are only outgoing lines. This is so because $\Omega_5$ is the lowest Weyl element and it
corresponds to the universal past infinity point for generic flows.
On the contrary in the vertex $\Omega_8$ there only incoming lines. This is so because $\Omega_8$ is the highest Weyl element
and it corresponds to the universal future infinity for
generic flows. The other vertices have both incoming and outgoing lines.}
\label{sp4graphus}
 \iffigs
 \hskip 1.5cm \unitlength=1.1mm
 \end{center}
  \fi
\end{figure}

For simplicity we use the $4\times 4$ representation and we find
\begin{equation}
  \mathcal{O} \, = \, \left( \begin{array}{cccc}
    O_{11} & O_{12} & O_{13} & O_{14} \\
    O_{21} & O_{22} & O_{23} & O_{24} \\
    O_{31} & O_{32} & O_{33} & O_{34} \\
    O_{41} & O_{42} & O_{43} & O_{44} \
  \end{array}\right)
\label{matraOsp4}
\end{equation}
where
\begin{equation}
  \begin{array}{ccl}
    O_{11} & = & \cos  \theta _1  \cos  \theta
   _3  \cos  \theta _4 -\sin
    \theta _1  \sin   \theta
   _3  \sin  \left(\theta _2-\theta
   _4 \right)~,  \\
    O_{12} & = & \cos  \theta _2  \cos  \theta
   _3  \sin  \theta _1~,  \\
    O_{13} & = & \cos  \theta _3  \sin  \theta
   _1  \sin  \theta _2 +\cos
    \theta _1  \sin  \theta
   _3 ~, \\
    O_{14} & = & \cos  \theta _2  \cos  \theta
   _4  \sin  \theta _1  \sin
    \theta _3 +\left(\sin
    \theta _1  \sin  \theta
   _2  \sin  \theta _3 -\cos
    \theta _1  \cos  \theta
   _3 \right) \sin  \theta _4~,  \\
    \end{array}
    \end{equation}
    \begin{equation}
    \begin{array}{ccl}
    O_{21} & = & -\cos  \theta _3  \cos  \theta
   _4  \sin  \theta _1 -\cos
    \theta _1  \sin  \theta
   _3  \sin  \left(\theta _2-\theta
   _4 \right)~,  \\
    O_{22} & = & \cos  \theta _1  \cos  \theta
   _2  \cos  \theta _3~,  \\
    O_{23} & = & \cos  \theta _1  \cos  \theta
   _3  \sin  \theta _2 -\sin
    \theta _1  \sin  \theta
   _3 ~, \\
    O_{24} & = & \cos  \theta _1  \cos  \theta
   _2-\theta _4  \sin  \theta
   _3 +\cos  \theta _3  \sin
    \theta _1  \sin  \theta
   _4 ~, \\
    \end{array}
    \end{equation}
    \begin{equation}
    \begin{array}{ccl}
    O_{31} & = & -\cos  \theta _1  \cos  \theta
   _2-\theta _4  \sin  \theta
   _3 -\cos  \theta _3  \sin
    \theta _1  \sin  \theta
   _4 ~, \\
    O_{32} & = & \sin  \theta _1  \sin  \theta
   _3 -\cos  \theta _1  \cos
    \theta _3  \sin  \theta
   _2 ~, \\
    O_{32} & = & \cos  \theta _1  \cos  \theta
   _2  \cos  \theta _3~,  \\
    O_{33} & = & \cos  \theta _1  \cos  \theta
   _2  \cos  \theta _3 ~, \\
    O_{34} & = & -\cos  \theta _3  \cos  \theta
   _4  \sin  \theta _1 -\cos
    \theta _1  \sin  \theta
   _3  \sin  \left(\theta _2-\theta
   _4 \right)~, \\
    \end{array}
    \end{equation}
    \begin{equation}
    \begin{array}{ccl}
    O_{41} & = & \left(\cos  \theta _1  \cos
    \theta _3 -\sin  \theta
   _1  \sin  \theta _2  \sin
    \theta _3 \right) \sin
    \theta _4 -\cos  \theta
   _2  \cos  \theta _4  \sin
    \theta _1  \sin  \theta
   _3 ~, \\
    O_{42} & = & -\cos  \theta _3  \sin  \theta
   _1  \sin  \theta _2 -\cos
    \theta _1  \sin  \theta
   _3~,  \\
    O_{43} & = & \cos  \theta _2  \cos  \theta
   _3  \sin  \theta _1~,  \\
    O_{44} & = & \cos  \theta _1  \cos  \theta
   _3  \cos  \theta _4 -\sin
    \theta _1  \sin  \theta
   _3  \sin  \left(\theta _2-\theta
   _4 \right)~. \
  \end{array}
\label{omatraelementi}
\end{equation}
\paragraph{Vertices}
Having parameterized in this way the $\mathrm{U(2)}$ group element
with the four Euler angles $\theta_i$, in a completely analogous way
to the case of $\mathrm{SL(3,\mathbb{R})}$, we can check that when
all of the $\theta_i$ take either the $0$ or the $\frac \pi 2$ value
then the corresponding matrix $\mathcal{O}$ becomes integer valued and its
similarity action on a Cartan subalgebra element (\ref{oppaCSA})
corresponds to the action of some Weyl group element on the
eigenvalues:
\begin{equation}
\mbox{If } \forall i  \,\, \theta_i \, = \, \left\{ \begin{array}{c}
  0 \\
  \mathrm{or} \\
  \frac {\pi}{2}
\end{array} \right. \quad , \quad  \exists \, \omega \, \in \, Weyl(C_2) \quad / \quad \mathcal{O}^T \,
\mathcal{C}\left( \{h_1,h_2\}\right)\, \mathcal{O} \, = \,\mathcal{C}\left( \omega\{h_1,h_2\}\right)~.
\label{vallettesp4}
\end{equation}
In this way the parameter space $\mathrm{U(2)}/\mathrm{{\cal W}}$ is reduced
to lie in a four dimensional hypercube and, using a notation analogous
to that of eq.(\ref{notazia}), the identification of the $16$ vertices
of the hypercube with Weyl group elements is displayed in table
\ref{sp4vertices}. As the reader can observe, in the chosen numbering
the odd-labeled Weyl group elements appear only once, while the
even-labeled appear three-times.
\begin{table}
{\scriptsize
\centering
$$
  \begin{array}{|c|c|c|c|}
  \hline
  \# &\mbox{vertex}&\mbox{Weyl group element}&\mbox{multiplicity of Weyl elem.}\\
  \hline
1 & \{0, 0, 0,
      0\}  &  \Omega_1& 1\\
2 & \{1, 0, 0,
      0\}  &  \Omega_5& 1\\
3 & \{0, 1, 0,
      0\}  &  \Omega_4& 3\\
4 & \{0, 0, 1,
      0\}  &  \Omega_8& 3\\
5 & \{0, 0, 0,
      1\}  &  \Omega_3& 1\\
6 & \{1, 1, 0,
      0\}  &  \Omega_6& 3\\
7 & \{1, 0, 1,
      0\}  &  \Omega_2& 3\\
8 & \{1, 0, 0,
      1\}  &  \Omega_7& 1\\
9 & \{0, 1, 1,
      0\}  &  \Omega_6& 3\\
10 & \{0, 1, 0,
      1\}  &  \Omega_2& 3\\
11 & \{0, 0, 1,
      1\}  &  \Omega_6& 3\\
12 & \{1, 1, 1,
      0\}  &  \Omega_4& 3\\
13 & \{1, 1, 0,
      1\}  &  \Omega_8& 3\\
14 & \{1, 0, 1,
      1\}  &  \Omega_4& 3\\
15 & \{0, 1, 1,
      1\}  &  \Omega_8& 3\\
16 & \{1, 1, 1,
      1\}  &  \Omega_2& 3\\
      \hline
\end{array}
$$
}
 \caption{The 16 vertices of the hypercubic parameter space for $\mathrm{Sp(4,\mathbb{R})/U(2)}$
  and their identification with Weyl group elements. }\label{sp4vertices}
\end{table}
\paragraph{Edges}
Using just the same strategy as in the previous case-study we can now
construct the $64$ oriented links connecting the $16$ vertices. These are all
the possible segments of straight lines in parameter space
connecting two vertices and by means of a computer programme we can
evaluate the orientation of the link, namely discover which of the
end-points (Weyl group element) corresponds to past infinity $t=-\infty$
and which to future infinity $t=+\infty$. As expected the orientation of all the links
is in the direction from lower to higher Weyl elements, according to the ordering of table \ref{Sp4Weylordo}. The result of these
computations is displayed in table \ref{sp4links} and summarized in
the flow diagram of fig.\ref{sp4graphus}.
\begin{table}
 {\scriptsize
 \centering
$$
\begin{array}{|c|ccc||c|cccc|}
\hline
\#
&\mbox{Vertex}&\mbox{Vertex}&\mbox{Flow}&\#&\mbox{Vertex}&\mbox{Vertex}&\mbox{Flow}&\\
\hline
1&\{0, 0, 0, 0\}& \{0, 0, 0,1\}&
\mbox{$\Omega_1$ $\mapsto $ $\Omega_3$} &\null& \null & \null &\null&\null\\ 2 &
\{0, 0, 0,
      0\}& \{0, 0, 1,
        0\}&
\mbox{$\Omega_1 $ $\mapsto $ $\Omega_8 $} &
3&\{0, 0, 0,
      0\}& \{0, 1, 0,
        0\}&
\mbox{$\Omega_1$ $\mapsto $ $\Omega_4$} &\\4&\
\{0, 0, 0,
      0\}& \{1, 0, 0,
        0\}&
\mbox{$\Omega_5$ $\mapsto $ $\Omega_1$} &
5&\{0, 0, 0,
      1\}& \{0, 0, 0,
        0\}&
\mbox{$\Omega_1$ $\mapsto $ $\Omega_3$} &\\6&\
\{0, 0, 0,
      1\}& \{0, 0, 1,
        1\}&
\mbox{$\Omega_6$ $\mapsto $ $\Omega_3$} &
7&\{0, 0, 0,
      1\}& \{0, 1, 0,
        1\}&
\mbox{$\Omega_3$ $\mapsto $ $\Omega_2$} &\\8&\
\{0, 0, 0,
      1\}& \{1, 0, 0,
        1\}&
\mbox{$\Omega_3$ $\mapsto $ $\Omega_7$} &
9&\{0, 0, 1,
      0\}& \{0, 0, 0,
        0\}&
\mbox{$\Omega_1$ $\mapsto $ $\Omega_8$} &\\10&\
\{0, 0, 1,
      0\}& \{0, 0, 1,
        1\}&
\mbox{$\Omega_6$ $\mapsto $ $\Omega_8$} &
11&\{0, 0, 1,
      0\}& \{0, 1, 1,
        0\}&
\mbox{$\Omega_6$ $\mapsto $ $\Omega_8$} &\\12&\
\{0, 0, 1,
      0\}& \{1, 0, 1,
        0\}&
\mbox{$\Omega_2$ $\mapsto $ $\Omega_8$} &
13&\{0, 0, 1,
      1\}& \{0, 0, 0,
        1\}&
\mbox{$\Omega_6$ $\mapsto $ $\Omega_3$} &\\14&\
\{0, 0, 1,
      1\}& \{0, 0, 1,
        0\}&
\mbox{$\Omega_6$ $\mapsto $ $\Omega_8$} &
15&\{0, 0, 1,
      1\}& \{0, 1, 1,
        1\}&
\mbox{$\Omega_6$ $\mapsto $ $\Omega_8$} &\\16&\
\{0, 0, 1,
      1\}& \{1, 0, 1,
        1\}&
\mbox{$\Omega_6$ $\mapsto $ $\Omega_4$} &
17&\{0, 1, 0,
      0\}& \{0, 0, 0,
        0\}&
\mbox{$\Omega_1$ $\mapsto $ $\Omega_4$} &\\18&\
\{0, 1, 0,
      0\}& \{0, 1, 0,
        1\}&
\mbox{$\Omega_4$ $\mapsto $ $\Omega_2$} &
19&\{0, 1, 0,
      0\}& \{0, 1, 1,
        0\}&
\mbox{$\Omega_6$ $\mapsto $ $\Omega_4$} &\\20&\
\{0, 1, 0,
      0\}& \{1, 1, 0,
        0\}&
\mbox{$\Omega_6$ $\mapsto $ $\Omega_4$} &
21&\{0, 1, 0,
      1\}& \{0, 0, 0,
        1\}&
\mbox{$\Omega_3$ $\mapsto $ $\Omega_2$} &\\22&\
\{0, 1, 0,
      1\}& \{0, 1, 0,
        0\}&
\mbox{$\Omega_4$ $\mapsto $ $\Omega_2$} &
23&\{0, 1, 0,
      1\}& \{0, 1, 1,
        1\}&
\mbox{$\Omega_2$ $\mapsto $ $\Omega_8$} &\\24&\
\{0, 1, 0,
      1\}& \{1, 1, 0,
        1\}&
\mbox{$\Omega_2$ $\mapsto $ $\Omega_8$} &
25&\{0, 1, 1,
      0\}& \{0, 0, 1,
        0\}&
\mbox{$\Omega_6$ $\mapsto $ $\Omega_8$} &\\26&\
\{0, 1, 1,
      0\}& \{0, 1, 0,
        0\}&
\mbox{$\Omega_6$ $\mapsto $ $\Omega_4$} &
27&\{0, 1, 1,
      0\}& \{0, 1, 1,
        1\}&
\mbox{$\Omega_6$ $\mapsto $ $\Omega_8$} &\\28&\
\{0, 1, 1,
      0\}& \{1, 1, 1,
        0\}&
\mbox{$\Omega_6$ $\mapsto $ $\Omega_4$} &
29&\{0, 1, 1,
      1\}& \{0, 0, 1,
        1\}&
\mbox{$\Omega_6$ $\mapsto $ $\Omega_8$} &\\30&\
\{0, 1, 1,
      1\}& \{0, 1, 0,
        1\}&
\mbox{$\Omega_2$] $\mapsto $ $\Omega_8$} &
31&\{0, 1, 1,
      1\}& \{0, 1, 1,
        0\}&
\mbox{$\Omega_6$ $\mapsto $ $\Omega_8$} &\\32&\
\{0, 1, 1,
      1\}& \{1, 1, 1,
        1\}&
\mbox{$\Omega_2$ $\mapsto $ $\Omega_8$} &
33&\{1, 0, 0,
      0\}& \{0, 0, 0,
        0\}&
\mbox{$\Omega_5$ $\mapsto $ $\Omega_1$} &\\34&\
\{1, 0, 0,
      0\}& \{1, 0, 0,
        1\}&
\mbox{$\Omega_5$ $\mapsto $ $\Omega_7$} &
35&\{1, 0, 0,
      0\}& \{1, 0, 1,
        0\}&
\mbox{$\Omega_5$ $\mapsto $ $\Omega_2$} &\\36&\
\{1, 0, 0,
      0\}& \{1, 1, 0,
        0\}&
\mbox{$\Omega_5$ $\mapsto $ $\Omega_6$} &
37&\{1, 0, 0,
      1\}& \{0, 0, 0,
        1\}&
\mbox{$\Omega_3$ $\mapsto $ $\Omega_7$} &\\38&\
\{1, 0, 0,
      1\}& \{1, 0, 0,
        0\}&
\mbox{$\Omega_5$ $\mapsto $ $\Omega_7$} &
39&\{1, 0, 0,
      1\}& \{1, 0, 1,
        1\}&
\mbox{$\Omega_4$ $\mapsto $ $\Omega_7$} &\\40&\
\{1, 0, 0,
      1\}& \{1, 1, 0,
        1\}&
\mbox{$\Omega_7$ $\mapsto $ $\Omega_8$} &
41&\{1, 0, 1,
      0\}& \{0, 0, 1,
        0\}&
\mbox{$\Omega_2$ $\mapsto $ $\Omega_8$} &\\42&\
\{1, 0, 1,
      0\}& \{1, 0, 0,
        0\}&
\mbox{$\Omega_5$ $\mapsto $ $\Omega_2$} &
43&\{1, 0, 1,
      0\}& \{1, 0, 1,
        1\}&
\mbox{$\Omega_4$ $\mapsto $ $\Omega_2$} &\\44&\
\{1, 0, 1,
      0\}& \{1, 1, 1,
        0\}&
\mbox{$\Omega_4$ $\mapsto $ $\Omega_2$} &
45&\{1, 0, 1,
      1\}& \{0, 0, 1,
        1\}&
\mbox{$\Omega_6$ $\mapsto $ $\Omega_4$} &\\46&\
\{1, 0, 1,
      1\}& \{1, 0, 0,
        1\}&
\mbox{$\Omega_4$ $\mapsto $ $\Omega_7$} &
47&\{1, 0, 1,
      1\}& \{1, 0, 1,
        0\}&
\mbox{$\Omega_4$ $\mapsto $ $\Omega_2$} &\\48&\
\{1, 0, 1,
      1\}& \{1, 1, 1,
        1\}&
\mbox{$\Omega_4$ $\mapsto $ $\Omega_2$} &
49&\{1, 1, 0,
      0\}& \{0, 1, 0,
        0\}&
\mbox{$\Omega_6$ $\mapsto $ $\Omega_4$} &\\50&\
\{1, 1, 0,
      0\}& \{1, 0, 0,
        0\}&
\mbox{$\Omega_5$ $\mapsto $ $\Omega_6$} &
51&\{1, 1, 0,
      0\}& \{1, 1, 0,
        1\}&
\mbox{$\Omega_6$ $\mapsto $ $\Omega_8$} &
\\52&\
\{1, 1, 0,
      0\}& \{1, 1, 1,
        0\}&
\mbox{$\Omega_6$ $\mapsto $ $\Omega_4$} &
53&\{1, 1, 0,
      1\}& \{0, 1, 0,
        1\}&
\mbox{$\Omega_2$ $\mapsto $ $\Omega_8$} &\\54&\
\{1, 1, 0,
      1\}& \{1, 0, 0,
        1\}&
\mbox{$\Omega_7$ $\mapsto $ $\Omega_8$} &
55&\{1, 1, 0,
      1\}& \{1, 1, 0,
        0\}&
\mbox{$\Omega_6$ $\mapsto $ $\Omega_8$} &\\56&\
\{1, 1, 0,
      1\}& \{1, 1, 1,
        1\}&
\mbox{$\Omega_2$ $\mapsto $ $\Omega_8$} &
57&\{1, 1, 1,
      0\}& \{0, 1, 1,
        0\}&
\mbox{$\Omega_6$ $\mapsto $ $\Omega_4$} &\\58&\
\{1, 1, 1,
      0\}& \{1, 0, 1,
        0\}&
\mbox{$\Omega_4$ $\mapsto $ $\Omega_2$} &
59&\{1, 1, 1,
      0\}& \{1, 1, 0,
        0\}&
\mbox{$\Omega_6$ $\mapsto $ $\Omega_4$} &\\60&\
\{1, 1, 1,
      0\}& \{1, 1, 1,
        1\}&
\mbox{$\Omega_4$ $\mapsto $ $\Omega_2$} &
61&\{1, 1, 1,
      1\}& \{0, 1, 1,
        1\}&
\mbox{$\Omega_2$ $\mapsto $ $\Omega_8$} &\\62&\
\{1, 1, 1,
      1\}& \{1, 0, 1,
        1\}&
\mbox{$\Omega_4$ $\mapsto $ $\Omega_2$} &
63&\{1, 1, 1,
      1\}& \{1, 1, 0,
        1\}&
\mbox{$\Omega_2$ $\mapsto $ $\Omega_8$} &\\64&\
\{1, 1, 1,
      1\}& \{1, 1, 1,
        0\}&
\mbox{$\Omega_4$ $\mapsto $ $\Omega_2$} &\null& \null & \null &\null&\null\\
\hline
      \end{array}
$$
}
 \caption{A $4$-dimensional hypercube has $32$ edges which amount to 64 edges if
 we consider also the possible orientation. Here are displayed the $64$ oriented links
 of the hypercubic parameter space for $\mathrm{Sp(4,\mathbb{R})/U(2)}$
 flows and the corresponding oriented links from Weyl group elements to Weyl group elements.}\label{sp4links}
\end{table}
\begin{figure}[!hbt]
\begin{center}
\iffigs
 \includegraphics[height=95mm]{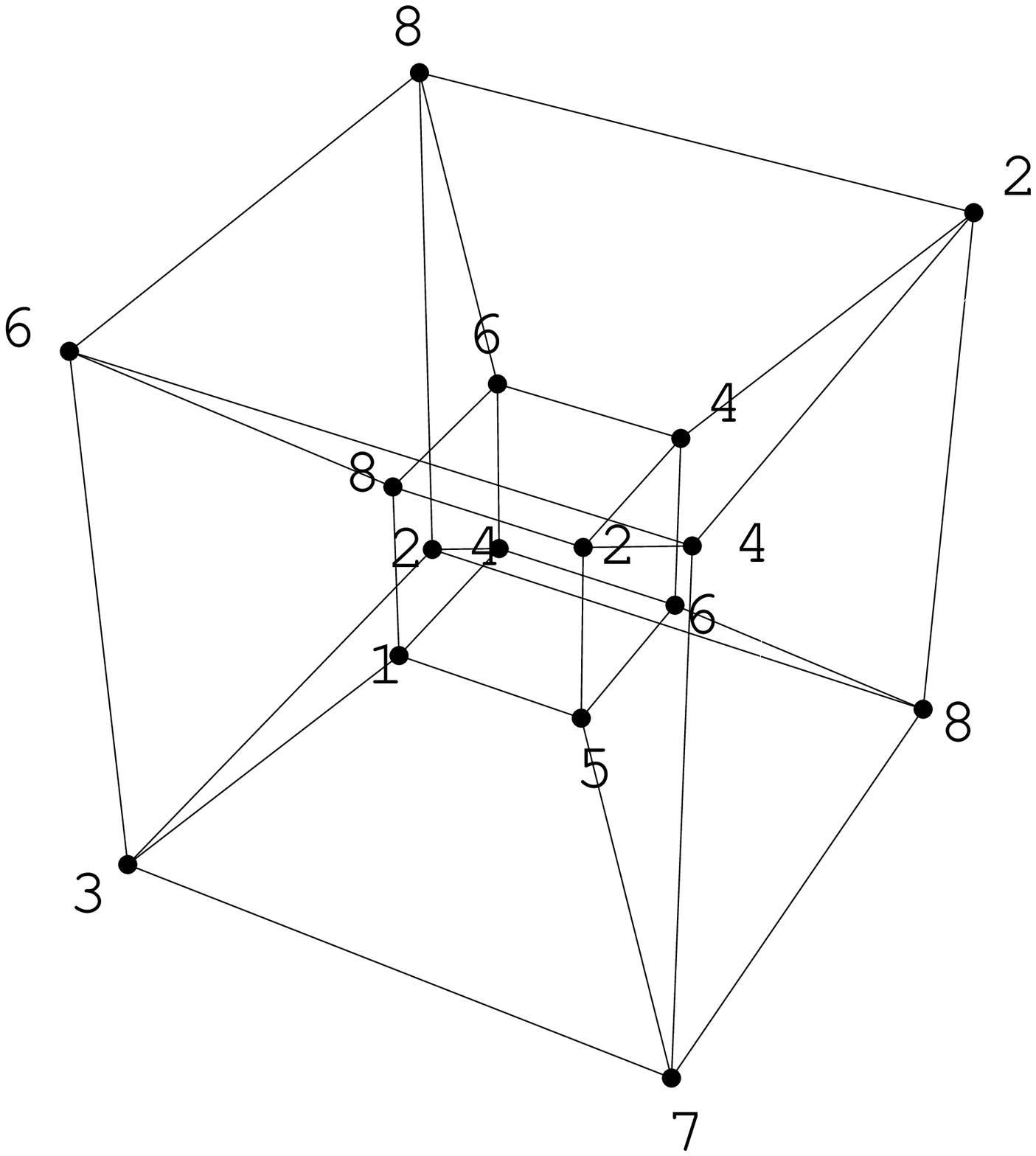}
\else
\end{center}
 \fi
\caption{\it Stereographic projection of the hypercubic parameter space for
$\mathrm{Sp(4,\mathbb{R})}/\mathrm{U(2)}$ motions.}
\label{hypcube1}
 \iffigs
 \hskip 1.5cm \unitlength=1.1mm
 \end{center}
  \fi
\end{figure}
A four dimensional hypercube cannot be drawn in three dimension but a
standard way to visualize it is provided by presenting its stereographic
projection. Indeed if we shift the origin of the coordinate system to
the point $\{\ft 12 ,\ft 12 ,\ft 12 , \ft 12 \}$ then all the $16$
vertices of the  hypercube are located on the standard
three-sphere, namely, as $4$-component vectors they have unit norm. So we
can consider their stereographic projection from $\mathbb{S}^3$ to $\mathbb{R}^3$ and
connecting them with segments we obtain the visualization of the
hypercube displayed in fig.\ref{hypcube1}.
\paragraph{Trapped hypersurfaces}
The study of trapped hypersurfaces can now be performed
once again in complete analogy with the case of
$\mathrm{SL(3,\mathbb{R})}$. We just have to calculate all the
relevant minors and impose their vanishing. In this way we determine
equations on the parameters that have to be solved within the
hypercubic range. If solutions within the hypercube exist, then we
have a trapped surface. Otherwise we just have a Weyl replica of an already existing
surface. In our case there are just
14 relevant minors distributed in the following way: 4 of rank 1, 4
of rank 3 and 6 of rank 2. Explicitly we can define the following
candidate trapped surfaces:
{\scriptsize
\begin{equation}
  \begin{array}{rclcl}
\Sigma _1  & : &  O_{1,
      1}  & = & 0~, \\
\Sigma _2  & : &  O_{2,
      1}  & = & 0~, \\
\Sigma _3  & : &  O_{3,
      1}  & = & 0~, \\
\Sigma _4  & : &  O_{4,
      1}  & = & 0~, \\
\Sigma _5  & : &  O_{1,
          3} \left(O_{2, 1} O_{3, 2} - O_{2, 2} O_{3, 1}\right) +
    O_{1, 2} \left(O_{2, 3} O_{3, 1} - O_{2, 1} O_{3, 3}\right) +
    O_{1, 1} \left(O_{2, 2} O_{3, 3} -
          O_{2, 3} O_{3, 2}\right)  & = & 0~, \\
\Sigma _6  & : &  O_{1,
          3} \left(O_{2, 1} O_{4, 2} - O_{2, 2} O_{4, 1}\right) +
    O_{1, 2} \left(O_{2, 3} O_{4, 1} - O_{2, 1} O_{4, 3}\right) +
    O_{1, 1} \left(O_{2, 2} O_{4, 3} -
          O_{2, 3} O_{4, 2}\right)  & = & 0 ~,\\
\Sigma _7  & : &  O_{1,
          3} \left(O_{3, 1} O_{4, 2} - O_{3, 2} O_{4, 1}\right) +
    O_{1, 2} \left(O_{3, 3} O_{4, 1} - O_{3, 1} O_{4, 3}\right) +
    O_{1, 1} \left(O_{3, 2} O_{4, 3} -
          O_{3, 3} O_{4, 2}\right)  & = & 0 ~,\\
\end{array}
\label{criticonesp4U}
\end{equation}
\begin{equation}
  \begin{array}{rclcl}
\Sigma _8  & : &  O_{2,
          3} \left(O_{3, 1} O_{4, 2} - O_{3, 2} O_{4, 1}\right) +
    O_{2, 2} \left(O_{3, 3} O_{4, 1} - O_{3, 1} O_{4, 3}\right) +
    O_{2, 1} \left(O_{3, 2} O_{4, 3} -
          O_{3, 3} O_{4, 2}\right)  & = & 0 ~,\\
\Sigma _9  & : &  O_{1, 1} O_{2, 2} -
    O_{1, 2} O_{2, 1}  & = & 0 ~,\\
\Sigma _{10}  & : &  O_{1, 1} O_{3, 2} -
    O_{1, 2} O_{3, 1}  & = & 0~, \\
\Sigma _{11}  & : &  O_{1, 1} O_{4, 2} -
    O_{1, 2} O_{4, 1}  & = & 0 ~,\\
\Sigma _{12}  & : &  O_{2, 1} O_{3, 2} -
    O_{2, 2} O_{3, 1}  & = & 0 ~,\\
\Sigma _{13}  & : &  O_{2, 1} O_{4, 2} -
    O_{2, 2} O_{4, 1}  & = & 0 ~,\\
\Sigma _{14}  & : &  O_{3, 1} O_{4, 2} -
    O_{3, 2} O_{4, 1}  & = & 0 \\
\end{array}
\label{criticonesp4W}
\end{equation}
}
where the explicit form of the equation can be obtained by
substituting the values of the $\mathrm{U(2)}$ matrix elements as given in
eq.s (\ref{matraOsp4}--\ref{omatraelementi}). A full-fledged analysis of all the
trapped surfaces is beyond the scope of the present paper
which aims at illustrating the general principles and at explaining the
method. What we can do without any analytic study of the trapped
surfaces is to determine the accessible Weyl group elements for each of them and
in this way single out the corresponding past infinity and future
infinity states.
The result is shown in table \ref{14sigma}.
\begin{table}
  \centering
  $$
  \begin{array}{|c|l|c|l|}
  \hline
  \mbox{Surf} & \mbox{Accessible Weyl el.} & \mbox{Flow}& \mbox{Type}\\
  \hline
   \Sigma _1 & \left\{w_5, w_6, w_3, w_2, w_7,
      w_8\right\} & w_5 \, \
\mapsto \, w_8 & \mbox{critical}\\
\Sigma _2 & \left\{w_1, w_3, w_4, w_2, w_7,
      w_8\right\} & w_1 \, \
\mapsto \, w_8 &\mbox{super-critical}\\
\Sigma _3 & \left\{w_5, w_6, w_1, w_3, w_4,
      w_2\right\} & w_5 \, \
\mapsto \, w_2 & \mbox{super-critical}\\
\Sigma _4 & \left\{w_5, w_6, w_1, w_4, w_7,
      w_8\right\} & w_5 \, \
\mapsto \, w_8 & \mbox{critical}\\
\Sigma _5 & \left\{w_5, w_6, w_3, w_2, w_7,
      w_8\right\} & w_5 \, \
\mapsto \, w_8 & \mbox{critical} \\
\Sigma _6 & \left\{w_1, w_3, w_4, w_2, w_7,
      w_8\right\} & w_1 \, \
\mapsto \, w_8 & \mbox{super-critical}\\
\Sigma _7 & \left\{w_5, w_6, w_1, w_3, w_4,
      w_2\right\} & w_5 \, \
\mapsto \, w_2 & \mbox{super-critical}\\
\Sigma _8 & \left\{w_5, w_6, w_1, w_4, w_7,
      w_8\right\} & w_5 \, \
\mapsto \, w_8 & \mbox{critical}\\
\Sigma _9 & \left\{w_6, w_3, w_4, w_2, w_7,
      w_8\right\} & w_6 \, \
\mapsto \, w_8 & \mbox{super-critical}\\
\Sigma _{10} & \left\{w_5, w_6, w_1, w_3,
      w_2, w_8\right\} & w_5\
 \, \mapsto \, w_8 & \mbox{critical}\\
\Sigma _{11} & \left\{w_5, w_6, w_1, w_3,
      w_4, w_2, w_7,
      w_8\right\} & w_5 \, \
\mapsto \, w_8 & \mbox{trapped non crit.}\\
\Sigma _{12} & \left\{w_5, w_6, w_1, w_3,
      w_4, w_2, w_7,
      w_8\right\} & w_5 \, \
\mapsto \, w_8 & \mbox{trapped non crit.}\\
\Sigma _{13} & \left\{w_5, w_1, w_4, w_2,
      w_7, w_8\right\} & w_5\
 \, \mapsto \, w_8 & \mbox{critical}\\
\Sigma _{14} & \left\{w_5, w_6, w_1, w_3,
      w_4, w_7\right\} & w_5\
 \, \mapsto \, w_7 & \mbox{super-critical}\\
  \hline
  \end{array}
  $$
  \caption{Accessible Weyl elements on the $14$ trapped surfaces of $\mathrm{Sp(4,\mathbb{R})/U(2)}$ flows.
  By inspecting the list $\mathcal{A}_{\Sigma}$ of accessible Weyl elements we easily
  deduce the character of the surface. If there are missing Weyl elements it is critical. It is
  super-critical if one of the missing elements is either $\Omega_{min} \, = \, w_5$ and/or
  $\Omega_{max} \, = \, w_8$. When no Weyl element is missing in $\mathcal{A}_{\Sigma}$, the surface
  is just only trapped. It would be critical inside the bigger group $\mathrm{SL(4,\mathbb{R})}$. }
  \label{14sigma}
\end{table}
\newpage
\par
As it is evident by inspection of this table the possible flows
realized on critical hypersurfaces of parameter space are just a very
small number, namely the following five:
\begin{equation}
  \begin{array}{ccc}
    1 & : & w_1 \, \mapsto \, w_8~, \\
    2 & : & w_5 \, \mapsto \, w_2~, \\
    3 & : & w_5 \, \mapsto \, w_7~, \\
    4 & : & w_5 \, \mapsto \, w_8~, \\
    5 & : & w_6 \, \mapsto \, w_8 ~.\
  \end{array}
\label{fiveflows}
\end{equation}
As we are going to see this is a property shared by the entire
universality class of manifolds that have the same Tits Satake
projection. The Weyl group, the flow diagram on the links and the possible
flows realized on critical surfaces do not depend on the
representative inside the class but are just a property of the class.
\par
In the next section we just focus on the detailed discussion  of a few examples of flows for
this maximally split manifold.
\subsection{Examples for $\sym(4,\mathbb{R})$}
In this section, as we just announced, we consider three examples of
$\mathrm{Sp(4,\mathbb{R})}$-flows. One will be in the bulk the other
two will be located on two different critical surfaces. We analyse
these cases in detail both to show the  relation
between the asymptotic states and the structure of the orthogonal
group element $\mathcal{O}$ and to illustrate the billiard phenomenon.
In the plots of the Cartan fields we will be able to observe the multiple
bouncing of the cosmic ball on the hyperplanes orthogonal to some of
the roots.
\subsubsection{An example of flow in the bulk of parameter space: $\Omega_5 \, \Rightarrow \, \Omega_8$}
\begin{figure}[!hbt]
\begin{center}
\iffigs
 \includegraphics[height=65mm]{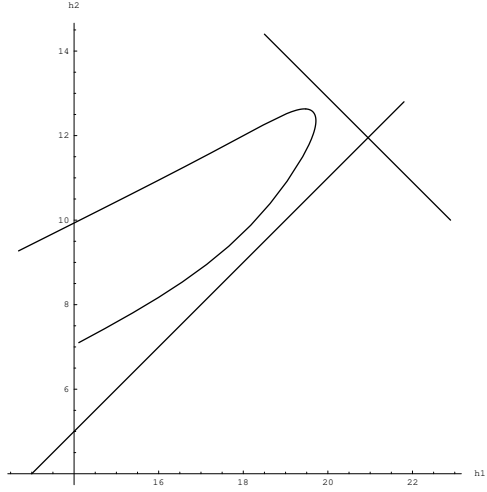}
\else
\end{center}
 \fi
\caption{\it Motion of the cosmic ball on the CSA billiard table of $\mathrm{Sp(4,\mathbb{R})}$ in a generic
bulk case. The choice of the angles is  $\theta_1 =\ft{\pi}{6}\, , \, \theta_2 = \ft{\pi}{4}\, ,
\,\theta_3 = \ft{\pi}{6}\, , \,\theta_4 =\ft{\pi}{3}$.
According to theory this motion realizes the smooth reflection $\Omega_2$ from the universal primordial Kasner era $\Omega_5$ to
the universal remote future Kasner era $\Omega_8$. This motion involves just two bounces on the wall respectively orthogonal to the root $\alpha_3 = (1,-1)$
and $\alpha_1 = (1,1)$. In this picture
the straight lines represent the walls orthogonal to $\alpha_1$ and $\alpha_3$, respectively. }
\label{Cbmotiongen2}
 \iffigs
 \hskip 1.5cm \unitlength=1.1mm
 \end{center}
  \fi
\end{figure}

\begin{figure}[!hbt]
\begin{center}
\iffigs
 \includegraphics[height=31mm]{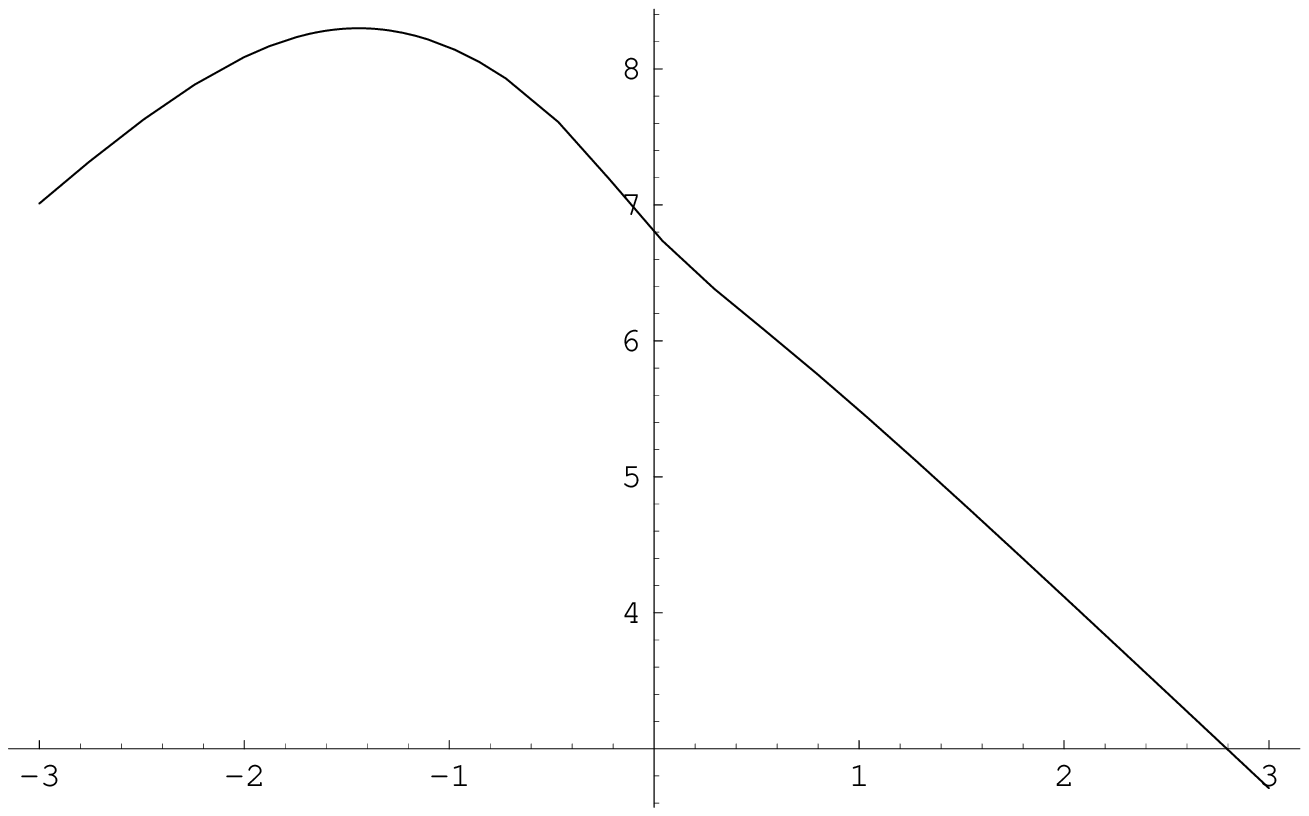}
 \includegraphics[height=31mm]{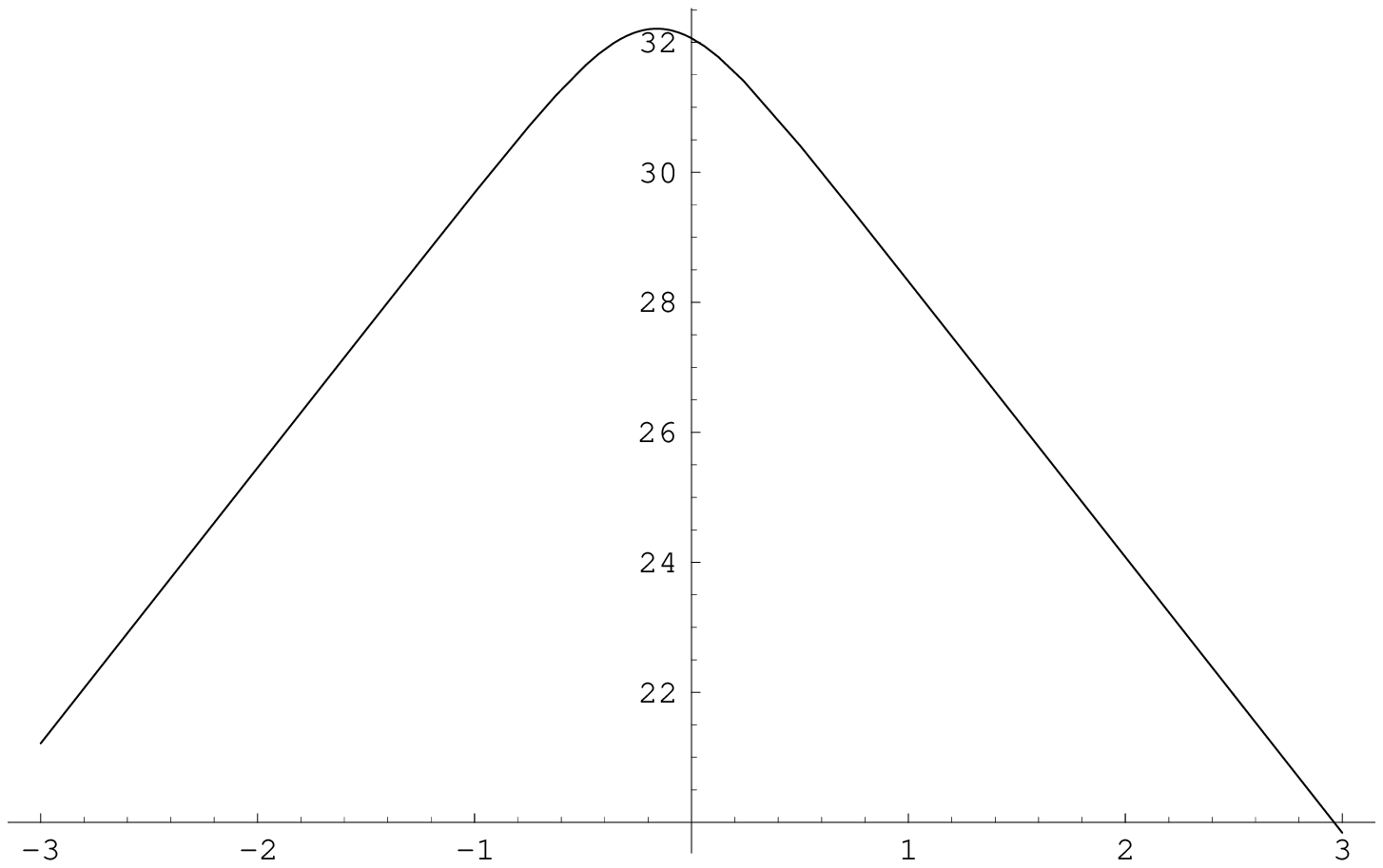}
\else
\end{center}
 \fi
\caption{\it Plot of ${\alpha}_{1,3} \cdot h$ projections for the $\mathrm{Sp(4,\mathbb{R})}$   generic
bulk flow generated by the parameter choice $\theta_1 =\ft{\pi}{6}\, , \, \theta_2 = \ft{\pi}{4}\, , \,\theta_3 = \ft{\pi}{6}\, ,
\,\theta_4 =\ft{\pi}{3}$which connects  the primordial Kasner era $\Omega_5$ to
the remote future Kasner era $\Omega_8$. There are two bounces in this flow because the projections ${\alpha_{1,3}} \cdot h$
have maxima at different instant of times. }
\label{a12gensp4max}
 \iffigs
 \hskip 1.5cm \unitlength=1.1mm
 \end{center}
  \fi
\end{figure}

If, as initial data we choose the following element of the compact
subgroup $\mathrm{U(2)} \subset \mathrm{Sp(4,\mathbb{R})}$:
\begin{eqnarray}
&& \mathrm{U(2)} \, \ni \, \mathcal{O} \, = \, \exp[\ft{\pi}{6}\, J_1]\,\, \exp[\ft{\pi}{4}\, J_2]\,
\, \exp[\ft{\pi}{6}\, J_3]\,\, \exp[\ft{\pi}{3}\, J_4]\, \,
\label{Hgenerica}
\end{eqnarray}
{\scriptsize
\begin{eqnarray}
  &&=\left(
\begin{array}{llll}
 \frac{1}{16} \left(6-\sqrt{2}+\sqrt{6}\right) &
   \frac{\sqrt{\frac{3}{2}}}{4} & \frac{1}{8}
   \sqrt{3} \left(2+\sqrt{2}\right) & \frac{1}{16}
   \left(\sqrt{2}-6 \sqrt{3}+\sqrt{6}\right) \\
 \frac{1}{16} \left(3 \sqrt{2}-2
   \sqrt{3}-\sqrt{6}\right) & \frac{3}{4 \sqrt{2}} &
   \frac{1}{8} \left(-2+3 \sqrt{2}\right) &
   \frac{1}{16} \left(6+3 \sqrt{2}+\sqrt{6}\right)
   \\
 \frac{1}{16} \left(-6-3 \sqrt{2}-\sqrt{6}\right) &
   \frac{1}{8} \left(2-3 \sqrt{2}\right) &
   \frac{3}{4 \sqrt{2}} & \frac{1}{16} \left(3
   \sqrt{2}-2 \sqrt{3}-\sqrt{6}\right) \\
 \frac{1}{16} \left(-\sqrt{2}+6
   \sqrt{3}-\sqrt{6}\right) & -\frac{1}{8} \sqrt{3}
   \left(2+\sqrt{2}\right) &
   \frac{\sqrt{\frac{3}{2}}}{4} & \frac{1}{16}
   \left(6-\sqrt{2}+\sqrt{6}\right)\nonumber\\
\end{array}
\right)
\end{eqnarray}
}
we are just in the bulk. Indeed as it is evident by inspection of
eq.(\ref{Hgenerica}), the
$4 \times 4$ matrix representing $\mathcal{O}$ has no minor with vanishing
determinant. Hence according to the advocated theorems we expect
asymptotic sorting of the eigenvalues. Indeed this is what happens.
Implementing by numerical evaluation the integration formula on a
computer we discover that the asymptotic form of the Lax operator at $t=-\infty$
corresponds to the Weyl group element $\Omega_5$
\begin{equation}
  \lim_{t \,\rightarrow \, -\infty} \, L(t) \, = \, \left(\begin{array}{cccc}
    2 & 0 & 0 & 0 \\
    0 & 1 & 0 & 0 \\
    0 & 0 & -1 & 0 \\
    0 & 0 & 0 & -2 \
  \end{array} \right) \, \Leftrightarrow \, \Omega_5~.
\label{minfgenersp4}
\end{equation}
Similarly at asymptotically late times the limit of the Lax operator
is that corresponding to the Weyl group element $\Omega_8$
\begin{equation}
  \lim_{t \,\rightarrow \, +\infty} \, L(t) \, = \, \left(\begin{array}{cccc}
    -2 & 0 & 0 & 0 \\
    0 & -1 & 0 & 0 \\
    0 & 0 & 1 & 0 \\
    0 & 0 & 0 & 2 \
  \end{array} \right) \, \Leftrightarrow \, \Omega_8~.
\label{pinfgenersp4}
\end{equation}
Algebraically we have $\Omega_8 \, = \, \Omega_2 \, \Omega_5$, so
that all generic flows in the bulk that avoid touching critical
surfaces are a smooth realization of the Weyl reflection $\Omega_2 \,
\in \, \mathcal{W}$. The particular smooth realization of this
reflection provided by the present choice of parameters is
illustrated in fig.\ref{Cbmotiongen2} which displays the motion
of the cosmic ball on the two dimensional billiard table whose
axes are the Cartan fields $h_1,h_2$.
This motion involves two bounces as it becomes evident by plotting
the projection of the Cartan vector $\overrightarrow{h} =\left(h_1,h_2
\right)$ along the two  roots $\alpha_1 = (1,-1) $ and
$\alpha_3=(1,1)$. These plots are displayed in fig.\ref{a12gensp4max}.
\subsubsection{An example of flow on the super-critical surface $\Sigma_9$:
$\Omega_6 \, \Rightarrow \, \Omega_8$}
As a next example we consider a flow confined on a super-critical surface.
If we set $\theta_2=\frac{\pi}{2}$ the $\mathrm{U(2)}$ matrix takes the
following form:
{\scriptsize
\begin{eqnarray}
  \mathcal{O}&=&\left(
\begin{array}{llll}
 \cos \left(\theta _1+\theta _3\right) \cos \left(\theta _4\right) & 0 & \sin \left(\theta _1+\theta
   _3\right) & -\cos \left(\theta _1+\theta _3\right) \sin \left(\theta _4\right) \\
 -\cos \left(\theta _4\right) \sin \left(\theta _1+\theta _3\right) & 0 & \cos \left(\theta _1+\theta
   _3\right) & \sin \left(\theta _1+\theta _3\right) \sin \left(\theta _4\right) \\
 -\sin \left(\theta _1+\theta _3\right) \sin \left(\theta _4\right) & -\cos \left(\theta _1+\theta
   _3\right) & 0 & -\cos \left(\theta _4\right) \sin \left(\theta _1+\theta _3\right) \\
 \cos \left(\theta _1+\theta _3\right) \sin \left(\theta _4\right) & -\sin \left(\theta _1+\theta
   _3\right) & 0 & \cos \left(\theta _1+\theta _3\right) \cos \left(\theta _4\right)
\end{array}
\right)\nonumber\\
\label{Hplaq134}
\end{eqnarray}
}
which has two notable properties. The first is that it has vanishing some
principal minors, the second is that it
actually depends on two variables only, namely $\theta_1 \, + \,
\theta_3$ and $\theta_4$.
\par
\begin{figure}[!hbt]
\begin{center}
\iffigs
 \includegraphics[height=70mm]{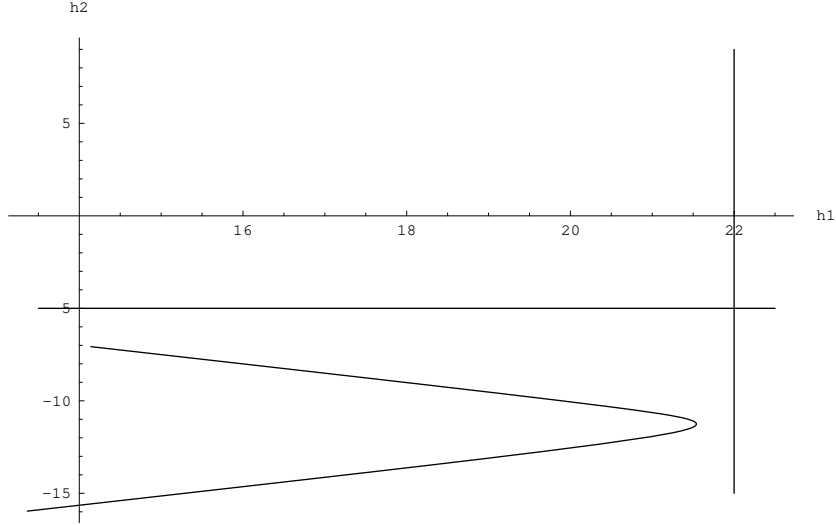}
\else
\end{center}
 \fi
\caption{\it Motion of the cosmic ball on the CSA billiard table of
$\mathrm{Sp(4,\mathbb{R})}$ in a
super-critical surface case. The choice of the angles is  $\theta_1 =\ft{\pi}{3}\, ,
\, \theta_2 = \ft{\pi}{2}\,
 , \,\theta_3 = \ft{\pi}{3}\, , \,\theta_4 =\ft{\pi}{3}$.
This motion realizes the smooth reflection $\Omega_3$ from the  Kasner era
$\Omega_6$ at $t=-\infty$ to
the  Kasner era $\Omega_8$ at $t=+\infty$. The two straight lines appearing in the picture are
the walls orthogonal to the roots $\alpha_2 =(0,2)$ and $\alpha_4=(2,0)$, respectively. As one sees the cosmic ball just bounces once the
on the $\alpha_4$ wall.}
\label{Cbmotionspec68}
 \iffigs
 \hskip 1.5cm \unitlength=1.1mm
 \end{center}
  \fi
\end{figure}
\begin{figure}[!hbt]
\begin{center}
\iffigs
 \includegraphics[height=35mm]{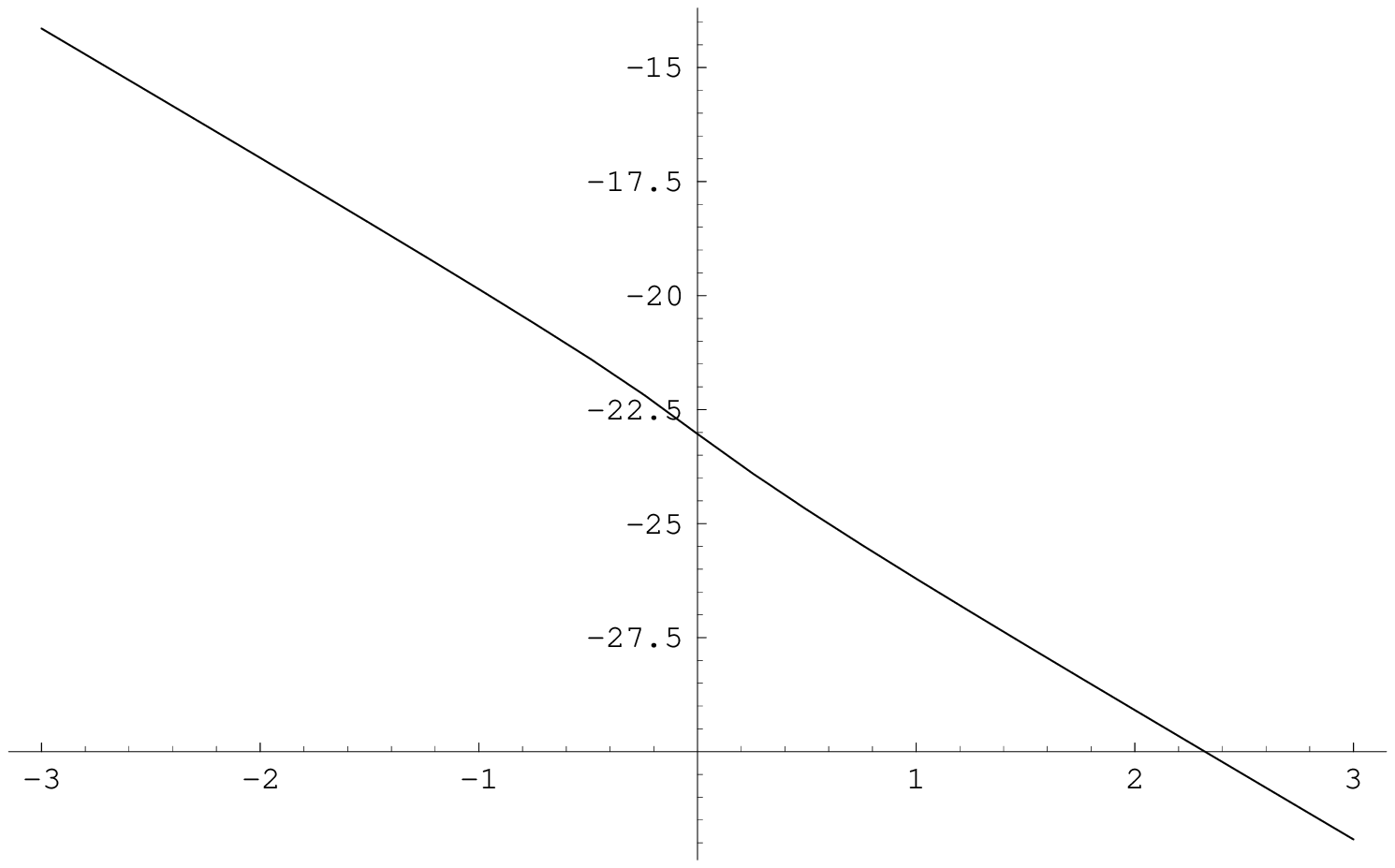}
 \includegraphics[height=35mm]{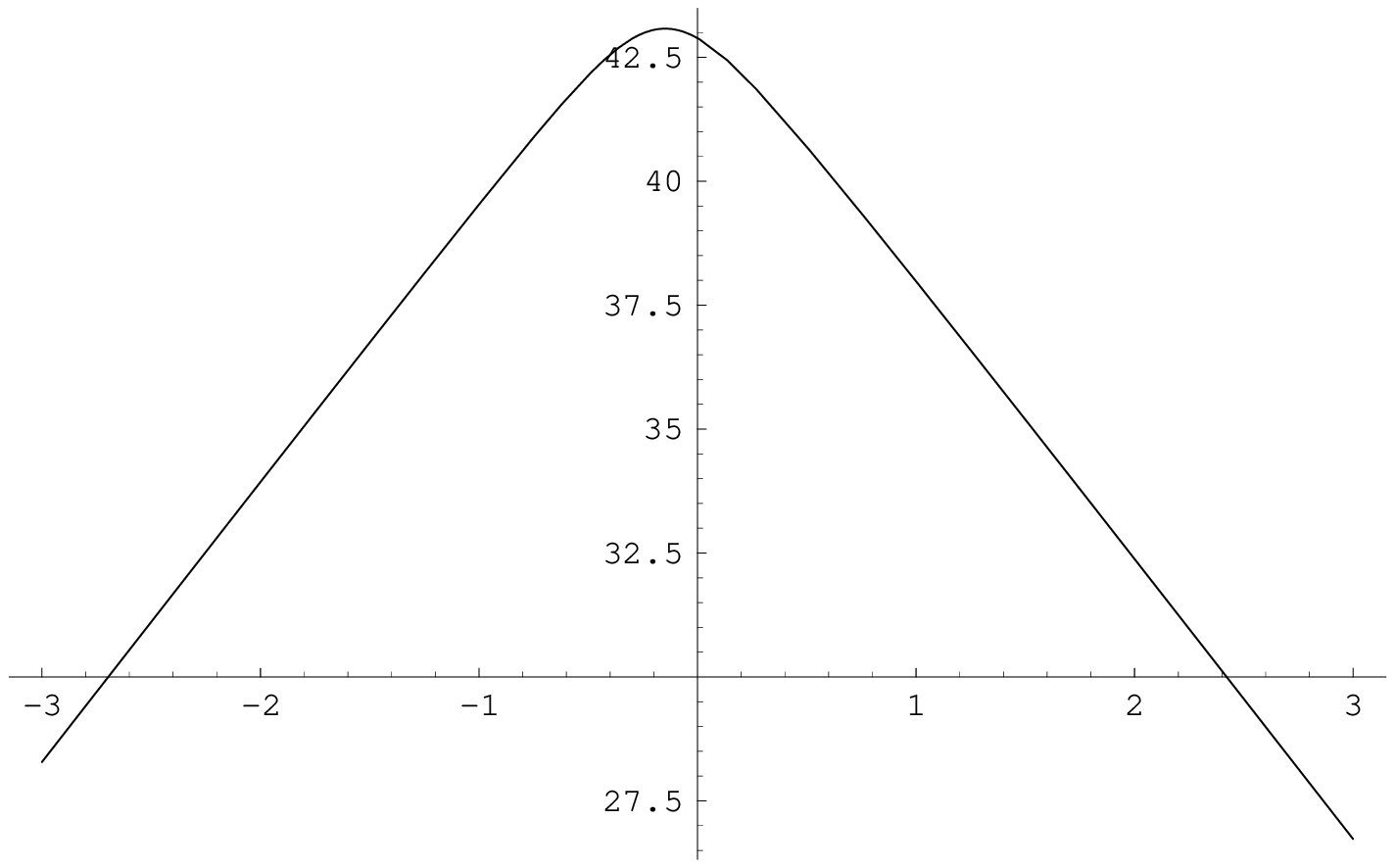}
\else
\end{center}
 \fi
\caption{\it Plot of ${\alpha}_{2,4} \cdot h$ projections for the $\mathrm{Sp(4,\mathbb{R})}$
flow on a super-critical surface generated by the parameter choice $\theta_1 =\ft{\pi}{3}\, , \,
\theta_2 = \ft{\pi}{2}\, , \,\theta_3 = \ft{\pi}{3}\, , \,\theta_4 =\ft{\pi}{3}$
which connects  the  past Kasner era $\Omega_6$ to
the  future Kasner era $\Omega_8$. There is just one bounce in this flow and this occurs on the
$\alpha_{4}$ wall. }
\label{a12spec68max}
 \iffigs
 \hskip 1.5cm \unitlength=1.1mm
 \end{center}
  \fi
\end{figure}

If we consider the equation for the critical super-surface $\Sigma_9$ we
find
\begin{equation}
  \cos \left(\theta _2\right) \cos ^2\left(\theta _3\right) \cos \left(\theta
   _4\right) \, = \, 0
\label{sigma9equa}
\end{equation}
so that the hyperplane $\theta_2=\frac{\pi}{2}$ where we have chosen our group
element is just one of the three components of $\Sigma_9$.
Furthermore, because of what we just observed, on this hyperplane
all points of the following form have to be identified:
\begin{equation}
\forall \, \phi \, \in \left[ 0, \ft{\pi}{2}\right] \, ;  \, \left(
\theta_1+\phi,\frac{\pi}{2},\theta_3-\phi,\theta_4\right) \, \sim \,\left(
\theta_1,\frac{\pi}{2},\theta_3,\theta_4\right)~.
\label{identifica}
\end{equation}
Having chosen initial data on a trapped surface, namely $\Sigma_9$,  we
do not expect full asymptotic sorting of the eigenvalues: actually, according to table \ref{14sigma} we expect flows
from $\Omega_6$ to $\Omega_8$. Indeed this is what happens.
We verify it in one example. For instance, as initial data we choose the following element of the compact
subgroup $\mathrm{U(2)} \subset \mathrm{Sp(4,\mathbb{R})}$, which lies in
the considered hypersurface:
\begin{eqnarray}
 \mathrm{U(2)} \, \ni \, \mathcal{O} & = & \exp[\ft{\pi}{3}\, J_1]\,  \exp[\ft{\pi}{2}\, J_2]\,
\, \exp[\ft{\pi}{3} \, J_3]\, \exp[\ft{\pi}{3} \, J_4]\nonumber\\
&=&\left(
\begin{array}{llll}
 -\frac{1}{4} & 0 & \frac{\sqrt{3}}{2} &
   \frac{\sqrt{3}}{4} \\
 -\frac{\sqrt{3}}{4} & 0 & -\frac{1}{2} &
   \frac{3}{4} \\
 -\frac{3}{4} & \frac{1}{2} & 0 &
   -\frac{\sqrt{3}}{4} \\
 -\frac{\sqrt{3}}{4} & -\frac{\sqrt{3}}{2} & 0 &
   -\frac{1}{4}
\end{array}
\right)~.
\label{Hspec}
\end{eqnarray}
Implementing by numerical evaluation the integration formula on a
computer we discover that the asymptotic form of the Lax operator at $t=-\infty$
corresponds to the Weyl group element $\Omega_6$ which implies no decreasing
sorting of the eigenvalues
\begin{equation}
  \lim_{t \,\rightarrow \, -\infty} \, L(t) \, = \, \left(\begin{array}{cccc}
    2 & 0 & 0 & 0 \\
    0 & -1 & 0 & 0 \\
    0 & 0 & 1 & 0 \\
    0 & 0 & 0 & -2 \
  \end{array} \right) \, \Leftrightarrow \, \Omega_6~.
\label{minf68}
\end{equation}
On the other hand the limit of the Lax operator at asymptotically late times
is that corresponding to the Weyl group element $\Omega_8$ which is
the same occurring in generic flows and yields increasing sorting of
the eigenvalues
\begin{equation}
  \lim_{t \,\rightarrow \, +\infty} \, L(t) \, = \, \left(\begin{array}{cccc}
    -2 & 0 & 0 & 0 \\
    0 & -1 & 0 & 0 \\
    0 & 0 & 1 & 0 \\
    0 & 0 & 0 & 2 \
  \end{array} \right) \, \Leftrightarrow \, \Omega_8~.
\label{pinf68}
\end{equation}
Algebraically we have $\Omega_8 \, = \, \Omega_3 \, \Omega_6$, so
that the  flows occurring on this super-critical
surface are smooth realizations of the Weyl reflection $\Omega_3 \,
\in \, \mathcal{W}$. The particular smooth realization of this
reflection encoded in this flow is
illustrated in fig.\ref{Cbmotionspec68} which displays the motion
of the cosmic ball on the two dimensional billiard table
This motion involves just one bounce on the wall orthogonal to the  root $\alpha_4$ as it becomes evident
by inspecting fig.\ref{a12spec68max}.
\subsubsection{An example of flow on the super-critical surface $\Sigma_2$: $\Omega_1 \, \Rightarrow \, \Omega_8$}
\label{critsigma2}
In the example considered below the initial
state is different  from that appearing in a generic bulk flow, namely there is
not decreasing sorting of the eigenvalues at past infinity which rather corresponds to the Weyl element $\Omega_1$.
Yet the end point at
$t=+\infty$ coincides with the universal one $\Omega_8$, namely there is  increasing
sorting at future infinity. We realize this situation
by choosing initial data on one of the critical surfaces, namely the surface
$\Sigma_2$.
\par
The equation for the trapped surface $\Sigma_2$, as defined in (\ref{criticonesp4U})
reads as follows:
\begin{equation}
  0 \, = \, \cos \left(\theta _3\right) \cos \left(\theta
   _4\right) \sin \left(\theta _1\right)+\cos
   \left(\theta _1\right) \sin \left(\theta
   _3\right) \sin \left(\theta _2-\theta _4\right)
\label{equaSig6}
\end{equation}
and it can be solved within the hypercube by expressing $\theta_1$ in
terms of the remaining three Euler angles as it follows:
\begin{equation}
  \theta_1 \, = \, \arccos\left(\frac{\cos \left(\theta _3\right)
   \cos \left(\theta _4\right)}{\sqrt{\cos
   ^2\left(\theta _3\right) \cos ^2\left(\theta
   _4\right)+\sin ^2\left(\theta _3\right) \sin
   ^2\left(\theta _2-\theta _4\right)}}\right)~.
\label{sigma2Solva}
\end{equation}
On the hypersurface $\Sigma_2$ we choose the particularly nice point
\begin{equation}
  \left\{ \frac{\pi}{3},\frac{\pi}{6},\frac{\pi}{3},\frac{\pi}{3}
  \right\} \, \in \, \Sigma_2
\label{specialpoint}
\end{equation}
which is easily seen to verify the defining equation (\ref{equaSig6})
and which leads to a quite simple form of the matrix $\mathcal{O}$.
With these values of the Euler angles
we obtain the following element of the maximally compact subgroup $\mathrm{U(2)} \subset
\mathrm{Sp(4,\mathbb{R})}$:
\begin{eqnarray}
 \mathrm{U(2)} \, \ni \, H_{sp} & = & \exp[\ft{\pi}{3}\, J_1]\,  \exp[\ft{\pi}{6}\, J_2]\,
\, \exp[\ft{\pi}{3} \, J_3]\, \exp[\ft{\pi}{3} \, J_4]\nonumber\\
&=&
\left(
\begin{array}{llll}
 \frac{1}{2} & \frac{3}{8} & \frac{3 \sqrt{3}}{8} &
   \frac{\sqrt{3}}{4} \\
 0 & \frac{\sqrt{3}}{8} & -\frac{5}{8} &
   \frac{3}{4} \\
 -\frac{3}{4} & \frac{5}{8} & \frac{\sqrt{3}}{8} &
   0 \\
 -\frac{\sqrt{3}}{4} & -\frac{3 \sqrt{3}}{8} &
   \frac{3}{8} & \frac{1}{2}
\end{array}
\right)
\label{Hspec2}
\end{eqnarray}
which indeed has vanishing $O_{2,1}$ matrix element as it is required by the definition
of the $\Sigma_2$ trapped surface. Hence
according to table \ref{14sigma} we expect a flow from $\Omega_1$ to $\Omega_8$.
Before proceeding to the integration of the Lax equation it is
interesting to consider the $\so(2,3)$ 5-dimensional representation of the same
$\mathrm{U(2)}$ group element. It is explicitly given by the following matrix:
\begin{equation}
H_{so} \, = \,  \left(
\begin{array}{lllll}
 \frac{\sqrt{3}}{16} & \frac{5}{16} & -\frac{3}{4
   \sqrt{2}} & -\frac{3}{16} & -\frac{7
   \sqrt{3}}{16} \\
 -\frac{19}{32} & \frac{11 \sqrt{3}}{32} & \frac{3
   \sqrt{\frac{3}{2}}}{8} & -\frac{5 \sqrt{3}}{32}
   & -\frac{3}{32} \\
 \frac{3 \sqrt{\frac{3}{2}}}{16} & \frac{15}{16
   \sqrt{2}} & -\frac{1}{8} & \frac{15}{16
   \sqrt{2}} & \frac{3 \sqrt{\frac{3}{2}}}{16} \\
 -\frac{3}{32} & -\frac{5 \sqrt{3}}{32} & \frac{3
   \sqrt{\frac{3}{2}}}{8} & \frac{11 \sqrt{3}}{32}
   & -\frac{19}{32} \\
 -\frac{7 \sqrt{3}}{16} & -\frac{3}{16} &
   -\frac{3}{4 \sqrt{2}} & \frac{5}{16} &
   \frac{\sqrt{3}}{16}
\end{array}
\right)~.
\label{so23repre}
\end{equation}
Since all the properties of the flows are intrinsic properties of the
group and cannot depend on the chosen representation it follows that
also the matrix (\ref{so23repre}) should be critical namely some of
its relevant minors (those obtained by intesecting the first
$k$-columns with $k$ arbitrary rows should vanish. Although not
evident at first sight, this is indeed true. Calculating the   minors we find that there are three relevant
$2 \times 2$ minors whose determinant vanishes, namely
\begin{eqnarray}
\mbox{Det} \, \left(
\begin{array}{ll}
 \frac{\sqrt{3}}{16} & \frac{5}{16} \\
 \frac{3 \sqrt{\frac{3}{2}}}{16} & \frac{15}{16 \sqrt{2}}
\end{array}
\right) & = & 0 \quad ; \quad \, \mbox{Det} \, \left(
\begin{array}{ll}
 \frac{\sqrt{3}}{16} & \frac{5}{16} \\
 -\frac{3}{32} & -\frac{5 \sqrt{3}}{32}
\end{array}
\right)\, = \, 0 ~,\\
 \mbox{Det}\left(
\begin{array}{ll}
 \frac{3 \sqrt{\frac{3}{2}}}{16} & \frac{15}{16 \sqrt{2}} \\
 -\frac{3}{32} & -\frac{5 \sqrt{3}}{32}
\end{array}
\right)&  = & 0~.
\label{minorini}
\end{eqnarray}
Hence the criticality condition is indeed intrinsic to the choice of
the group element and not to its specific representation as a matrix.
\par
\begin{figure}[!hbt]
\begin{center}
\iffigs
 \includegraphics[height=70mm]{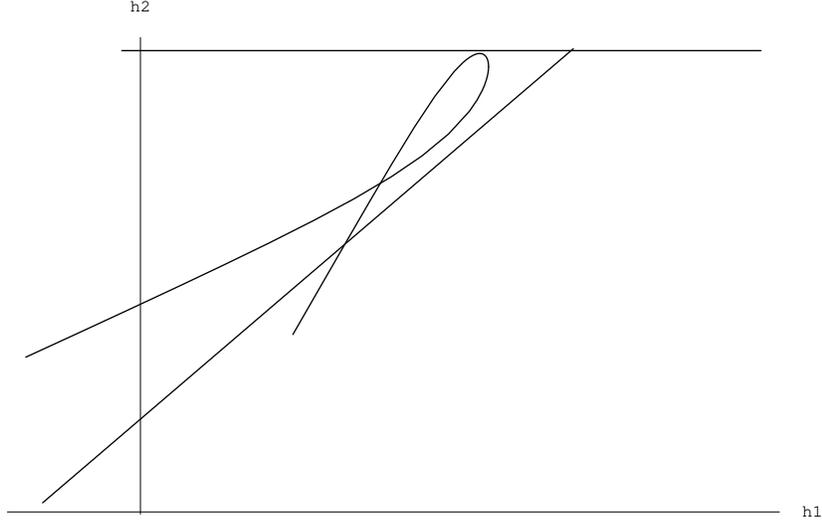}
\else
\end{center}
 \fi
\caption{\it Motion of the cosmic ball on the CSA billiard table of
$\mathrm{Sp(4,\mathbb{R})}$ in a
super-critical surface case. The choice of the angles is
$\theta_1 =\ft{\pi}{3}\, , \,
\theta_2 = \ft{\pi}{6}\,
 , \,\theta_3 = \ft{\pi}{3}\, , \,\theta_4 =\ft{\pi}{3}$ which lie on
 the trapped and super-critical surface $\Sigma_2$.
This motion realizes the smooth reflection $\Omega_8$ from the  Kasner era
$\Omega_1$ at $t=-\infty$ to
the  Kasner era $\Omega_8$ at $t=+\infty$. The peculiar knot appearing in this picture
implies the existence
of two bounces on the same root wall. The two straight lines displayed in the figure are the walls orthogonal
to the two simple roots $\alpha_1 = (1,-1)$ and $\alpha_2 = (0,2)$.
The ball bounces twice on the $\alpha_1$ wall.}
\label{Cbmotionspec52}
 \iffigs
 \hskip 1.5cm \unitlength=1.1mm
 \end{center}
  \fi
\end{figure}
\begin{figure}[!hbt]
\begin{center}
\iffigs
 \includegraphics[height=35mm]{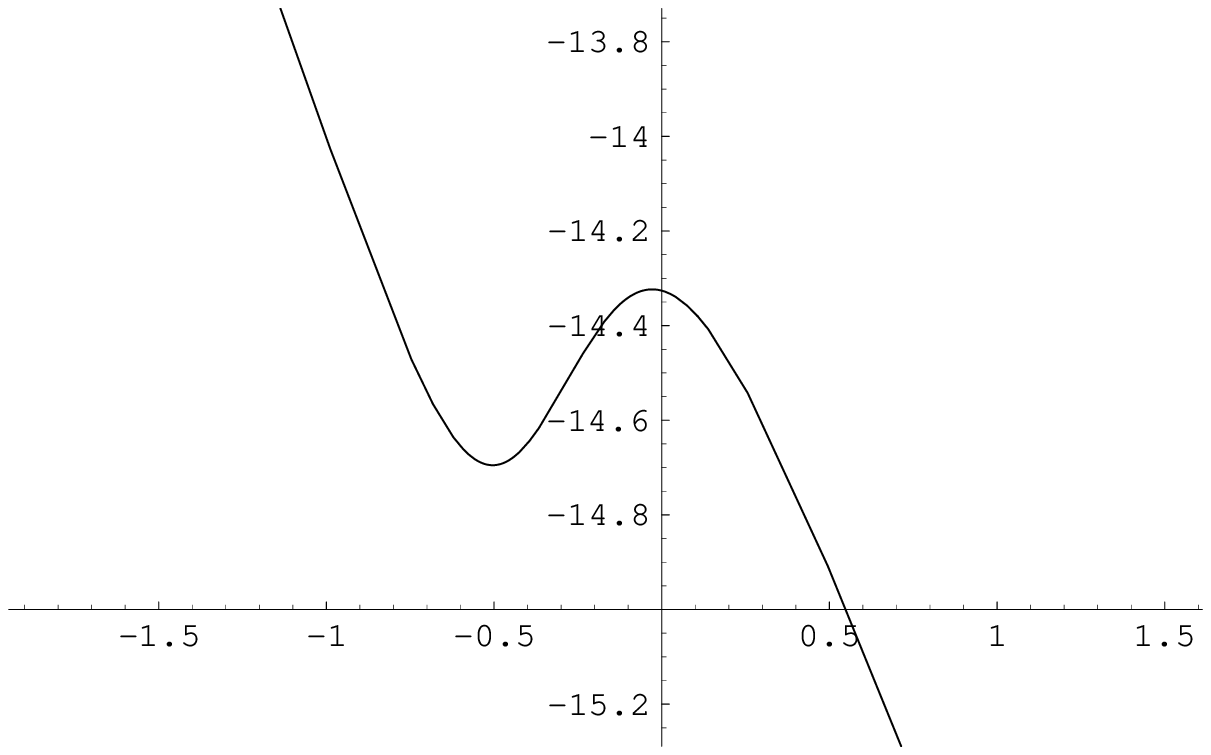}
 \includegraphics[height=35mm]{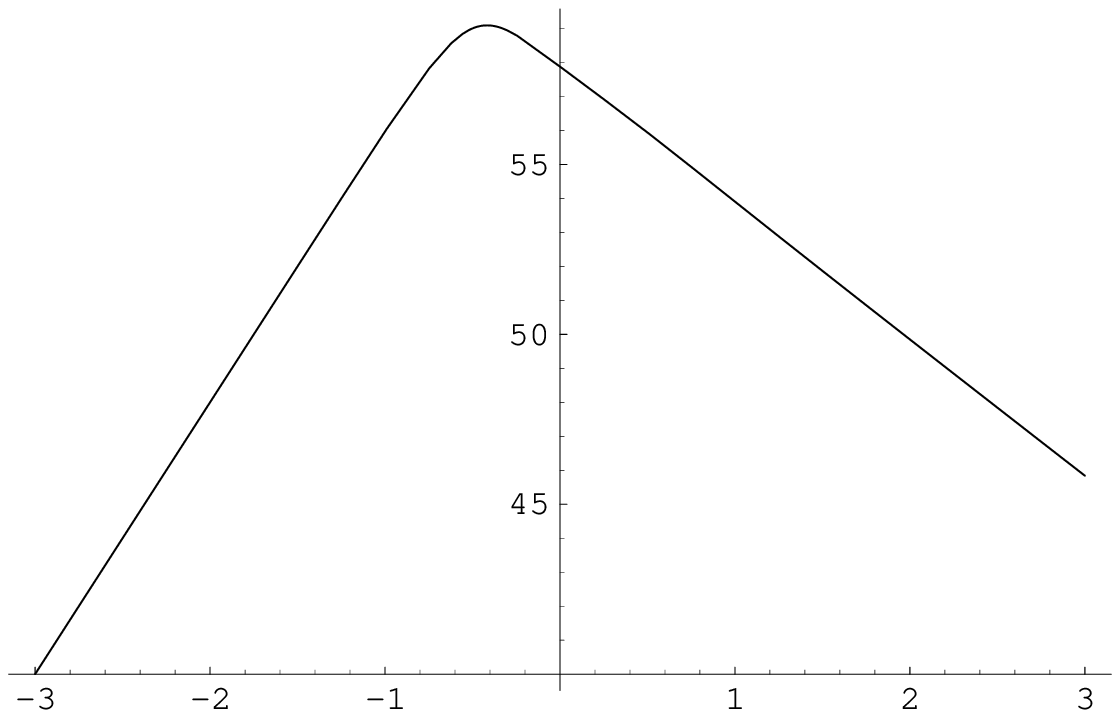}
\else
\end{center}
 \fi
\caption{\it Plot of ${\alpha}_{1,2} \cdot h$ projections for the $\mathrm{Sp(4,\mathbb{R})}$
flow on a super-critical surface generated by the parameter choice
$\theta_1 =\ft{\pi}{3}\, , \, \theta_2 = \ft{\pi}{6}\, , \,\theta_3 = \ft{\pi}{3}\, , \,\theta_4 =\ft{\pi}{3}$
which connects  the  past Kasner era $\Omega_1$ to
the  future Kasner era $\Omega_8$. The two bounces are clearly
visible in the maxima and minima of the first graph. }
\label{a12spec52max}
 \iffigs
 \hskip 1.5cm \unitlength=1.1mm
 \end{center}
  \fi
\end{figure}

Implementing by numerical evaluation the integration formula on a
computer we discover that the asymptotic form of the Lax operator at $t=-\infty$
corresponds to the Weyl group element $\Omega_1$ as expected:
\begin{equation}
  \lim_{t \,\rightarrow \, -\infty} \, L(t) \, = \, \left(\begin{array}{cccc}
    1 & 0 & 0 & 0 \\
    0 & 2 & 0 & 0 \\
    0 & 0 & -2 & 0 \\
    0 & 0 & 0 & -12 \
  \end{array} \right) \, \Leftrightarrow \, \Omega_1
\label{minf52}
\end{equation}
while the limit at asymptotically late times
is that corresponding to the Weyl group element $\Omega_8$ as we also expected
\begin{equation}
  \lim_{t \,\rightarrow \, +\infty} \, L(t) \, = \, \left(\begin{array}{cccc}
    -2 & 0 & 0 & 0 \\
    0 & -1 & 0 & 0 \\
    0 & 0 & 1 & 0 \\
    0 & 0 & 0 & 2 \
  \end{array} \right) \, \Leftrightarrow \, \Omega_8~.
\label{pinf52}
\end{equation}
Algebraically we have $\Omega_8 \, = \, \Omega_8 \, \Omega_1$, so
that the  flows occurring on this super-critical
surface are smooth realizations of the Weyl reflection $\Omega_8 \,
\in \, \mathcal{W}$. The  smooth realization of this
reflection encoded in the flow with these initial data
is illustrated in fig.\ref{Cbmotionspec52} which displays the motion
of the cosmic ball in the $h_1,h_2$ Cartan subalgebra plane.
This motion involves three bounces two on the wall orthogonal to the simple root $\alpha_1$
and one on the wall orthogonal to $\alpha_2$. This is clearly visible
by inspection of fig.\ref{a12spec52max}.
\section{The case of the $\so(r,r+2s)$ algebra}
\label{sorr2salgebra}
We are interested in considering the sigma model on the symmetric non compact coset manifold
\begin{equation}
  \mathcal{M}_{(r,2s)}\, = \, \frac{\mathrm{SO(r,r+2s)}}{\mathrm{SO(r)} \times \mathrm{SO(r+2s)}}~.
\label{p2qman}
\end{equation}
For $r=4$ the above manifold is quaternionic and corresponds to the
family of special geometries $L(0,P=2s)$.
\begin{figure}
\caption{\it The Dynkin diagram of the  $D_\ell$ Lie algebra.
\label{Dell}}
\centering
\begin{picture}(60,130)
\put (-70,65){$D_\ell$}
\put (-20,70){\circle {10}}
\put (-23,55){$\alpha_1$}
\put (-15,70){\line (1,0){20}}
\put (10,70){\circle {10}}
\put (7,55){$\alpha_2$}
\put (15,70){\line (1,0){20}}
\put (40,70){\circle {10}}
\put (37,55){$\alpha_3$}
\put (47,70){$\dots$}
\put (70,70){\circle {10}}
\put (67,55){$\alpha_{\ell-3}$}
\put (75,70){\line (1,0){25}}
\put (105,70){\circle {10}}
\put (100,55){$\alpha_{\ell-2}$}
\put (110,70){\line (1,1){20}}
\put (110,70){\line (1,-1){20}}
\put (133.2,93.2){\circle {10}}
\put (133.2,46.8){\circle {10}}
\put (143.2,93.2){$\alpha_{\ell-1}$}
\put (143.2,43.8){$\alpha_\ell$}
\end{picture}
\end{figure}
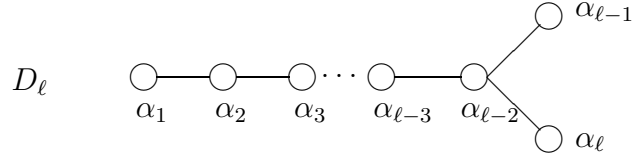
\subsection{The corresponding complex Lie algebra and root system}
The complex Lie algebra of which $\so(r,r+2s)$ is a non-compact real
section is just $D_\ell$ where
\begin{equation}
  \ell = r + s~.
\label{elldefi}
\end{equation}
The corresponding Dynkin
diagram is displayed in fig \ref{Dell} and the associated root system
 is realized by the following set of vectors in
$\mathbb{R}^\ell$:
\begin{equation}
  \Delta \equiv \left\{ \pm \,\epsilon ^A \, \pm \epsilon ^B \right\} \quad ;
  \quad \mbox{card} \, \Delta \, = \, 2\left( \ell^2 \, - \, \ell \right)
\label{roots}
\end{equation}
where $\epsilon ^A$ denotes an orthonormal basis of unit
vectors. The set of positive roots is then easily defined as follows:
\begin{equation}
  \alpha \, > \, 0 \quad \Rightarrow \, \alpha \, \in \, \Delta_+ \,
  \equiv \, \left\{ \epsilon ^A \, \pm \epsilon ^B \right\} \quad ( A < B )~.
\label{Deltapiu}
\end{equation}
A standard basis of simple roots
representing the Dynkin diagram \ref{Dell} is given by
\begin{eqnarray}
  \alpha _1 & = & \epsilon _1 \, - \, \epsilon _2~, \nonumber\\
  \alpha _2 & = & \epsilon _2 \, - \, \epsilon _3~, \nonumber\\
  \dots & \dots & \dots~, \nonumber\\
   \alpha _{\ell-1} & = & \epsilon _{\ell -1} \, - \, \epsilon _\ell~, \nonumber\\
   \alpha _{\ell} & = & \epsilon _{\ell -1} \, + \, \epsilon _\ell~.
\label{simplerute}
\end{eqnarray}
The maximally split real form of the $D_\ell$ Lie algebra is
$\so(\ell,\ell)$ and it is explicitly realized by the following
$2\ell \times 2\ell$ matrices. Let $e_{A,B}$ denote the $2\ell \times 2\ell$
matrix whose entries are all zero except the entry $A,B$ which is
equal to one. Then the Cartan generators $\mathcal{H}_A$ and the
positive root step operators $E^\alpha$ are represented as follows:
\begin{eqnarray}
\mathcal{H}_A & = & e_{A,A} \, - \, e_{A+\ell,A+\ell}~, \nonumber\\
E^{\epsilon _A \, - \, \epsilon _B} & = & e_{B,A} \, - \,
e_{A+\ell,B+\ell}~, \nonumber\\
E^{\epsilon _A \, + \, \epsilon _B} & = & e_{A+\ell,B} \, - \,
e_{B+\ell,A}~.
\label{trogut}
\end{eqnarray}
The solvable algebra of the maximally split coset
\begin{equation}
  \mathcal{M}_{(\ell,0)}\, = \, \frac{\mathrm{SO(\ell,\ell)}}{\mathrm{SO(\ell)} \times \mathrm{SO(\ell)}}
\label{ell2man}
\end{equation}
has therefore a very simple form in terms of matrices. Following the
general constructive principles $Solv_{(\ell,\ell)}$ is just the algebraic
span of all the matrices (\ref{trogut}) so that
\begin{equation}
  Solv_{(\ell,\ell)} \, \ni \, M \, \Leftrightarrow \, M \, = \, \left(
  \begin{array}{c|c}
    T & B \\
    \hline
    0 & - T^T \
  \end{array} \right) \quad ; \quad \left \{\begin{array}{rcl}
  T & = & \mbox{upper triangular}~, \\
  B & = & - \, B^T \quad \mbox{antisymmetric.}
\end{array} \right.
\label{solvellell}
\end{equation}
The matrices of the form (\ref{solvellell}) clearly form a subalgebra
of the  $\so(\ell,\ell)$ algebra which, in this representation, is
defined as the set of matrices $\Lambda$ fulfilling the following
condition:
\begin{equation}
  \Lambda^T \, \left(
  \begin{array}{c|c}
    0 & \mathbf{1}_{l} \\
    \hline
    \mathbf{1}_{l} & 0 \
  \end{array} \right) \, + \, \left(
  \begin{array}{c|c}
    0 & \mathbf{1}_{l} \\
    \hline
    \mathbf{1}_{l} & 0 \
  \end{array} \right)\Lambda \, = \, 0~.
\label{fertilina}
\end{equation}
\subsection{The real form $\so(r,r+2s)$ of the $D_{r+s}$ Lie algebra}
The main point in order to apply to the coset manifold (\ref{p2qman}) the general integration algorithm of
the Lax equation devised for the case
$\mathrm{SL(2\ell)}/\mathrm{SO(2\ell)}$ consists of introducing a
convenient basis of generators of the Lie algebra $\so(r,r+2s)$
where, in the fundamental representation, all elements of the
solvable Lie algebra associated with the coset under study turn out
to be given by upper triangular matrices. With some ingenuity such a
basis can be found by defining the $\so(r,r+2s)$ Lie algebra as the set
of matrices $\Lambda_t$ satisfying the following constraint:
\begin{equation}
  \Lambda_t^T \, \eta_t \, + \, \eta_t \, \Lambda_t \, = \, 0
\label{Lambdatdefi}
\end{equation}
where the symmetric invariant metric $\eta_t$ with $r+2s$ positive
eigenvalues $(+1)$
and $r$ negative ones $(-1)$ is given by the following matrix.
\begin{equation}
  \eta_t \, = \, \left(\begin{array}{c|c|c}
    0 & 0 & \varpi_r \\
    \hline
    0 & \mathbf{1}_{2s} & 0 \\
    \hline
    \varpi_r & 0 & 0 \
  \end{array} \right)~.
\label{etaT}
\end{equation}
In the above equation the symbol $\varpi_r$ denotes the completely
anti-diagonal $r \times r$ matrix which follows:
\begin{equation}
  \varpi_r \, = \, \left. \underbrace{\left(\begin{array}{cccccc}
    0 & 0 & \dots & \dots & 0 & 1 \\
    0 & 0 & \dots & \dots & 1 & 0 \\
    0 & 0 & \dots & 1 & 0 & 0 \\
    \dots & \dots & \dots & \dots & \dots & \dots \\
    0 & 1 & 0 & \dots & \dots & 0 \\
    1 & 0 & 0 & \dots & \dots & 0 \
  \end{array} \right)}_{r} \right \} \, r~.
\label{varpimatra}
\end{equation}
Obviously there is a simple orthogonal transformation which maps the
metric $\eta_t$ into the standard block diagonal metric $\eta_b$
written below
\begin{equation}
  \eta_b \, = \, \left( \begin{array}{c|c|c}
    \mathbf{1}_r & 0 & 0 \\
    \hline
    0 & \mathbf{1}_{2s} & 0 \\
    \hline
    0 & 0 & -\mathbf{1}_r \
  \end{array}\right)~.
\label{etab}
\end{equation}
Indeed we can write
\begin{equation}
  \Omega^T \, \eta_b \, \Omega \, = \, \eta_t
\label{omegatransf}
\end{equation}
where the explicit form of the matrix $\Omega$ is the following:
\begin{equation}
  \Omega \, = \, \left(\begin{array}{c|c|c}
    0 & \mathbf{1}_{2s} & 0 \\
    \hline
    \ft {1}{\sqrt{2}} \,\mathbf{1}_{r}  & 0 & \ft {1}{\sqrt{2}} \varpi_r \\
    \hline
        \ft {1}{\sqrt{2}} \,\mathbf{1}_{r} & 0 & - \ft {1}{\sqrt{2}} \varpi_r \
  \end{array} \right)~.
\label{Omega}
\end{equation}
Correspondingly the orthogonal transformation $\Omega$ maps the Lie
algebra and group elements of $\so(r,r+2s)$ from the standard basis
where the invariant metric is $\eta_b$ to the basis where it is
$\eta_t$
\begin{equation}
  \Lambda_t \, = \,  \Omega^T \, \Lambda_b \, \Omega \,.
\label{Lambdatb}
\end{equation}
In the $t$-basis the general form of an element of the solvable Lie
algebra which generates the coset manifold (\ref{p2qman}) has the
following appearance:
\begin{eqnarray}
  Solv\left(\frac{\mathrm{SO(r,r+2s)}}{\mathrm{SO(r)}\times \mathrm{SO(r+2s)}}\right)  \, \ni
  \, \Lambda_t & = & \left(\begin{array}{c|c|c}
    T & X & B \\
    \hline
    0 & 0 & X^T \, \varpi_r \\
    \hline
        0 & 0 & -  \varpi_r \, T^T \, \varpi_r \
  \end{array} \right)
\label{Solmatra}
\end{eqnarray}
where
\begin{eqnarray}
T & = &  \left(\begin{array}{cccccc}
    T_{1,1} & T_{1,2} & \dots & \dots & T_{1,r-1} & T_{1,r} \\
    0 & T_{2,2} & \dots & \dots & T_{2,r-1} & T_{2,r} \\
    0 & 0 & T_{3,3} & \dots & \dots & T_{3,r} \\
    \dots & \dots & \dots & \dots & \dots & \dots \\
    0 & 0 & 0 & \dots & T_{r-1,r-1} & T_{r-1,r} \\
    0 & 0 & 0 & \dots & \dots & T_{r,r} \
  \end{array} \right)  \, \quad \,\mbox{upper triangular $r \times r$~,} \nonumber\\
  B&=&-B^T \, \quad \, \mbox{antisymmetric $r \times r$~,}\nonumber\\
  X & = & \mbox{arbitrary $r \times 2s$}
\label{condizie}
\end{eqnarray}
while an element of the maximal compact subalgebra has instead the
following appearance:
\begin{eqnarray}
  \so(r) \, \oplus \, \so(r+2s) \, \ni
  \, \Lambda_t & = & \left(\begin{array}{c|c|c}
    Z & Y & C \, \varpi_r \\
    \hline
    -Y^T & Q & - \, Y^T \, \varpi_r \\
    \hline
        \varpi_r \, C & \varpi_r Y & -  \varpi_r \, Z^T \, \varpi_r \
  \end{array} \right)
\label{sosomatra}
\end{eqnarray}
where
\begin{eqnarray}
Z & = & - \, Z^T \,  \mbox{antisymmetric $r \times r$}~, \nonumber\\
C & = & - \, C^T \,  \mbox{antisymmetric $r \times r$}~, \nonumber\\
Q & = & - \, Q^T \,  \mbox{antisymmetric $2s \times 2s$}~. \nonumber\\
Y & = & \mbox{arbitrary $r \times 2s$}
\label{condesoso}
\end{eqnarray}
Having clarified the structure of the matrices representing Lie
algebra elements in this basis well adapted to the Tits Satake
projection, we can now discuss a basis of generators also well
adapted to the same projection.
To this effect, let us denote by $\mathcal{I}_{ij}$ the
$r \times r$ matrices whose only non vanishing entry is
the $ij$-th one which is equal to $1$
\begin{equation}
  \mathcal{I}_{ij} \, = \, \left(\begin{array}{ccccccl}
    0 & 0 & \dots & \dots & \dots & 0 &\null\\
    0 & 0 & \dots & \dots & \dots & 0 &\null\\
    \dots & \dots & \dots & \dots & \dots & \dots & \null\\
    0 & 0 & \dots & 1 & \dots & 0 &\} \, \mbox{i-th row}\\
     0 & 0 & \dots & \dots & \dots & 0 \\
    0 & 0 & \dots & \underbrace{\dots}_{\mbox{j-th column}} & \dots & 0 &\null \
  \end{array} \right)~.
\label{Imatri}
\end{equation}
Using this notation the $r$ non-compact Cartan generators are given by
\begin{equation}
  \mathcal{H}_i \, = \, \left(\begin{array}{c|c|c}
    \mathcal{I}_{ii} & 0 & 0 \\
    \hline
    0 & 0 & 0\\
    \hline
        0 & 0 & -  \varpi_r \, \mathcal{I}_{ii} \, \varpi_r \
  \end{array} \right) \,
  \quad ; \quad \, (i=1,\dots , r)~.
\label{cartani1}
\end{equation}
Next we introduce the coset generators associated with the long roots
of type: $\alpha = \epsilon^i - \epsilon^j$.
\begin{eqnarray}
\begin{array}{c}
   \mbox{\scriptsize{$\alpha = \epsilon^i - \epsilon^j$ }} \\
 \mbox{\scriptsize{ $i<j =1,\dots , r$}}
\end{array}
  & \Rightarrow & K^{ij}_-
  =  \ft {1}{\sqrt{2}} \,\left( E^\alpha + E^{-\alpha} \right)  =
  \ft {1}{\sqrt{2}} \left(\begin{array}{c|c|c}
    \mathcal{I}_{ij}  +  \mathcal{I}_{ji} & 0 & 0 \\
    \hline
    0 & 0 & 0\\
    \hline
        0 & 0 & -  \varpi_r \, \left( \mathcal{I}_{ij}  +  \mathcal{I}_{ji}\right)  \, \varpi_r \
  \end{array} \right)\nonumber\\
\label{Kijm}
\end{eqnarray}
and the coset generators associated with the long roots of type $\alpha = \epsilon^i +
\epsilon^j$:
\begin{eqnarray}
\begin{array}{c}
   \mbox{\scriptsize{$\alpha = \epsilon^i + \epsilon^j$ }} \\
 \mbox{\scriptsize{ $i<j =1,\dots , r$}}
\end{array}
  & \Rightarrow & K^{ij}_+
  =  \ft {1}{\sqrt{2}} \,\left( E^\alpha + E^{-\alpha} \right)  =
  \ft {1}{\sqrt{2}} \left(\begin{array}{c|c|c}
    0 & 0 & \left( \mathcal{I}_{ij}  -  \mathcal{I}_{ji}\right) \varpi_r \\
    \hline
    0 & 0 & 0\\
    \hline
       \varpi_r \left( \mathcal{I}_{ji}  -  \mathcal{I}_{ij}\right) & 0 & 0 \
  \end{array} \right)~.\nonumber\\
\label{Kijp}
\end{eqnarray}
The short roots, after the Tits-Satake projection, are just $r$,
namely $\epsilon^i$. Each of them, however, appears with multiplicity
$2s$, due to the paint group. We introduce a $2s$-tuple of coset
generators associated to each of the short roots in such a way that
such $2s$-tuple transforms in the fundamental representation of
$\mathbf{G}_{paint} \, = \, \so(2s)$. To this effect let us define
the rectangular $r\times 2s$ matrices$ \mathcal{J}_{im}$ analogous to the
square matrices $\mathcal{I}_{ij}$, namely
\begin{equation}
  \mathcal{J}_{im} \, = \, \left(\begin{array}{ccccccccl}
    0 & 0 & \dots & \dots & \dots &  \dots & \dots & 0 &\null\\
    0 & 0 & \dots & \dots & \dots & \dots & \dots & 0 &\null\\
    \dots & \dots & \dots & \dots & \dots & \dots \dots & \dots & \null\\
    0 & 0 & \dots & 1 & \dots &\dots & \dots & 0 &\} \, \mbox{i-th row}\\
     0 & 0 & \dots & \dots & \dots &\dots & \dots & 0 \\
    0 & 0 & \dots & \underbrace{\dots}_{\mbox{$m$-th column}} & \dots & \dots & \dots & 0 &\null \
  \end{array} \right)~.
\label{Jmatri}
\end{equation}
Then we introduce the following coset generators:
\begin{eqnarray}
\begin{array}{c}
   \mbox{\scriptsize{$\alpha = \epsilon^i $ }} \\
 \mbox{\scriptsize{ $i=1,\dots , r$}}\\
 \mbox{\scriptsize{ $m=1,\dots , 2s$}}\\
\end{array}
  & \Rightarrow & K^{i}_m
  =
  \ft {1}{\sqrt{2}} \left(\begin{array}{c|c|c}
    0 & \mathcal{J}_{im} & 0 \\
    \hline
    \mathcal{J}_{im}^T & 0 & - \mathcal{J}_{im}^T \, \varpi_r\\
    \hline
        0 & - \varpi_r \,\mathcal{J}_{im}  & 0 \
  \end{array} \right)~.\nonumber\\
\label{Kim}
\end{eqnarray}
The remaining generators of the $\so(r,r+2s)$ algebra are all compact
and span the subalgebra $\so(r)\oplus \so(r+2s)\subset \so(r,r+2s)$.
According to the nomenclature of eq.(\ref{sosomatra}) we introduce
four sets of generators. The first set is associated with the long
roots of type $\alpha = \epsilon ^i -\epsilon ^j$ and is defined as
follows:
\begin{equation}
  Z^{ij} \, = \, \ft {1}{\sqrt{2}} \,\left( E^\alpha - E^{-\alpha} \right)  =
  \ft {1}{\sqrt{2}} \left(\begin{array}{c|c|c}
    \mathcal{I}_{ij}  -  \mathcal{I}_{ji} & 0 & 0 \\
    \hline
    0 & 0 & 0\\
    \hline
        0 & 0 & -  \varpi_r \, \left( \mathcal{I}_{ij}  -  \mathcal{I}_{ji}\right)  \, \varpi_r \
  \end{array} \right)~.
\label{Zij}
\end{equation}
The second set is associated with the long roots of type $\alpha = \epsilon ^i +\epsilon
^j$ and is defined as follows:
\begin{equation}
  C^{ij}
  =  \ft {1}{\sqrt{2}} \,\left( E^\alpha - E^{-\alpha} \right)  =
  \ft {1}{\sqrt{2}} \left(\begin{array}{c|c|c}
    0 & 0 & \left( \mathcal{I}_{ij}  -  \mathcal{I}_{ji}\right) \varpi_r \\
    \hline
    0 & 0 & 0\\
    \hline
      - \varpi_r \left( \mathcal{I}_{ji}  -  \mathcal{I}_{ij}\right) & 0 & 0 \
  \end{array} \right)~.
\label{Cij}
\end{equation}
The third group of compact generators spans the compact coset
\begin{equation}
  \frac{\mathrm{SO(r+2s)}}{\mathrm{SO(r)} \times \mathrm{SO(2s)}}
\label{Kcoset}
\end{equation}
and it is given by
\begin{equation}
  Y^{i}_{m}
  =
  \ft {1}{\sqrt{2}} \left(\begin{array}{c|c|c}
    0 & \mathcal{J}_{im} & 0 \\
    \hline
    -\mathcal{J}_{im}^T & 0 & - \mathcal{J}_{im}^T \, \varpi_r\\
    \hline
        0 &  \varpi_r \,\mathcal{J}_{im}  & 0 \
  \end{array} \right)~.
\label{Yim}
\end{equation}
The fourth set of compact generators spans the paint group Lie algebra
$\so(2s)$ and is given by
\begin{equation}
  Q_{mn} \, = \, \left(\begin{array}{c|c|c}
    0 & 0 & 0 \\
    \hline
    0 & \mathcal{Q}_{mn} -\mathcal{Q}_{nm} & 0\\
    \hline
     0 & 0 & 0 \
  \end{array} \right)
\label{Qmn}
\end{equation}
where $\mathcal{Q}_{mn}$ denotes the analogue of the
$\mathcal{I}_{ij}$ in $2s$ rather than in $r$ dimensions.
\par
By performing the change of basis to the block diagonal form of the
matrices we can verify that $C_{ij} - Z_{ij}$ generate the $\so(r)$
subalgebra while $C_{ij} + Z_{ij}$ together with $Q_{mn}$ and
$Y_{im}$ generate the subalgebra $\so(r+2s)$.
\par
The full set of generators is ordered in the following way:
\begin{equation}
  T_\Lambda \, = \, \left\{ \underbrace{\mathcal{H}_i}_{r} \, , \, \underbrace{K^{ij}_- }_{\ft 12 r(r-1)} \, , \, \underbrace{K^{ij}_+}_{\ft 12 r(r-1)} \, , \,
  \underbrace{K^i_m}_{2rs} \, , \, \underbrace{Z^{ij}}_{\ft 12 r(r-1)} \, , \, \underbrace{C^{ij}}_{\ft 12 r(r-1)} \, , \, \underbrace{Y^i_{m}}_{2rs} \, ,
  \, \underbrace{Q_{mn}}_{s(2s-1)} \right\}
\label{allgenera}
\end{equation}
and satisfy the trace relation:
\begin{eqnarray}
  \mbox{Tr} \,\left ( T_\Lambda \, T_\Sigma \right) & = & g_{\Lambda \Sigma}~, \nonumber\\
g_{\Lambda \Sigma
  } & = & 2 \, \mbox{diag} \left(\underbrace{+,+,\dots, +}_{r(r+2s)},
  \underbrace{-,-,\dots,-}_{r^2-r + 2rs + 2s^2 -s} \right)~.
\label{tracciagenera}
\end{eqnarray}
In this way we have obtained the needed and detailed construction of
the embedding (\ref{embeddatone}) which is necessary to apply the
integration algorithm. In the next section we make a detailed study
of the case $r=2,s=1$.
\section{A case study for the Tits Satake projection: $\mathrm{SO(2,4)}$}
The simplest example of \textbf{not maximally split manifold} inside the series defined by
eq.(\ref{p2qman})  corresponds to the choice: $r=2, \, s=1$, namely
\begin{equation}
  \mathcal{M}_{2,2} \, \equiv \,
  \frac{\mathrm{SO(2,4)}}{\mathrm{SO(2) \times SO(4)}}~.
\label{so24defi}
\end{equation}
The Tits Satake projection yields the manifold studied at length in
section \ref{sectsp4}
\begin{equation}
  \Pi_{TS} \, : \, \frac{\mathrm{SO(2,4)}}{\mathrm{SO(2) \times
  SO(4)}} \, \mapsto \, \frac{\mathrm{SO(2,3)}}{\mathrm{SO(2) \times
  SO(3)}} \, \sim \, \frac{\mathrm{Sp(4,\mathbb{R})}}{\mathrm{U(2)}}
\label{Tsprojetta24}
\end{equation}
and the paint group is the simplest possible group
\begin{equation}
  \mathrm{G}_{paint} \, = \, \mathrm{SO(2)}~.
\label{soPaint}
\end{equation}
This manifold will be the target of our case study in order to illustrate the
bearing of the Tits Satake projection and the features of the Tits
Satake universality classes.
\par
Following the discussion of section \ref{sorr2salgebra}
we can organize the roots in a well adapted way for the Tits Satake
projection and introduce a basis where the solvable Lie algebra of
the coset is represented by upper triangular matrices.
\par
The root system associated with $\so(2,4)$ is actually that of
$D_3\sim A_3$ described by the Dynkin diagram which follows:
\begin{equation}
  \begin{picture}(60,130)
\put (-70,65){$D_3$}
\put (-20,70){\circle {10}}
\put (-23,55){$\alpha_1$}
\put (-15,70){\line (1,0){20}}
\put (10,70){\circle {10}}
\put (7,55){$\alpha_2$}
\put (15,70){\line (1,0){20}}
\put (40,70){\circle {10}}
\put (37,55){$\alpha_3$}
\end{picture}
\label{ornitoro}
\end{equation}
There are $6$ positive roots that are vectors in $\mathbb{R}^3$ and can be
organized as it follows:
\begin{equation}
\begin{array}{rclclcl}
\alpha_{1,1} & = & \epsilon ^2 \, - \, \epsilon^3 &
\stackrel{\Pi_{TS}}{\longrightarrow} & \epsilon^2 & \equiv & \alpha_1~, \\
\alpha_{1,2} & = & \epsilon ^2 \, + \, \epsilon^3 &
\stackrel{\Pi_{TS}}{\longrightarrow} & \epsilon^2 & \equiv & \alpha_1 ~,\\
\alpha_{2} & = & \epsilon ^1 \, - \, \epsilon^2 &
\stackrel{\Pi_{TS}}{\longrightarrow} & \epsilon ^1 \, - \, \epsilon^2 & \equiv &
\alpha_2~,\\
\alpha_{3,1} & = & \epsilon ^1 \, - \, \epsilon^3 &
\stackrel{\Pi_{TS}}{\longrightarrow} & \epsilon ^1  & \equiv &
\alpha_1+ \alpha_2~, \\
\alpha_{3,2} & = & \epsilon ^1 \, + \, \epsilon^3 &
\stackrel{\Pi_{TS}}{\longrightarrow} & \epsilon ^1  & \equiv &
\alpha_1+ \alpha_2 ~,\\
\alpha_{4} & = & \epsilon ^1 \, + \, \epsilon^2 &
\stackrel{\Pi_{TS}}{\longrightarrow} & \epsilon ^1 \, + \, \epsilon^2  & \equiv &
2 \alpha_1+ \alpha_2 ~. \
\end{array}
\label{so24rutte}
\end{equation}
In the above formulae the last three columns describe the Tits-Satake
projection of the root system which, in this case, is simply given by the
geometrical projection of the three--vectors onto the plane $\{12\}$. In
this way the correspondence with the $\sym(4,\mathbb{R})$ root system
becomes explicit (compare with fig.\ref{c2b2rutte}).
We have $2s=2$ preimages of each of the  short roots $\alpha_1$ and
$\alpha_1+\alpha_2$ while the long roots $\alpha_2$ and
$2\alpha_1+\alpha_2$ have a single preimage. In complete analogy with
eq.(\ref{so23fundrep}) we can define the appropriate basis for the
realization of the considered Lie algebra by giving the  explicit expression of
the most general element of the solvable Lie algebra $Solv\left(
\mathrm{SO(2,4)/SO(2)\times SO(4)}\right) $. Abstractly this is given by
\begin{eqnarray}
\mathcal{T} & = & h_1 \, \mathcal{H}_1 \, + \, h_2 \, \mathcal{H}_2
\, + \, e_{1,1} \, E^{\alpha_{1,1}} \, + \, e_{1,2} \, E^{\alpha_{1,2}} \,
\nonumber\\
\null  & \null & + \, e_{2} \, E^{\alpha_{2}} \, + \, e_{3,1} \, E^{\alpha_{3,1}} \, +
\, e_{3,2} \, E^{\alpha_{3,2}} \, +\, e_4 \, E^{\alpha_{4}}~.
\label{T24}
\end{eqnarray}
We define the form of all Cartan and step generators
by writing the same Lie algebra element (\ref{T24}) as a $6 \times 6$ matrix
 \begin{equation}
\mathcal{T} \, = \,  \left(
\begin{array}{llllll}
 h_1+h_2 & -\sqrt{2} e_2 & -\sqrt{2} e_{3,1} &
   -\sqrt{2} e_{3,2} & -\sqrt{2} e_4 & 0 \\
 0 & h_1-h_2 & -\sqrt{2} e_{1,1} & -\sqrt{2} e_{1,2}
   & 0 & \sqrt{2} e_4 \\
 0 & 0 & 0 & 0 & \sqrt{2} e_{1,1} & \sqrt{2} e_{3,1}
   \\
 0 & 0 & 0 & 0 & \sqrt{2} e_{1,2} & \sqrt{2} e_{3,2}
   \\
 0 & 0 & 0 & 0 & h_2-h_1 & \sqrt{2} e_2 \\
 0 & 0 & 0 & 0 & 0 & -h_1-h_2
\end{array}
\right)
\label{T24explicit}
\end{equation}
which satisfies the condition (\ref{Lambdatdefi}) with the metric
$\eta_t$ defined in eq.(\ref{etaT}).
\par
Then in full analogy with eq.s (\ref{Kgenerisp4},\ref{HHgenerisp4})
we can construct a basis for the subspace $\mathbb{K}$ and for the subalgebra
$\mathbb{H}$ by writing
\begin{eqnarray}
\mathrm{K}_1 & = & \mathcal{H}_1 \nonumber~,\\
\mathrm{K}_2 & = & \mathcal{H}_2 \nonumber~,\\
\mathrm{K}_3 & = & \ft{1}{\sqrt{2}}\left( E^{\alpha_{1,1}} \, + \, (E^{\alpha_{1,1}})^T\right)~,  \nonumber\\
\mathrm{K}_4 & = & \ft{1}{\sqrt{2}}\left( E^{\alpha_{1,2}} \, + \, (E^{\alpha_{1,2}})^T\right)~,  \nonumber\\
\mathrm{K}_5 & = & \ft{1}{\sqrt{2}}\left( E^{\alpha_2} \, + \, (E^{\alpha_2})^T\right)~, \nonumber\\
\mathrm{K}_6 & = & \ft{1}{\sqrt{2}}\left( E^{\alpha_{3,1}} \, + \, (E^{\alpha_{3,1}})^T\right)~, \nonumber\\
\mathrm{K}_7 & = & \ft{1}{\sqrt{2}}\left( E^{\alpha_{3,2}} \, + \, (E^{\alpha_{3,2}})^T\right)~, \nonumber\\
\mathrm{K}_8 & = & \ft{1}{\sqrt{2}}\left( E^{\alpha_{4}} \, + \, (E^{\alpha_{4}})^T\right)
\label{Kgeneriso24}
\end{eqnarray}
and
\begin{eqnarray}
\mathrm{J}_1 & = & \ft{1}{\sqrt{2}}\left( E^{\alpha_{1,1}} \, - \, (E^{\alpha_{1,1}})^T\right)~,  \nonumber\\
\mathrm{J}_2 & = & \ft{1}{\sqrt{2}}\left( E^{\alpha_{1,2}} \, - \, (E^{\alpha_{1,2}})^T\right)~,  \nonumber\\
\mathrm{J}_3 & = & \ft{1}{\sqrt{2}}\left( E^{\alpha_2} \, - \, (E^{\alpha_2})^T\right)~, \nonumber\\
\mathrm{J}_4 & = & \ft{1}{\sqrt{2}}\left( E^{\alpha_{3,1}} \, - \, (E^{\alpha_{3,1}})^T\right)~, \nonumber\\
\mathrm{J}_5 & = & \ft{1}{\sqrt{2}}\left( E^{\alpha_{3,2}} \, - \, (E^{\alpha_{3,2}})^T\right) \nonumber\\
\mathrm{J}_6 & = & \ft{1}{\sqrt{2}}\left( E^{\alpha_{4}} \, - \, (E^{\alpha_{4}})^T\right)~.
\label{HHgeneriso2}
\end{eqnarray}
In this way we have constructed $8+6=14$ generators. One is still missing to
complete a $15$-dimensional basis for the Lie algebra $\so(2,4)$. The
missing item is $Q$, namely the generator of the paint group $\mathrm{SO(2)}$
\begin{equation}
  Q=\left(\begin{array}{cc|cc|cc}
    0 & 0 & 0 & 0 & 0 & 0 \\
    0 & 0 & 0 & 0 & 0 & 0 \\
    \hline
    0 & 0 & 0 & 1 & 0 & 0 \\
    0 & 0 & -1 & 0 & 0 & 0 \\
    \hline
    0 & 0 & 0 & 0 & 0 & 0 \\
    0 & 0 & 0 & 0 & 0 & 0 \
  \end{array} \right)~.
\label{Paintgenera}
\end{equation}
Naively one might think that the six generators $\mathrm{J}_i$ defined
in (\ref{HHgeneriso2}) close the Lie algebra of $\so(4)$, while $Q$
generates the factor $\so(2)$ in the denominator group of our
manifold $\frac{\mathrm{SO(2,4)}}{\mathrm{SO(2) \times SO(4)}}$. From
the discussion of the previous section \ref{sorr2salgebra} we know
that this is not the case. Indeed the paint group is inside the
factor $\mathrm{SO(r+2s)}$ so that the listed $\mathrm{J}_i$
constitute a tangent basis for the coset manifold
\begin{equation}
  \widetilde{\mathcal{P}} \, = \, \mathrm{\mathrm{SO(2)}} \, \times \, \frac{\mathrm{\mathrm{SO(4)}}}{\mathrm{SO(2)}_{paint}}
\label{Paraspazio}
\end{equation}
which is the universal covering of the true parameter space for the integration of our Lax
equation. The actual $\mathcal{P}$ is obtained from
$\widetilde{\mathcal{P}}$ by modding out the generalized Weyl group
as stated in equation (\ref{bunduparametru}).
\par
Having established these notations we can just proceed to the
construction of the initial data in the usual way. The Cartan
subalgebra element is given by the following $6 \times 6$ matrix:
\begin{equation}
  \mathcal{C} \, = \, \left(\begin{array}{cc|cc|cc}
    h_1+h_2 & 0 & 0 & 0 & 0 & 0 \\
    0 &  h_1 -h_2 &0 & 0 & 0 & 0 \\
    \hline
    0 & 0 & 0 & 0 & 0 & 0 \\
    0 & 0 & 0 & 0 & 0 & 0 \\
    \hline
    0 & 0 & 0 & 0 &  -h_1+h_2 & 0 \\
    0 & 0 & 0 & 0 & 0 & -h_1 -h_2 \
  \end{array} \right)
\label{CSA24}
\end{equation}
while the orthogonal matrix $\mathcal{O} \, \in \, \mathcal{P}$ can
be defined in complete analogy to eq.(\ref{Omatra44}):
\begin{eqnarray}
  &&\mathcal{O}\left(\theta_1,\,\dots,\theta_6\right) \, = \, \nonumber\\
  &&\exp \left[\sqrt{2} \,\theta_1 \, \mathrm{J_1} \right] \, \exp \left[\sqrt{2} \,\theta_2 \, \mathrm{J_2} \right]\,
  \exp \left[\theta_3 \, \mathrm{J_3}  \right] \,
  \exp \left[ \sqrt{2} \,\theta_4 \, \mathrm{J_4} \right] \,
  \exp \left[ \sqrt{2} \,\theta_5 \, \mathrm{J_5} \right]\, \exp \left[\theta_6 \, \mathrm{J_6}
  \right] \, \,\nonumber\\
&&\quad\quad\quad\quad\quad\quad\quad\quad = \left(\begin{array}{cccccc}
  O_{11} & O_{12} & O_{13} & O_{14} & O_{15} & O_{16} \\
  O_{21} & O_{22} & O_{23} & O_{24} & O_{25} & O_{26} \\
  O_{31} & O_{32} & O_{33} & O_{34} & O_{35} & O_{36} \\
  O_{41} & O_{42} & O_{43} & O_{44} & O_{45} & O_{46} \\
  O_{51} & O_{52} & O_{53} & O_{54} & O_{55} & O_{56} \\
  O_{61} & O_{62} & O_{63} & O_{64} & O_{65} & O_{66}\
\end{array} \right)~.
\label{Omatra66}
\end{eqnarray}
We do not write the explicit functional form of the $36$ entries
because it takes too much space yet it is clear that they are
uniquely defined by the above equation and by the explicit form of
the generators. We just go over to discuss the Weyl group.
\subsection{The generalized Weyl group for $\mathrm{SO(2,4)}$}
Applying the procedure of definition \ref{genWeyl}, we
introduce six generators for the generalized Weyl group corresponding to the
reflections with respect to the $6$ roots. These can be represented
as the rotation matrices
\begin{equation}
  \gamma_i \, = \, \mathcal{O}\left(
  \underbrace{0,\dots,0,}_{i-1}\frac{\pi}{2},\underbrace{0,\dots,0}_{6-i}\right)
  \quad ; \quad i=1,\dots,6
\label{Wgenera}
\end{equation}
which are integer valued. Considering all products and all relations
among these generators we obtain the finite group $\mathcal{W}\left( \so(2,4) \right)$
which has 32 elements. The group $\mathcal{W}\left( \so(2,4) \right)$ has a normal
subgroup
\begin{equation}
  \mathbb{Z}_2 \, \times  \, \mathbb{Z}_2 \, \sim \, \mathrm{N}\left( \so(2,4) \right) \,
  \subset \, \mathcal{W}\left( \so(2,4) \right)
\label{normalsubgroup}
\end{equation}
given by the following four diagonal matrices:
\begin{equation}
  \begin{array}{ccccccc}
    n_1 & = & \left( \begin{array}{cccccc}
      1 & 0 & 0 & 0 & 0 & 0 \\
      0 & 1 & 0 & 0 & 0 & 0 \\
      0 & 0 & 1 & 0 & 0 & 0 \\
      0 & 0 & 0 & 1 & 0 & 0 \\
      0 & 0 & 0 & 0 & 1 & 0 \\
      0 & 0 & 0 & 0 & 0 & 1 \\
    \end{array}\right)  & ; & n_2 & = & \left( \begin{array}{cccccc}
      -1 & 0 & 0 & 0 & 0 & 0 \\
      0 & -1 & 0 & 0 & 0 & 0 \\
      0 & 0 & 1 & 0 & 0 & 0 \\
      0 & 0 & 0 & 1 & 0 & 0 \\
      0 & 0 & 0 & 0 & -1 & 0 \\
      0 & 0 & 0 & 0 & 0 & -1 \\
    \end{array}\right)~, \\
    \null & \null & \null & \null & \null & \null \\
    n_3 & = & \left( \begin{array}{cccccc}
      1 & 0 & 0 & 0 & 0 & 0 \\
      0 & 1 & 0 & 0 & 0 & 0 \\
      0 & 0 & -1 & 0 & 0 & 0 \\
      0 & 0 & 0 & -1 & 0 & 0 \\
      0 & 0 & 0 & 0 & 1 & 0 \\
      0 & 0 & 0 & 0 & 0 & 1 \\
    \end{array}\right) & ; & n_4 & = & \left( \begin{array}{cccccc}
      -1 & 0 & 0 & 0 & 0 & 0 \\
      0 & -1 & 0 & 0 & 0 & 0 \\
      0 & 0 & -1 & 0 & 0 & 0 \\
      0 & 0 & 0 & -1 & 0 & 0 \\
      0 & 0 & 0 & 0 & -1 & 0 \\
      0 & 0 & 0 & 0 & 0 & -1 \\
    \end{array}\right)~. \
  \end{array}
\label{Nullgroup}
\end{equation}
The normal subgroup $\mathrm{N}\left( \so(2,4) \right)$ when acting by similarity
transformation on a   Cartan subalgebra element of the form
(\ref{CSA24}) leaves it invariant
\begin{equation}
  \forall \, n \, \in \, \mathrm{N}\left( \so(2,4) \right) \, : \, n^T \, \mathcal{C} \, n \, = \, \mathcal{C}~.
\label{stabiloN}
\end{equation}
The order $8$ factor group obtained by modding $\mathcal{W}\left( \so(2,4) \right)$ with respect
to $\mathrm{N}\left( \so(2,4) \right)$ is isomorphic to the Weyl group of the Tits Satake
subalgebra $\mathrm{Weyl}\left( \sym(4)\right) $ and has the same action on the eigenvalues
$h_1,h_2$
\begin{equation}
  \frac{\mathcal{W}\left( \so(2,4) \right)}{\mathrm{N}\left( \so(2,4) \right)} \, \simeq \,
  \mathrm{Weyl}\left( \sym(4)\right)~.
\label{logicamente}
\end{equation}
A representative for each of the eight equivalence classes can be
easily written. We find
\begin{equation}
  \begin{array}{ccccccc}
    \Lambda_1 & = & \left(
\begin{array}{llllll}
 1 & 0 & 0 & 0 & 0 & 0 \\
 0 & 1 & 0 & 0 & 0 & 0 \\
 0 & 0 & 1 & 0 & 0 & 0 \\
 0 & 0 & 0 & 1 & 0 & 0 \\
 0 & 0 & 0 & 0 & 1 & 0 \\
 0 & 0 & 0 & 0 & 0 & 1
\end{array}
\right) & ; & \Lambda_2 & = & \left(
\begin{array}{llllll}
 0 & 0 & 0 & 0 & 0 & 1 \\
 0 & 0 & 0 & 0 & 1 & 0 \\
 0 & 0 & 1 & 0 & 0 & 0 \\
 0 & 0 & 0 & 1 & 0 & 0 \\
 0 & 1 & 0 & 0 & 0 & 0 \\
 1 & 0 & 0 & 0 & 0 & 0
\end{array}
\right)~, \\
    \Lambda_3 & = & \left(
\begin{array}{llllll}
 0 & 0 & 0 & 0 & 0 & 1 \\
 0 & -1 & 0 & 0 & 0 & 0 \\
 0 & 0 & 1 & 0 & 0 & 0 \\
 0 & 0 & 0 & -1 & 0 & 0 \\
 0 & 0 & 0 & 0 & -1 & 0 \\
 1 & 0 & 0 & 0 & 0 & 0
\end{array}
\right) & ; & \Lambda_4 & = & \left(
\begin{array}{llllll}
 1 & 0 & 0 & 0 & 0 & 0 \\
 0 & 0 & 0 & 0 & -1 & 0 \\
 0 & 0 & 1 & 0 & 0 & 0 \\
 0 & 0 & 0 & -1 & 0 & 0 \\
 0 & -1 & 0 & 0 & 0 & 0 \\
 0 & 0 & 0 & 0 & 0 & 1
\end{array}
\right)~, \\
    \Lambda_5 & = & \left(
\begin{array}{llllll}
 0 & 1 & 0 & 0 & 0 & 0 \\
 -1 & 0 & 0 & 0 & 0 & 0 \\
 0 & 0 & 1 & 0 & 0 & 0 \\
 0 & 0 & 0 & 1 & 0 & 0 \\
 0 & 0 & 0 & 0 & 0 & -1 \\
 0 & 0 & 0 & 0 & 1 & 0
\end{array}
\right) & ; & \Lambda_6 & = & \left(
\begin{array}{llllll}
 0 & 1 & 0 & 0 & 0 & 0 \\
 0 & 0 & 0 & 0 & 0 & 1 \\
 0 & 0 & 1 & 0 & 0 & 0 \\
 0 & 0 & 0 & -1 & 0 & 0 \\
 1 & 0 & 0 & 0 & 0 & 0 \\
 0 & 0 & 0 & 0 & 1 & 0
\end{array}
\right)~, \\
    \Lambda_7 & = & \left(
\begin{array}{llllll}
 0 & 0 & 0 & 0 & 1 & 0 \\
 1 & 0 & 0 & 0 & 0 & 0 \\
 0 & 0 & 1 & 0 & 0 & 0 \\
 0 & 0 & 0 & -1 & 0 & 0 \\
 0 & 0 & 0 & 0 & 0 & 1 \\
 0 & 1 & 0 & 0 & 0 & 0
\end{array}
\right) & ; & \Lambda_8 & = & \left(
\begin{array}{llllll}
 0 & 0 & 0 & 0 & 1 & 0 \\
 0 & 0 & 0 & 0 & 0 & -1 \\
 0 & 0 & 1 & 0 & 0 & 0 \\
 0 & 0 & 0 & 1 & 0 & 0 \\
 -1 & 0 & 0 & 0 & 0 & 0 \\
 0 & 1 & 0 & 0 & 0 & 0
\end{array}
\right)~. \
  \end{array}
\label{representative8}
\end{equation}
Assembling the information presented above we come to a stronger
conclusion. Not only the factor group is isomorphic to the Weyl
group of the Tits Satake projection but even the generalized Weyl
group is isomorphic. Indeed we have found
\begin{equation}
  \mathcal{W}\left(\so(2,4) \right) \, \sim \, \mathcal{W}\left(\sym(4) \right)~.
\label{strongerconclusion}
\end{equation}
We have not proved so far that this is true in general but it is an
attractive conjecture to postulate that
\begin{equation}
  \mathcal{W}\left(\mathbb{U}\right) \, \sim \, \mathcal{W}\left(\mathbb{U}_{\mathrm{TS}}\right)~.
\label{congettura}
\end{equation}
We leave the proof of such a conjecture to future publications.
\subsection{Vertices, edges and trapped surfaces}
\begin{figure}[!hbt]
\begin{center}
\iffigs
 \includegraphics[height=60mm]{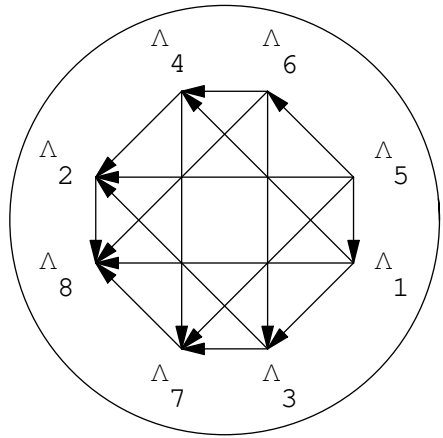}
\else
\end{center}
 \fi
\caption{\it The oriented phase diagram of the $\mathrm{SO(2,4)}/\mathrm{SO(2)\times SO(4)}$ flows. The Lie algebra $\so(2,4)$ is
\textbf{not maximally split} and its Tits Satake subalgebra is $\sym(4,\mathbb{R}) \sim \so(2,3) \, \subset \, \so(2,4)  $.
The relevant Weyl group is that of the Tits Satake subalgebra and the flow diagram for the integration of the Lax equation on this space
just coincides with that of $\sym(4,\mathbb{R})$. At fixed value of the Cartan fields if $h_1,h_2,-h_1,-h_2$ are the
eigenvalues of the $\sym(4,\mathbb{R})$ Lax operator, those of the
$\so(2,4)$ Lax operator are $h_1+h_2,h_1-h_2,0,0,-h_1-h_2,-h_1+h_2$.
Using $\{h_1+h_2,h_1-h_2\}$ as coordinates to identify the Weyl
element we obtain the presented flow graph.}
\label{so24graphus}
 \iffigs
 \hskip 1.5cm \unitlength=1.1mm
 \end{center}
  \fi
\end{figure}

By means of a computer programme we can now study the vertices, the
links, the critical surfaces and the accessible vertices on each
critical surface. We summarize the results.
\paragraph{Vertices}
Our parameter space is now a $6$ dimensional hypercube that has $64$
vertices and $192$ edges. On each of the $64$ vertices we find one of
the $8$ Weyl elements which obviously reappears several times. Each
of the odd elements $\Lambda_{1,3,5,7}$ appears $4$ times, while each
of the even elements $\Lambda_{2,4,6,8}$ appears $12$ times so that we
have $4 \times 4 \, + \, 4 \times 12 \, = \, 64$. The $64$ vertices
with their Weyl element correspondence are listed in table
\ref{so24verticini}.
\paragraph{Edges}
The one dimensional links connecting the $64$ vertices are $192$ and
each of them represents a flow from one lower Weyl element to a
higher one. A priori we might expect that the lines connecting the
$8$ Weyl elements could now be more numerous than in the case of the Tits
Satake projected manifold. However calculating all
these links on a computer  we find that the independent lines are just $16$ and the
same $16$ appearing in  the Tits Satake projection. This is made
evident by the flow diagram displayed in fig. \ref{so24graphus}
\paragraph{Trapped surfaces and accessible vertices}
The trapped hypersurfaces in parameter space are obtained by
equating to zero the minors obtained by intersecting the first $k$
columns of $\mathcal{O}$ with an equal number of arbitrarily chosen
rows. In this way we generate a total of $62$ trapped surfaces. They
are enumerated as follows:
\begin{equation}
\begin{array}{|c|c|}
\hline
  \mbox{Order of the minor} & \mbox{Number}  \\
  \hline
  5 & 6  \\
  4 & 15 \\
  3 & 20  \\
  2 & 15  \\
  1 & 6  \\
  \hline
  \null & 62 \\
  \hline
\end{array}
\label{conteggi}
\end{equation}
{\scriptsize
\begin{table}
  \centering
  $$
  \begin{array}{|c|c|}
  \hline
  \begin{array}{ccccc}
  1 ) & ; &  \{0, 0, 0, 0, 0,
      0\}  & = &  \Lambda_1 \
      \\
2 ) & ; &  \{1, 0, 0, 0, 0,
      0\}  & = &  \Lambda_5 \
 \\
3 ) & ; &  \{0, 1, 0, 0, 0,
      0\}  & = &  \Lambda_5 \
      \\
4 ) & ; &  \{0, 0, 1, 0, 0,
      0\}  & = &  \Lambda_4 \ \\
5 ) & ; &  \{0, 0, 0, 1, 0,
      0\}  & = &  \Lambda_8 \
 \\
6 ) & ; &  \{0, 0, 0, 0, 1,
      0\}  & = &  \Lambda_8 \
 \\
7 ) & ; &  \{0, 0, 0, 0, 0,
      1\}  & = &  \Lambda_3 \
 \\
8 ) & ; &  \{1, 1, 0, 0, 0,
      0\}  & = &  \Lambda_1 \
 \\
9 ) & ; &  \{1, 0, 1, 0, 0,
      0\}  & = &  \Lambda_6 \
\\
10 ) & ; &  \{1, 0, 0, 1, 0,
      0\}  & = &  \Lambda_2 \
\\
11 ) & ; &  \{1, 0, 0, 0, 1,
      0\}  & = &  \Lambda_2 \
 \\
12 ) & ; &  \{1, 0, 0, 0, 0,
      1\}  & = &  \Lambda_7 \
 \\
13 ) & ; &  \{0, 1, 1, 0, 0,
      0\}  & = &  \Lambda_6 \
 \\
14 ) & ; &  \{0, 1, 0, 1, 0,
      0\}  & = &  \Lambda_2 \
 \\
15 ) & ; &  \{0, 1, 0, 0, 1,
      0\}  & = &  \Lambda_2 \
 \\
16 ) & ; &  \{0, 1, 0, 0, 0,
      1\}  & = &  \Lambda_7 \
 \\
17 ) & ; &  \{0, 0, 1, 1, 0,
      0\}  & = &  \Lambda_6 \
 \\
18 ) & ; &  \{0, 0, 1, 0, 1,
      0\}  & = &  \Lambda_6 \
\\
19 ) & ; &  \{0, 0, 1, 0, 0,
      1\}  & = &  \Lambda_2 \
 \\
20 ) & ; &  \{0, 0, 0, 1, 1,
      0\}  & = &  \Lambda_1 \
 \\
21 ) & ; &  \{0, 0, 0, 1, 0,
      1\}  & = &  \Lambda_6 \
 \\
22 ) & ; &  \{0, 0, 0, 0, 1,
      1\}  & = &  \Lambda_6 \
 \\
23 ) & ; &  \{1, 1, 1, 0, 0,
      0\}  & = &  \Lambda_4 \
 \\
24 ) & ; &  \{1, 1, 0, 1, 0,
      0\}  & = &  \Lambda_8 \
 \\
25 ) & ; &  \{1, 1, 0, 0, 1,
      0\}  & = &  \Lambda_8 \
 \\
26 ) & ; &  \{1, 1, 0, 0, 0,
      1\}  & = &  \Lambda_3 \
 \\
27 ) & ; &  \{1, 0, 1, 1, 0,
      0\}  & = &  \Lambda_4 \
 \\
28 ) & ; &  \{1, 0, 1, 0, 1,
      0\}  & = &  \Lambda_4 \
 \\
29 ) & ; &  \{1, 0, 1, 0, 0,
      1\}  & = &  \Lambda_8 \
 \\
30 ) & ; &  \{1, 0, 0, 1, 1,
      0\}  & = &  \Lambda_5 \
 \\
31 ) & ; &  \{1, 0, 0, 1, 0,
      1\}  & = &  \Lambda_4 \
 \\
32 ) & ; &  \{1, 0, 0, 0, 1,
      1\}  & = &  \Lambda_4 \
 \
\end{array} &
\begin{array}{ccccc}
33 ) & ; &  \{0, 1, 1, 1, 0,
      0\}  & = &  \Lambda_4 \
 \\
34 ) & ; &  \{0, 1, 1, 0, 1,
      0\}  & = &  \Lambda_4 \
 \\
35 ) & ; &  \{0, 1, 1, 0, 0,
      1\}  & = &  \Lambda_8 \
 \\
36 ) & ; &  \{0, 1, 0, 1, 1,
      0\}  & = &  \Lambda_5 \
 \\
37 ) & ; &  \{0, 1, 0, 1, 0,
      1\}  & = &  \Lambda_4 \
 \\
38 ) & ; &  \{0, 1, 0, 0, 1,
      1\}  & = &  \Lambda_4 \
 \\
39 ) & ; &  \{0, 0, 1, 1, 1,
      0\}  & = &  \Lambda_4 \
 \\
40 ) & ; &  \{0, 0, 1, 1, 0,
      1\}  & = &  \Lambda_8 \
 \\
41 ) & ; &  \{0, 0, 1, 0, 1,
      1\}  & = &  \Lambda_8 \
 \\
42 ) & ; &  \{0, 0, 0, 1, 1,
      1\}  & = &  \Lambda_3 \
 \\
43 ) & ; &  \{1, 1, 1, 1, 0,
      0\}  & = &  \Lambda_6 \
 \\
44 ) & ; &  \{1, 1, 1, 0, 1,
      0\}  & = &  \Lambda_6 \
 \\
45 ) & ; &  \{1, 1, 1, 0, 0,
      1\}  & = &  \Lambda_2 \
 \\
46 ) & ; &  \{1, 1, 0, 1, 1,
      0\}  & = &  \Lambda_1 \
 \\
47 ) & ; &  \{1, 1, 0, 1, 0,
      1\}  & = &  \Lambda_6 \
 \\
48 ) & ; &  \{1, 1, 0, 0, 1,
      1\}  & = &  \Lambda_6 \
 \\
49 ) & ; &  \{1, 0, 1, 1, 1,
      0\}  & = &  \Lambda_6 \
 \\
50 ) & ; &  \{1, 0, 1, 1, 0,
      1\}  & = &  \Lambda_2 \
 \\
51 ) & ; &  \{1, 0, 1, 0, 1,
      1\}  & = &  \Lambda_2 \
 \\
52 ) & ; &  \{1, 0, 0, 1, 1,
      1\}  & = &  \Lambda_7 \
 \\
53 ) & ; &  \{0, 1, 1, 1, 1,
      0\}  & = &  \Lambda_6 \
 \\
54 ) & ; &  \{0, 1, 1, 1, 0,
      1\}  & = &  \Lambda_2 \
 \\
55 ) & ; &  \{0, 1, 1, 0, 1,
      1\}  & = &  \Lambda_2 \
 \\
56 ) & ; &  \{0, 1, 0, 1, 1,
      1\}  & = &  \Lambda_7 \
 \\
57 ) & ; &  \{0, 0, 1, 1, 1,
      1\}  & = &  \Lambda_2 \
 \\
58 ) & ; &  \{1, 1, 1, 1, 1,
      0\}  & = &  \Lambda_4 \
 \\
59 ) & ; &  \{1, 1, 1, 1, 0,
      1\}  & = &  \Lambda_8 \
 \\
60 ) & ; &  \{1, 1, 1, 0, 1,
      1\}  & = &  \Lambda_8 \
 \\
61 ) & ; &  \{1, 1, 0, 1, 1,
      1\}  & = &  \Lambda_3 \
 \\
62 ) & ; &  \{1, 0, 1, 1, 1,
      1\}  & = &  \Lambda_8 \
 \\
63 ) & ; &  \{0, 1, 1, 1, 1,
      1\}  & = &  \Lambda_8 \
 \\
64 ) & ; &  \{1, 1, 1, 1, 1,
      1\}  & = &  \Lambda_2 \
 \\
  \end{array}\\
  \hline
  \end{array}
  $$
  \caption{Vertices/Weyl group  correspondence for the case
  $\mathrm{SO(2,4)}$. }\label{so24verticini}
\end{table}
}
We can now calculate the set of accessible Weyl elements for each of these
$62$ surfaces and within the accessible set we can single out the
lowest and the highest Weyl elements which will correspond to the
initial and final end points of the flows confined on that
surface. The result is displayed in table \ref{flussoni}.
\begin{table}
  \centering
  $$
  \begin{array}{|c|c|}
  \hline
 \begin{array}{cc}
 \Sigma _1  &  \left\{w_6,
      w_8\right\} \\
\Sigma _2  &  \left\{w_5,
      w_8\right\} \\
\Sigma _3  &  \left\{w_5,
      w_8\right\} \\
\Sigma _4  &  \left\{w_5,
      w_8\right\} \\
\Sigma _5  &  \left\{w_5,
      w_8\right\} \\
\Sigma _6  &  \left\{w_5,
      w_7\right\} \\
\Sigma _7  &  \left\{w_5,
      w_8\right\} \\
\Sigma _8  &  \left\{w_5,
      w_8\right\} \\
\Sigma _9  &  \left\{w_5,
      w_8\right\} \\
\Sigma _{10}  &  \left\{w_5,
      w_8\right\} \\
\Sigma _{11}  &  \left\{w_5,
      w_8\right\} \\
\Sigma _{12}  &  \left\{w_5,
      w_8\right\} \\
\Sigma _{13}  &  \left\{w_1,
      w_8\right\} \\
\Sigma _{14}  &  \left\{w_5,
      w_8\right\} \\
\Sigma _{15}  &  \left\{w_5,
      w_8\right\} \\
\Sigma _{16}  &  \left\{w_5,
      w_8\right\} \\
\Sigma _{17}  &  \left\{w_5,
      w_8\right\} \\
\Sigma _{18}  &  \left\{w_5,
      w_2\right\} \\
\Sigma _{19}  &  \left\{w_5,
      w_8\right\} \\
\Sigma _{20}  &  \left\{w_5,
      w_8\right\} \\
\Sigma _{21}  &  \left\{w_5,
      w_8\right\} \\
\Sigma _{22}  &  \left\{w_5,
      w_8\right\} \\
\Sigma _{23}  &  \left\{w_5,
      w_8\right\} \\
\Sigma _{24}  &  \left\{w_5,
      w_8\right\} \\
\Sigma _{25}  &  \left\{w_5,
      w_8\right\} \\
\Sigma _{26}  &  \left\{w_5,
      w_8\right\} \\
\Sigma _{27}  &  \left\{w_1,
      w_8\right\} \\
\Sigma _{28}  &  \left\{w_5,
      w_8\right\} \\
\Sigma _{29}  &  \left\{w_5,
      w_8\right\} \\
\Sigma _{30}  &  \left\{w_5,
      w_8\right\} \\
\Sigma _{31}  &  \left\{w_5,
      w_8\right\} \\
      \end{array}
      &
      \begin{array}{cc}
\Sigma _{32}  &  \left\{w_5,
      w_8\right\} \\
\Sigma _{33}  &  \left\{w_5,
      w_8\right\} \\
\Sigma _{34}  &  \left\{w_5,
      w_2\right\} \\
\Sigma _{35}  &  \left\{w_5,
      w_8\right\} \\
\Sigma _{36}  &  \left\{w_5,
      w_8\right\} \\
\Sigma _{37}  &  \left\{w_5,
      w_8\right\} \\
\Sigma _{38}  &  \left\{w_5,
      w_8\right\} \\
\Sigma _{39}  &  \left\{w_5,
      w_8\right\} \\
\Sigma _{40}  &  \left\{w_5,
      w_8\right\} \\
\Sigma _{41}  &  \left\{w_5,
      w_8\right\} \\
\Sigma _{42}  &  \left\{w_5,
      w_8\right\} \\
\Sigma _{43}  &  \left\{w_5,
      w_8\right\} \\
\Sigma _{44}  &  \left\{w_5,
      w_8\right\} \\
\Sigma _{45}  &  \left\{w_1,
      w_8\right\} \\
\Sigma _{46}  &  \left\{w_5,
      w_8\right\} \\
\Sigma _{47}  &  \left\{w_5,
      w_8\right\} \\
\Sigma _{48}  &  \left\{w_5,
      w_8\right\} \\
\Sigma _{49}  &  \left\{w_5,
      w_8\right\} \\
\Sigma _{50}  &  \left\{w_5,
      w_2\right\} \\
\Sigma _{51}  &  \left\{w_5,
      w_8\right\} \\
\Sigma _{52}  &  \left\{w_5,
      w_8\right\} \\
\Sigma _{53}  &  \left\{w_5,
      w_8\right\} \\
\Sigma _{54}  &  \left\{w_5,
      w_8\right\} \\
\Sigma _{55}  &  \left\{w_5,
      w_8\right\} \\
\Sigma _{56}  &  \left\{w_5,
      w_8\right\} \\
\Sigma _{57}  &  \left\{w_6,
      w_8\right\} \\
\Sigma _{58}  &  \left\{w_5,
      w_8\right\} \\
\Sigma _{59}  &  \left\{w_5,
      w_8\right\} \\
\Sigma _{60}  &  \left\{w_5,
      w_8\right\} \\
\Sigma _{61}  &  \left\{w_5,
      w_8\right\} \\
\Sigma _{62}  &  \left\{w_5,
      w_7\right\} \\
\end{array}\\
\hline
\end{array}
  $$
  \caption{Initial and final endpoints of flows confined on trapped surfaces
  for the case $\mathrm{SO(2,4)}$.}\label{flussoni}
\end{table}
Inspection of this list reveals that the available flows, although
repeated on many different surfaces are just a small set of five
possibilities, exactly the same five possible flows appearing in the
the case of the Tits Satake projection that were shown in eq.(\ref{fiveflows})
\par
This concludes our discussion. As we have seen the vertices and the
possible flows on critical links or trapped hypersurfaces do not depend on
the chosen representative within a Tits Satake universality class
rather they depend only on the class. In other words the study of the
maximally split Tits Satake projection already provides us with a
complete picture of all possible flows. It is only the detailed
structure of bouncing which varies from one representative to the
other.
\subsection{Examples of flows for $\mathrm{SO(2,4)}$}
\begin{figure}[!hbt]
\begin{center}
\iffigs
 \includegraphics[height=45mm]{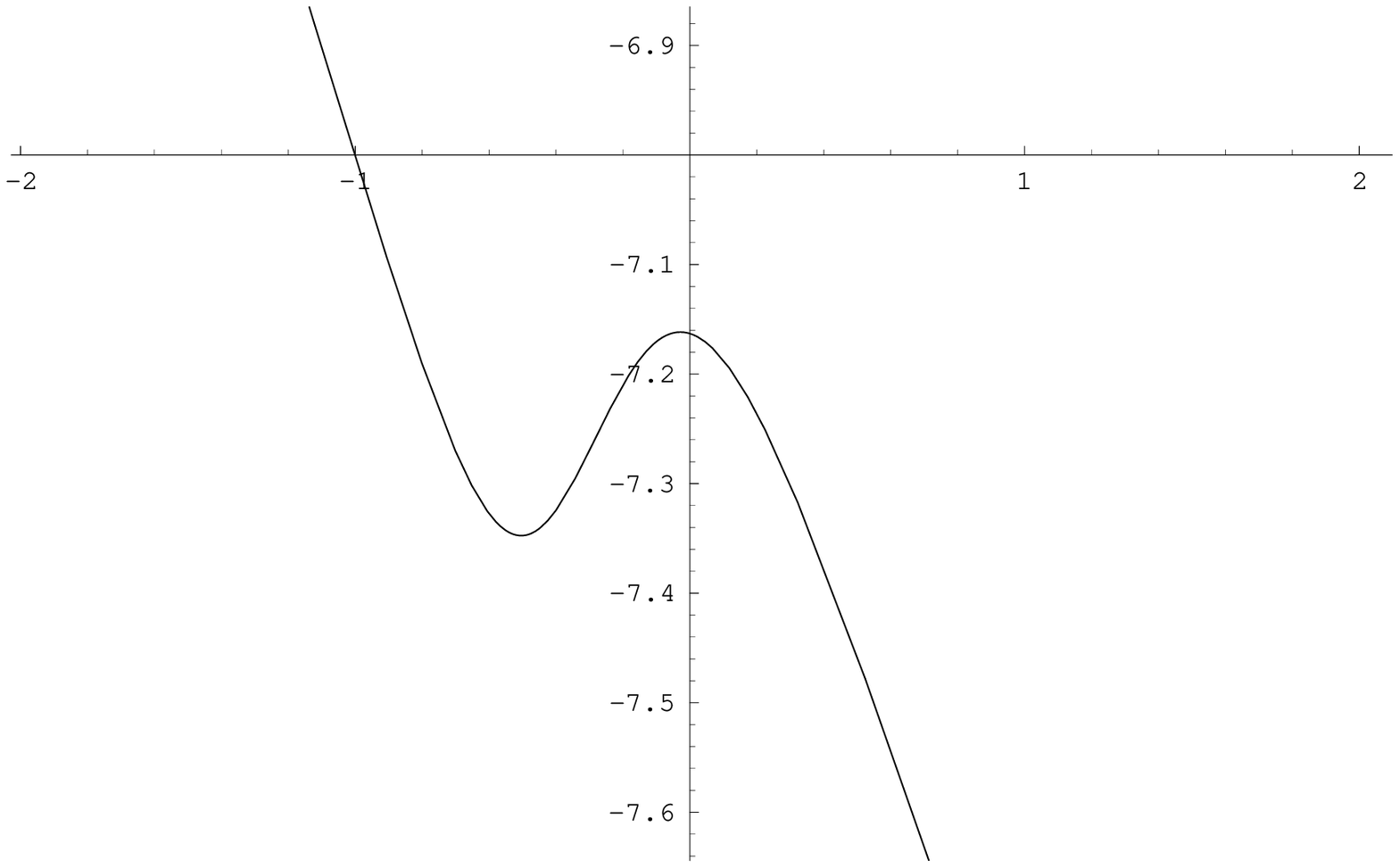}
 \includegraphics[height=50mm]{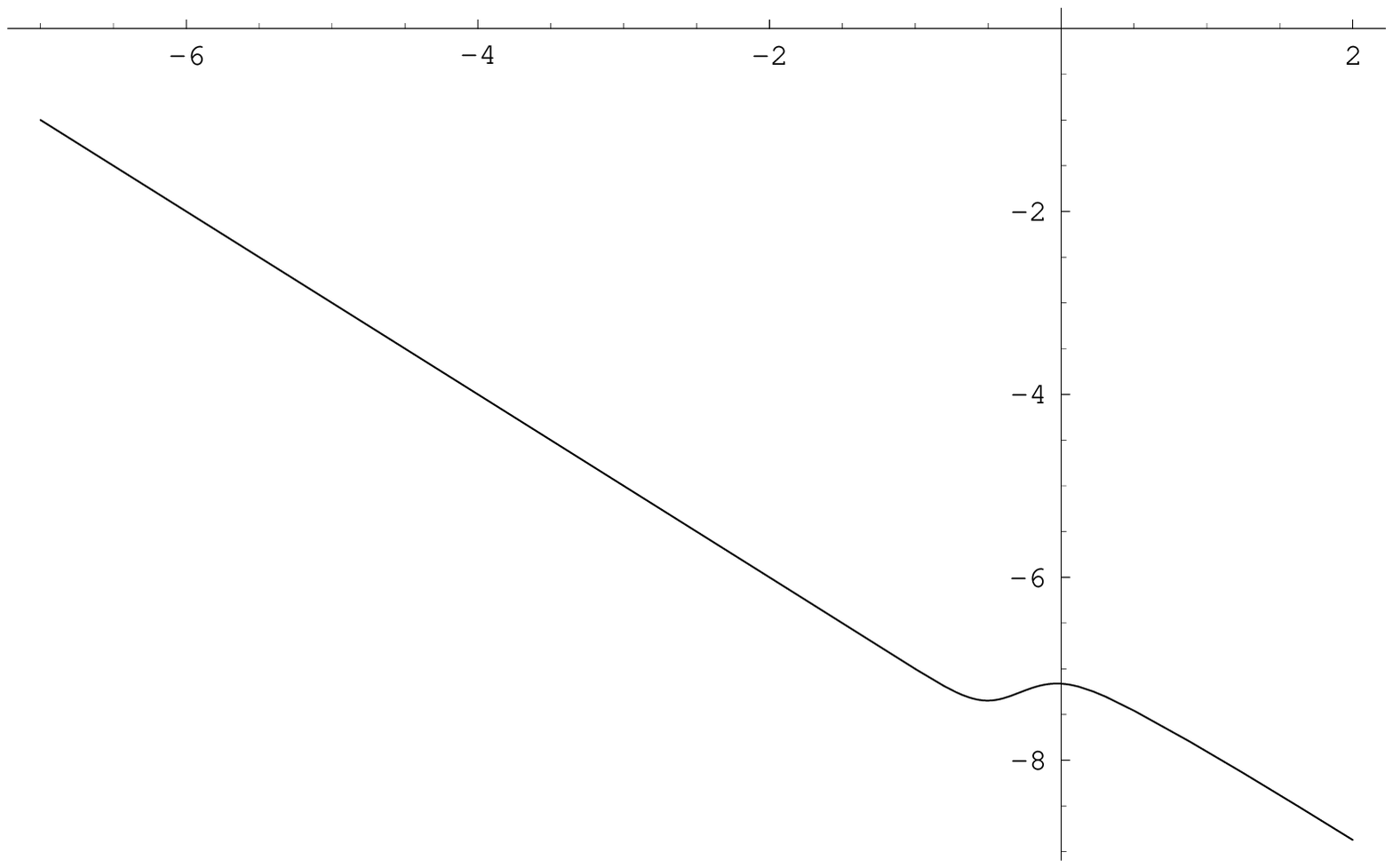}
\else
\end{center}
 \fi
\caption{\it Plot of the ${\alpha}_{1} \cdot h$ projection
 for the $\mathrm{SO(2,4)}$   flow
generated by the parameter choice $\left\{ \theta_1 , \, \theta_2 , \, \theta_3 \, , \theta_4 , \,
  \theta_5 , \, \theta_6 \right\}  \, = \,\left\{ \frac{\pi}{3},\,
  0,\, \frac{\pi}{6},\, \frac{\pi}{3},\,0\, , \frac{\pi}{3}\,
  \right\}$. This is actually a flow in the
  Tits Satake submanifold and corresponds
  to a super-critical surface. This super-critical flow connects  the primordial Kasner era $\Lambda_1$
  to the remote future Kasner era $\Lambda_8$. The plot on the left and on the right are the same.
The only difference is that on the right we have an enlargement of the time region around
$t=0$, while
on the left we consider a time range covering a much wider portion of the
early epochs.}
\label{a1so24np}
 \iffigs
 \hskip 1.5cm \unitlength=1.1mm
 \end{center}
  \fi
\end{figure}

\begin{figure}[!hbt]
\begin{center}
\iffigs
 \includegraphics[height=60mm]{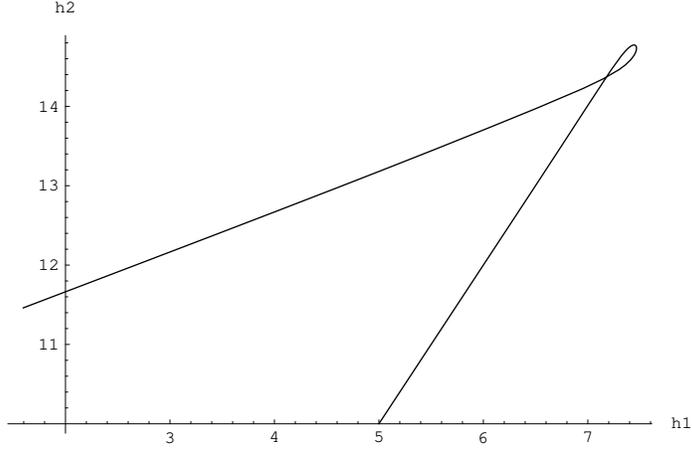}
\else
\end{center}
 \fi
\caption{\it Trajectory of the cosmic ball in the
  $\mathrm{SO(2,4)}$ flow
generated by the parameter choice $\left\{ \theta_1 , \, \theta_2 , \, \theta_3 \, , \theta_4 , \,
  \theta_5 , \, \theta_6 \right\}  \, = \,\left\{ \frac{\pi}{3},\,
  0,\, \frac{\pi}{6},\, \frac{\pi}{3},\,0\, , \frac{\pi}{3}\,
  \right\}$. This flow  is inside the
  Tits Satake submanifold and corresponds
  to a super-critical surface. It connects the primordial Kasner era
  $\Lambda_1$ to the remote future Kasner era $\Lambda_8$. }
\label{h1h2so24n}
 \iffigs
 \hskip 1.5cm \unitlength=1.1mm
 \end{center}
  \fi
\end{figure}

We come now to the analysis of two explicit examples of flows aiming
at illustrating three aspects:
\begin{description}
  \item[a] The embedding of the Tits Satake flows within the
  flows of the bigger coset manifold.
  \item[b] The role of the extra parameters not contained in the Tits
  Satake projection.
  \item[c] The instability of super-critical and in general of trapped surfaces.
\end{description}
To this effect we shall reconsider the case analyzed in section
\ref{critsigma2} of a flow on the critical surface $\Sigma_2$ for the
Tits Satake projection of $\mathrm{SO(2,4)}$, namely $\mathrm{Sp(4,\mathbb{R})} \sim
\mathrm{SO(2,3)}$.
\paragraph{The unperturbed super-critical flow}
The choice of the Euler angles is that of eq.(\ref{Hspec2}) which, in
the five dimensional representation $\mathrm{SO(2,3)}$ produces the matrix of eq.(\ref{so23repre}). This
latter has three vanishing minors, as
shown in eq.(\ref{minorini}). It is quite easy to embed this case and
the corresponding flow into the non maximally split representative
$\mathrm{SO(2,4)}$ of the same universality class. It suffices to
choose the six $\theta_i$ angles as follows:
\begin{equation}
  \left\{ \theta_1 , \, \theta_2 , \, \theta_3 \, , \theta_4 , \,
  \theta_5 , \, \theta_6 \right\}  \, = \,\left\{ \frac{\pi}{3},\,
  0,\, \frac{\pi}{6},\, \frac{\pi}{3},\,0\, , \frac{\pi}{3}\,
  \right\}
\label{tettangoli}
\end{equation}
since, according to eq.(\ref{so24rutte}) and (\ref{HHgeneriso2}), the
angles $\theta_2$ and $\theta_5$ correspond to the second copy of the
compact generators respectively associated with the first and the third of the
$\sym(4)$ roots. The result of this choice is the following matrix in $\mathrm{SO(2)} \times
\mathrm{SO(4)} \subset \mathrm{SO(2,4)}$:
\begin{equation}
  \mathcal{O}_{unp}\, = \, \left(
\begin{array}{llllll}
 \frac{\sqrt{3}}{16} & \frac{5}{16} & -\frac{3}{4
   \sqrt{2}} & 0 & -\frac{3}{16} & -\frac{7
   \sqrt{3}}{16} \\
 -\frac{19}{32} & \frac{11 \sqrt{3}}{32} & \frac{3
   \sqrt{\frac{3}{2}}}{8} & 0 & -\frac{5
   \sqrt{3}}{32} & -\frac{3}{32} \\
 \frac{3 \sqrt{\frac{3}{2}}}{16} & \frac{15}{16
   \sqrt{2}} & -\frac{1}{8} & 0 & \frac{15}{16
   \sqrt{2}} & \frac{3 \sqrt{\frac{3}{2}}}{16} \\
 0 & 0 & 0 & 1 & 0 & 0 \\
 -\frac{3}{32} & -\frac{5 \sqrt{3}}{32} & \frac{3
   \sqrt{\frac{3}{2}}}{8} & 0 & \frac{11
   \sqrt{3}}{32} & -\frac{19}{32} \\
 -\frac{7 \sqrt{3}}{16} & -\frac{3}{16} & -\frac{3}{4
   \sqrt{2}} & 0 & \frac{5}{16} & \frac{\sqrt{3}}{16}
\end{array}
\right)~.
\label{lenivai}
\end{equation}
As one sees, by deleting the 4th row and the 4th column one retrieves
the matrix of eq.(\ref{so23repre}). Indeed the matrix (\ref{lenivai})
is manifestly inside the Tits Satake subgroup $\mathrm{SO(2,3)} \subset
\mathrm{SO(2,4)}$. If we use $\mathcal{O}_{unp}$ as initial data for
our integration algorithm implemented on a computer we find that the
asymptotic limits are $\Lambda_1$ at past infinity and $\Lambda_8$ at
future infinity just as in the original case discussed in section
\ref{critsigma2}.
Consider fig.\ref{a1so24np}. It displays
the plot of the Cartan fields projected along the root $\alpha_1$
and clearly demonstrates that there are just two bounces on the wall
orthogonal to this root. Both of them occur in a narrow time range
around $t=0$. At very early and very late times there are no more
bounces and the result is the trajectory of the cosmic ball displayed
in fig.\ref{h1h2so24n}. The two asymptotic lines (incoming and
outgoing) are the Kasner epochs $\Lambda_1$ and $\Lambda_8$,
respectively.
\paragraph{Perturbing the super-critical flow with painted walls}
In order to illustrate both the nature of the Tits Satake projection
and the instability of super-critical surfaces we consider now a small
perturbation of the initial data used in the previous example. Keeping all
the other angles unchanged we shift from zero the angle $\theta_2$ by
a very small amount. Explicitly we choose
\begin{equation}
  \theta _1\, = \,  \frac{\pi }{3}\, , \, \theta _2\, = \,  \arcsin\left(\frac{1}{100}\right)\, , \, \theta _3\, = \,
   \frac{\pi }{6}\, , \, \theta _4\, = \,  \frac{\pi }{3}\, , \, \theta
   _5\, = \,  0\, , \, \theta _6\, = \,  \frac{\pi }{3}~.
\label{perturbatus}
\end{equation}
As explained $\theta_2$ is associated with the second copy of the
root $\alpha_1$. Hence introducing this small angle is equivalent to
creating a new $\alpha_1$ wall just painted with a different color.
This new wall is very very small and therefore it will produce very
little effects at finite times. Yet it is sufficient to remove us
from the super-critical surface and this necessarily changes asymptotics.
Instead of $\Lambda_1$ we expect now $\Lambda_5$ at past infinity. It
is interesting to analyze in detail how this happens.
\par
\begin{figure}[!hbt]
\begin{center}
\iffigs
 \includegraphics[height=45mm]{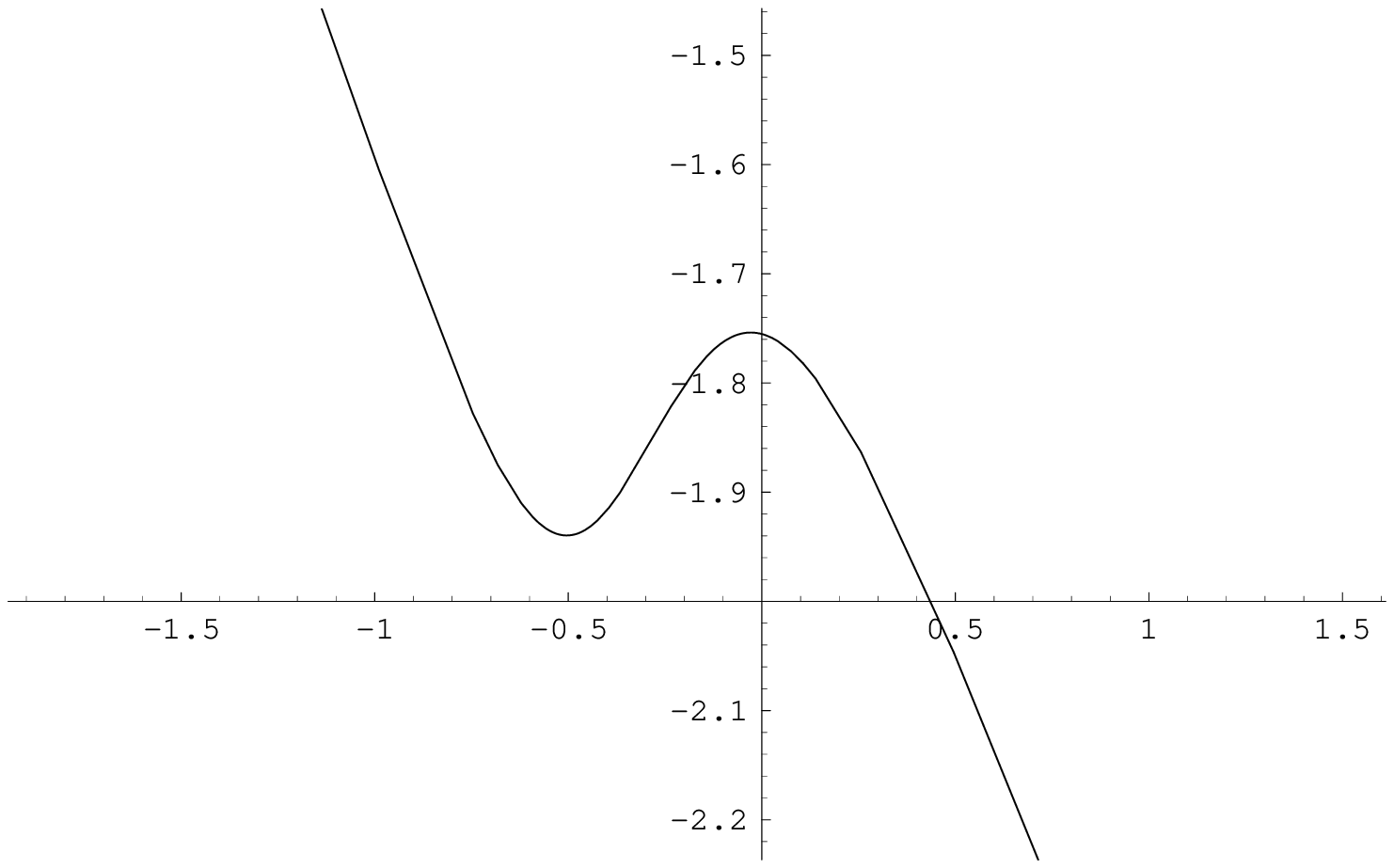}
 \includegraphics[height=50mm]{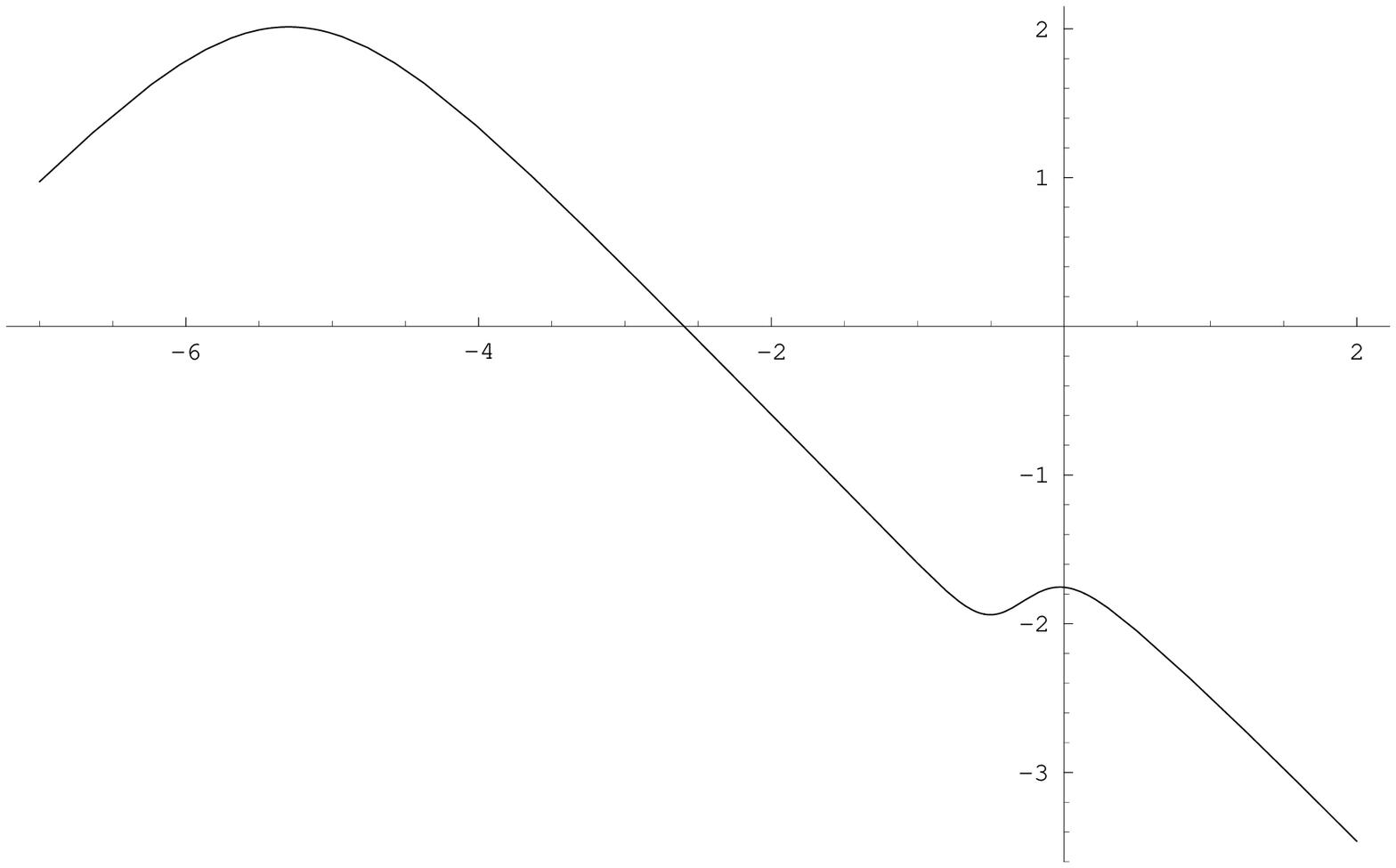}
\else
\end{center}
 \fi
\caption{\it Plot of the ${\alpha}_{1} \cdot h$ projection
 for the $\mathrm{SO(2,4)}$   flow
generated by the parameter choice
$\left\{ \theta_1 , \, \theta_2 , \, \theta_3 \, ,
\theta_4 , \,
  \theta_5 , \, \theta_6 \right\}  \, = \,\left\{ \frac{\pi}{3},\,
  \arcsin\left(\ft{1}{100} \right),\, \frac{\pi}{6},\, \frac{\pi}{3},\,0\, ,
  \frac{\pi}{3}\,
  \right\}$. This flow is a perturbation of a super-critical flow. The
  shift from $\Lambda_1$ to $\Lambda_5$ at past infinity occurs via
  an extra bump on the $\alpha_1$ wall which occurs at very early
  times. This bump is not visible in the plot on the
  right which is in the range near $t=0$ but it is evident
  in the plot on the left which goes further back in time.
  This picture is to be compared with fig.\ref{a1so24np}.}
\label{a1so24perturb}
 \iffigs
 \hskip 1.5cm \unitlength=1.1mm
 \end{center}
  \fi
\end{figure}
\begin{figure}[!hbt]
\begin{center}
\iffigs
 \includegraphics[height=60mm]{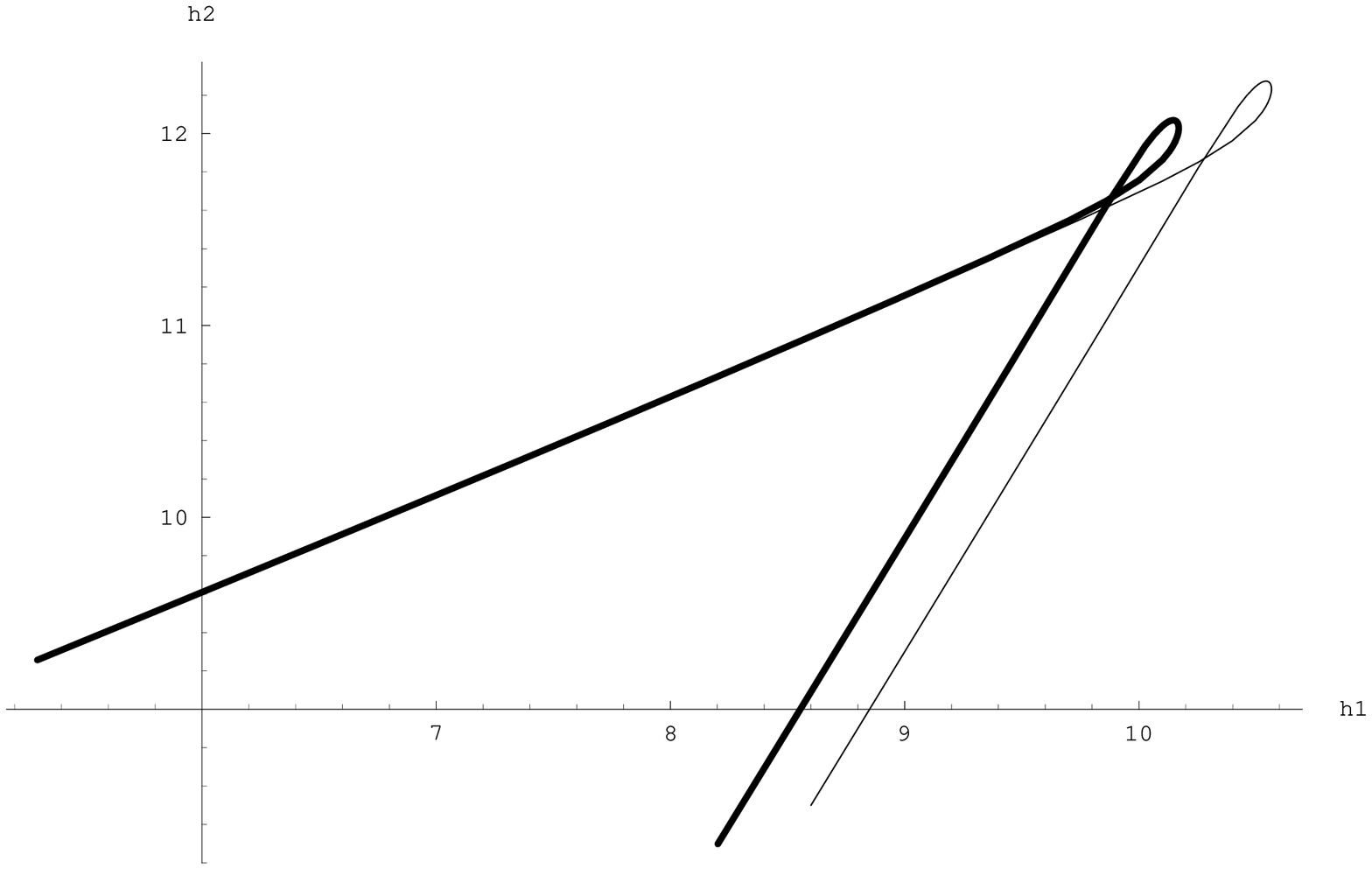}
 \includegraphics[height=60mm]{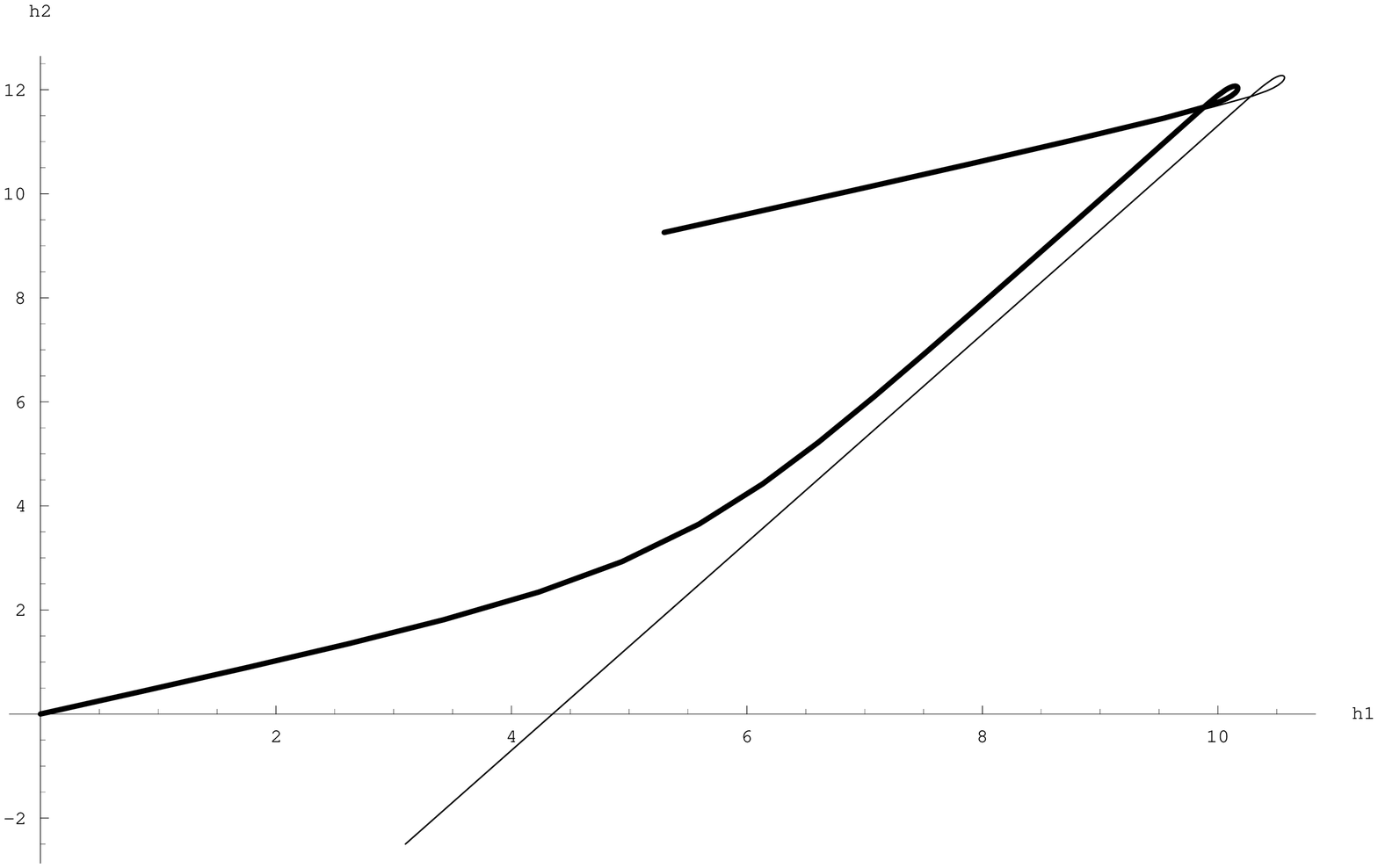}
\else
\end{center}
 \fi
\caption{\it Comparison of the trajectories of the cosmic ball in the
  $\mathrm{SO(2,4)}$   flow
generated by the parameter choice $\left\{ \theta_1 , \, \theta_2 , \, \theta_3 \, ,
\theta_4 , \,
  \theta_5 , \, \theta_6 \right\}  \, = \,\left\{ \frac{\pi}{3},\,
  0,\, \frac{\pi}{6},\, \frac{\pi}{3},\,0\, , \frac{\pi}{3}\,
  \right\}$ and in its perturbation by a small
  $\theta_2 = \arcsin \left ( \ft{1}{100} \right) $. The thin line is the
  unperturbed flow, the fatter line is the perturbed one. The first
  plot covers a time range around $t=0$ while the second plot extends
  much earlier in time. The additional bounce responsible for the
  changing of asymptotic is visible in the second plot. }
\label{h1h2so24pertu}
 \iffigs
 \hskip 1.5cm \unitlength=1.1mm
 \end{center}
  \fi
\end{figure}

If we name $\mathcal{O}_{pert}$ the matrix corresponding to the
choice of angles (\ref{perturbatus}) we can appreciate the
perturbation of initial data by writing $\mathcal{O}_{pert}$ in the following way:
\begin{eqnarray}
\mathcal{O}_{pert} & = & \mathcal{O}_{unp} \, + \, \epsilon_1 \,
\left(
\begin{array}{llllll}
 0 & 0 & 0 & 0 & 0 & 0 \\
 0 & 0 & 0 & -6 & 0 & 0 \\
 0 & 0 & 0 & 6 \sqrt{6} & 0 & 0 \\
 -\frac{15}{2} & -\frac{9 \sqrt{3}}{2} & 3 \sqrt{6} &
   0 & -\frac{9 \sqrt{3}}{2} & -\frac{15}{2} \\
 0 & 0 & 0 & -6 & 0 & 0 \\
 0 & 0 & 0 & 0 & 0 & 0
\end{array}
\right) \,\nonumber\\
&& + \, \epsilon_2 \, \left(
\begin{array}{llllll}
 0 & 0 & 0 & 0 & 0 & 0 \\
 -\frac{5}{16} & -\frac{3 \sqrt{3}}{16} &
   \frac{\sqrt{\frac{3}{2}}}{4} & 0 & -\frac{3
   \sqrt{3}}{16} & -\frac{5}{16} \\
 \frac{5 \sqrt{\frac{3}{2}}}{8} & \frac{9}{8
   \sqrt{2}} & -\frac{3}{4} & 0 & \frac{9}{8
   \sqrt{2}} & \frac{5 \sqrt{\frac{3}{2}}}{8} \\
 0 & 0 & 0 & 2 & 0 & 0 \\
 -\frac{5}{16} & -\frac{3 \sqrt{3}}{16} &
   \frac{\sqrt{\frac{3}{2}}}{4} & 0 & -\frac{3
   \sqrt{3}}{16} & -\frac{5}{16} \\
 0 & 0 & 0 & 0 & 0 & 0
\end{array}
\right)~,  \nonumber\\
\epsilon_1 & \simeq & 1.2 \, \times \, 10^{-3}~,\nonumber\\
\epsilon_2 & \simeq & 1.0 \, \times \, 10^{-4}~.
\label{perturbo0}
\end{eqnarray}
Then we can implement the integration algorithm on our computer and
calculate the asymptotic values of the Lax operator. Notwithstanding
the smallness of the perturbation, the past infinity regime jumps from
$\Lambda_1$ to $\Lambda_5$ as expected, while at future infinity it
remains $\Lambda_8$ which is already the highest possible Weyl
element. We can appreciate the mechanism which realizes this effect
by looking at fig.s \ref{a1so24perturb} and \ref{h1h2so24pertu}.
There is an extra bounce on the $\alpha_1$ wall as we expected which
corrects the trajectory and directs the cosmic ball to $\Lambda_5$
rather than $\Lambda_1$ when we go back in time. Since the
perturbation is small this bounce occurs at very early times so that
for most of the time the flow is almost on the critical surface. The
smaller the perturbation, the earlier the occurrence of the primeval
bounce. It should also be noted that we would have obtained exactly
the same effect if we had perturbed the $\theta_1$ angle instead of
the $\theta_2$. Indeed they are associated with the same root.
This is the meaning of the Tits Satake projection which captures all
the essential features of the dynamical processes for the entire
universality class.
\part{Perspectives}
\section{Summary of results}
In this paper we have made a few steps forward in developing the general
programme of supergravity billiards as a paradigm for superstring
cosmology. Our results are both of physical and mathematical nature.
\par
On the physical side, which for us means supergravity/superstring
theory, the essential points are the following ones:
\begin{description}
  \item[1)] We have shown that all supergravity billiards are
  completely integrable, irrespectively whether they are defined on a
  maximally split coset manifold $\mathrm{U/H}$ as it happens in the
  case of maximal susy or a non maximally split $\mathrm{U/H}$, as it happens in all lower
  supersymmetry cases. We have provided the explicit integration
  algorithm which just depends on the triangular embedding of the
  solvable Lie algebra $Solv(\mathrm{U/H})$ into that of
  $Solv(\mathrm{SL(N)/SO(N)})$.
  \item[2)] We have discovered a new principle of time orientation of
  the cosmic flow which relies on the natural ordering of the Weyl
  group elements (or of the permutations) according to their length
  $\ell_T$ in terms of transpositions. Cosmic evolution is always in
  the direction of increasing $\ell_T$ which plays the role of an
  entropy. There is a fascinating similarity, in this context between
  the laws of cosmic evolution and those of black hole
  thermodynamics.
  \item[3)] We have clarified the meaning of Tits Satake universality
  classes, introduced in \cite{contoine}, at least from the vantage
  point of cosmic billiards.  The asymptotic states, the type of
  available flows and the critical surfaces in parameter space are properties of the class and do
  not depend on the representative manifold in the class.
\end{description}
On the mathematical side the  highlights of our paper are the
following ones:
\begin{description}
  \item[1)] We have introduced the notion of generalized Weyl group
  for a non compact symmetric space $\mathrm{U/H}$ and shown that the
  factor group
  with respect to its normal subgroup is just the Weyl group of
  the Tits Satake subalgebra $\mathbb{U}_{TS} \, \subset \, \mathbb{U}.$
Moreover, we have demonstrated that not only
the factor group is isomorphic to the Weyl group of the Tits Satake projection
but even the generalized Weyl group is also isomorphic
$\mathcal{W}\left(\mathbb{U}\right) \, \sim \,
\mathcal{W}\left(\mathbb{U}_{\mathrm{TS}}\right)$. At least this is
true in the considered examples and we make the conjecture that it is
true in general.
  \item[2)] We have established a remarkable conjecture encoded in
  property \ref{kkminorprinc} of the main text: the constraints
  on minors of the diagonalizing orthogonal matrix for the Lax
  operator commute with the Toda flow.
  \item[3)] We have proposed a very simple efficient method of calculating
  the Toda flow asymptotics at $t=\pm \infty$ for the Lax operator of
  a $\sigma$-model  with target space any non compact-symmetric coset space ${\mathrm{U}}/{\mathrm{H}}$.
  Our algorithm requires only the knowledge of the corresponding Weyl group
  $\mathrm{Weyl}(\mathbb{U})$ as well as that of the small group $\mathrm{H}$.
  \item[4)] We have posed the question how the equations cutting out
  algebraic loci in compact group or coset manifolds and defined in
  terms of vanishing minors in the defining representation can be
  lifted to the abstract group level and extended to all irreducible
  representations.
\end{description}
\section{Open problems and directions to be pursued}
The results we have obtained are just steps ahead in a programme to
be further developed. They have solved some standing problems and
opened new directions of investigations which seem to us quite exciting. We just mention, as
conclusion, the milestones we would like to attain in the near future, evaluated from
the view-point which was generated by our present results:
\begin{description}
  \item[1)] Construction of all the triangular embeddings for all the
  solvable Lie algebras of all supergravity models and corresponding
  construction of the integration algorithm for all special
  homogeneous geometries.
  \item[2)] Oxidation and physical interpretation of the Toda flows
  we have just shown how to construct within supergravity models as those
  coming from string compactifications on manifolds of restricted holonomy.
  \item[3)] Extension of the integration algorithm to  sigma
  models with a potential emerging from flux compactifications and
  gauged supergravities in higher dimensions.
  \item[4)] Study of the integration algorithm in affine and
  hyperbolic Ka\v c-Moody extensions of the symmetric space
  $\mathrm{U/H}$, as they emerge from stepping down to $D=2$ and
  $D=1$ dimensions.
  \item[5)] Comparison between the law of increasing $\ell_T$ and the
  second principle of black hole mechanics in search of an adequate
  formulation of  cosmological thermodynamics and of a possible
  mapping between the attractor mechanism in black hole physics and
  Toda flows in cosmic billiards.
  \item[6)] More in depth study of the topology of parameter space
  $\mathrm{H}/\mathcal{W}(\mathbb{U})$ and in particular its
  partition in complex hulls that admit the trapped surfaces as walls.
  In this context an exciting open question is whether these hulls
  are completely closed or whether there is the possibility of going
  from one to the other avoiding the trapped walls. Clearly if the
  answer to this question is no then we have the notion of parallel
  disconnected universes.
\end{description}
{\bf Acknowledgements}
The authors acknowledge inspiring discussions and conversations with
their good friend and frequent collaborator Mario Trigiante.

\end{document}